\documentclass[fleqn,usenatbib]{mnras}

\usepackage{newtxtext,newtxmath}
\usepackage{mathtools, cuted}

\usepackage[T1]{fontenc}

\DeclareRobustCommand{\VAN}[3]{#2}
\let\VANthebibliography\thebibliography
\def\thebibliography{\DeclareRobustCommand{\VAN}[3]{##3}\VANthebibliography}

\usepackage{graphicx}	
\usepackage{amsmath}	
\usepackage{booktabs}

\usepackage{xcolor}
\usepackage[normalem]{ulem}
\definecolor{ochre}{rgb}{0.8, 0.47, 0.13}

\title[Evolution of CHE triples]{Stellar triples with chemically homogeneously evolving inner binaries}

\author[A. Dorozsmai et al.]{
Andris Dorozsmai$^{1}$\thanks{E-mail: andris@star.sr.bham.ac.uk},
Silvia Toonen$^{2}$,
Alejandro Vigna-G\'omez$^{3}$,
Selma E. de Mink$^{3,2}$ and
Floris Kummer$^{2}$
\\
$^{1}$Institute of Gravitational Wave Astronomy and School of Physics and Astronomy, University of Birmingham, Edgbaston, Birmingham B15 2TT, United Kingdom\\
$^{2}$ Astronomical Institute Anton Pannekoek, University of Amsterdam, Science Park 904, 1098 XH Amsterdam, The Netherlands\\
$^{3}$Max-Planck-Institut f\"ur Astrophysik, Karl-Schwarzschild-Str. 1, D-85748 Garching, Germany
}

\date{Accepted XXX. Received YYY; in original form ZZZ}

\begin{document}
\label{firstpage}
\pagerange{\pageref{firstpage}--\pageref{lastpage}}
\maketitle

\begin{abstract}
Observations suggest that massive stellar triples are common. However, their evolution is not yet fully understood.
We investigate the evolution of hierarchical triples in which the stars of the inner binary experience chemically homogeneous evolution (CHE), particularly to understand the role of the tertiary star in the  formation of gravitational-wave (GW) sources. We use the triple-star rapid population synthesis code \texttt{TRES} to determine the evolution of these systems at two representative metallicities: $Z = 0.005$ and $Z = 0.0005$. 
About half of all triples harbouring a CHE inner binary (CHE triples) experience tertiary mass transfer (TMT) episodes, an event which is  rare for classically evolving stars. 
In the majority of TMT episodes, the inner binary consists of two main-sequence stars (58-60 per cent) or two black holes (BHs, 24-31 per cent).  Additionally, we explore the role of von Zeipel-Lidov-Kozai (ZLK) oscillations for CHE triples.
 ZLK oscillations  can result in eccentric stellar mergers or lead to the formation of eccentric compact binaries in systems with initial outer pericenters smaller than $\sim$ 1200 $R_{\odot}$.  
Approximately $24$-$30$ per cent of CHE triples form GW sources,
and in 31 per cent of these, the tertiary star plays a significant role and leads to configurations that are not predicted for isolated binaries.
We conclude that the evolution of CHE binaries can be affected by a close tertiary companion, resulting in astronomical transients such as  BH-BH binaries that merge via GW emission orders of magnitude faster than their isolated binary counterparts and tertiary-driven 
massive stellar mergers. 
\end{abstract}




\section{Introduction}
An accurate and detailed understanding of the evolution of massive stars is essential for various important open questions in astrophysics, such as nucleosynthesis of heavy elements, the origin of supernova events, gamma-ray bursts, and GW sources (e.g. \citealt{Langer_2012}).
Observational evidence shows that the fraction of stars in hierarchical triples
or in higher-order multiple-stellar systems increases  with the mass of the primary star (\citealt{EvansRemage2011, Sana2014}). 
In particular, \citet{MoeDiStefano2017} showed that the majority of O-type stars reside either in triple or quadruple stellar systems.
This implies that in order to understand the evolution of massive stars, and to correctly interpret
the various astrophysical phenomena related to them, we need to consider stellar interactions in hierarchical triples.

The evolution of hierarchical triples involves a complex interplay between three-body dynamics, stellar evolution, and stellar interactions \citep[e.g.][]{Toonen2016}. 
Three-body interactions can result in e.g. ZLK oscillations \citep{1910AN....183..345V,Lidov1962, Kozai1962, Naoz2016}, a secular effect, where the eccentricity of the inner binary can be significantly enhanced as a result of dynamics.
ZLK oscillations coupled with various dissipative processes (e.g. tides, GWs) can shrink the orbit \citep[e.g.][]{ Mazeh+Shaham1979, FabryckyTremaine2007,Thompson2011} and prompt the merger of the inner binary \citep[e.g.][]{PeretsFabrycky2009, VignaGomez2022}.
These type of mergers can result in astronomical transient events such as Type Ia supernova \citep[e.g.][]{KatzDong2012,hamers2013,Toonen2018,Rajamuthukumar2022}
or double compact object mergers \citep[e.g.][]{Antognini2014,Antonini2017,RodriguezAntonini2018,HamersThompson2019,Fragione2019,Stegmann2022}.
Furthermore, stellar evolution can affect the orbital dynamics of the triple.
For example, radial expansion leading to a mass transfer event and mass loss can prompt ZLK oscillations or dynamical instabilities \citep{PeretsKratter2012, ShappeeThompson2013,MichaelyPerets2014, Toonen2022, HamersTEDI2022}.

Population synthesis studies of stellar triples show that the inner binaries in hierarchical triples  have increased stellar interactions compared to isolated binaries \citep[e.g.][]{Toonen2020, Stegman2022, Hamers2022}.
Similarly, tertiary-driven dynamics could play an essential role in double compact object mergers.
While GW sources detected by the LIGO/Virgo collaboration \citep[LVC, e.g.][]{Abbott_2019, Abbott2019_GWTC-1, Abbott2021_GWTC2, Abbott2021GWTC-3} have been studied in the context of stellar triples, this has been done so far only in a limited parameter space.
For example, for systems in which the inner binary is wide enough such that interaction between the two stars in the form of mass exchange can be neglected
\citep[e.g.][]{SilsbeeTremaine2017, Antonini2017, RodriguezAntonini2018,Fragione2019, VignaGomez2021, Martinez2022},
or in which the stars of the inner binary merge during the main sequence \citep{Stegmann2022}.
There are still major uncertainties and a need to explore and to understand the population of merging binary BHs from hierarchical triples.

In this paper, we focus on the evolution of hierarchical triples in which the stars of the inner binaries are chemically homogeneously evolving. CHE stars have been discussed in the context of rapidly-rotating stars \citep{Maeder1987, Yoon+2005, Yoon+2006, Brott2011, Kohler2015, Szecsi2015}, which can experience enhanced mixing during the MS stage. This mixing allows hydrogen-rich matter in the radiative envelope to be deposited into the convective core, where it is fused to helium. At the same time, helium is mixed throughout the star. This prevents the build-up of a chemical gradient inside the star and the classical core-envelope structure. As a result, the stars remain very compact over their lifetime. CHE has been proposed to occur in very close binaries where the tidal deformation of both stars is strong and they are forced to rotate rapidly \citep{deMink2009, Song2016}.
More recently, CHE binaries received renewed interest as they have been proposed as a new pathway to form BH binaries that can merge within the age of the universe \citep{de_Mink_2016, Mandel2016, Marchant2016, duBuisson2020,Hastings2020, Riley2021MNRAS.505..663R}. Recently, \citet{VignaGomez2021} studied triples with CHE inner binaries in the context of sequential merging BH-BHs with masses that fall in the pair-instability mass gap. Specifically, they considered sequential mergers of hierarchical co-planar triples, a simplified approach which neglected three-body dynamics. In this paper, we remove the constraints of co-planarity and explore, for the first time, the evolution of massive stellar triples with CHE inner binaries in the entire parameter space. As isolated CHE binaries are known to be promising GW progenitors, we will mostly focus on the role of the tertiary star in the evolution of the inner binary in the context of GW astronomy.

This paper is structured as follows.
In section \ref{sec:methodology}, we introduce \texttt{TRES}, the triple evolutionary code we use in this study, 
and the adaptations we have made to model CHE and contact binaries. In section \ref{sec:Results}, we discuss the results of our population synthesis in \texttt{TRES} and identify the most important evolutionary channels.
In section \ref{subsec:origin}, we show that the initial parameters of the tertiary star are sufficient to predict 
the evolutionary channel of each system. 
Finally, in section \ref{sec:GW}, we use analytical and numerical methods to explore our synthetic population of stellar triples in the context of GW sources.

\section{Methodology}
\label{sec:methodology}
We use \texttt{TRES} to simulate the evolution of our hierarchical triples \citep[see][for a detailed description of the code]{Toonen2016}. \texttt{TRES} couples secular dynamics of stellar triples with stellar evolution, and takes into account additional physical processes such as stellar interactions and dissipative processes.

\texttt{TRES} determines the evolution of each star by using the fitting formulae of \citet{Hurley2000} to the stellar tracks of \citet{Pols1998}, as implemented in the the rapid binary synthesis code \texttt{SeBa} \citep{Portegies_Zwart1996, Toonen_2012}, while interactions between the stars are determined by \texttt{TRES}. \texttt{TRES} treats three-body dynamics in the following way. For secular evolution, we include secular three body dynamics (subscript `3b') including quadrupole (\citealt{Harrington1968}) and octupole terms (\citealt{Ford_2004} with corrections of \citealt{Naoz2013}). Regarding the additional physical processes, we take into account: i) general relativistic effects (GR) and GW emission \citep[subscript `GR'][]{Peters64, Blaes2002}, ii) tidal friction \citep[subscript `TF'][]{Hurley2002}, iii)
the effects of stellar winds under the assumptions of fast, adiabatic wind \citep[see e.g.][]{Veras2011, DebesSigurdsson2002} with mass loss rates provided by \texttt{SeBa} (subscript `wind'), iv) precession due to ZLK, GR, tides \citep[subscript `tides'][]{SmeyersWillems2001} 
and intrinsic stellar rotation \citep[subscript `rotate'][]{FabryckyTremaine2007}, and v) the change in the stellar rotation due to stellar evolution based on spin angular momentum conservation (subscript `I'). This gives rise to a set of first-order ordinary differential equations, that are solved numerically.
These equations are:

\begin{eqnarray}
\left \{
\begin{array}{l c l}
 \dot{a}_{\rm in} &=& \dot{a}_{\rm in, GR} +\dot{a}_{\rm in, TF} +\dot{a}_{\rm in, wind} \\
 \dot{a}_{\rm out} &=& \dot{a}_{\rm out, GR} +\dot{a}_{\rm out, TF} +\dot{a}_{\rm out, wind} \\
 \dot{e}_{\rm in} &=& \dot{e}_{\rm in,3b} + \dot{e}_{\rm in,GR} +\dot{e}_{\rm in,TF} \\
 \dot{e}_{\rm out} &=& \dot{e}_{\rm out,3b} +\dot{e}_{\rm out,GR} + \dot{e}_{\rm out,TF} \\
 \dot{g}_{\rm in} &=& \dot{g}_{\rm in,3b} +  \dot{g}_{\rm in,GR} + \dot{g}_{\rm in,tides} + \dot{g}_{\rm in,rotate}\\
 \dot{g}_{\rm out} &=& \dot{g}_{\rm out, 3b} + \dot{g}_{\rm out,GR} + \dot{g}_{\rm out,tides} +\\ &&\dot{g}_{\rm out,rotate}\\
 \dot{h}_{\rm in} &=& \dot{h}_{\rm in, 3b}\\
 \dot{\theta} &=& \frac{-1}{J_{\rm b, in}J_{\rm b, out}} [\dot{J}_{\rm b, in}(J_{\rm b, in}+J_{\rm b, out}\theta) + \\
&& \dot{J}_{\rm b, out}(J_{\rm b, out}+ J_{\rm b, in}\theta)]\\
 \dot{\Omega}_{1} &=& \dot{\Omega}_{\rm 1, TF}  +\dot{\Omega}_{\rm 1, I} +\dot{\Omega}_{\rm 1, wind} \\
 \dot{\Omega}_{2} &=& \dot{\Omega}_{\rm  2, TF}  +\dot{\Omega}_{\rm 2, I}  +\dot{\Omega}_{\rm 2, wind}\\
 \dot{\Omega}_{3} &=& \dot{\Omega}_{\rm  3, TF}  +\dot{\Omega}_{\rm 3, I}  +\dot{\Omega}_{\rm 3, wind}
 \end{array}\right. 
\label{eq:ODE}
\end{eqnarray}
where $a$, $e$, $g$, $h$ and $J_b$ represent the semimajor axis, eccentricity, argument of pericenter, line of ascending nodes, and the orbital angular momentum for the inner (subscript `in') and outer (subscript `out') orbit. The dot represents the time derivatives. Lastly $\theta\equiv\cos(i)$, where $i$ is the mutual inclination between the inner and outer orbit, and $\Omega_1, \Omega_2, \Omega_3$ the spin frequency of the primary, secondary and tertiary star respectively. Per definition the primary and secondary stars are the stars in the inner binary, with the primary star initially more massive than the secondary star, and the tertiary star orbits the inner binary.

We highlight three aspects of the orbital evolution of hierarchical triples that is particularly relevant for the systems we study in this paper. Firstly, if the apsidal precession of the inner binary due to short range forces, such as tides ($\dot{g}_{\rm in, tides}$) and GR effects ($\dot{g}_{\rm in, GR}$) occurs on a much shorter timescale than the precession due to three-body dynamics ($\dot{g}_{\rm in, 3b}$), ZLK oscillations will be quenched \citep[see e.g.][]{Holman1997,Eggleton2001, Blaes2002, FabryckyTremaine2007, Thompson2011, Dong2014, Liu2015, Petrovich2015, Kassandra2017}. 
The timescale of ZLK oscilations can be approximated as \citep[e.g.][]{Innanen1997, Holman1997, KinoshitaNakai1999}:
\begin{equation}
\label{eq:zlk_timescale}
    t_{\rm ZLK} = \left(\frac{M_1 + M_2}{G M_{\rm out}^2}\right)^{1/2} \left(\frac{a_{\rm out}}{a_{\rm in}^{1/2}}\right)^{3}(1-e_{\rm out}^2)^{3/2}.
\end{equation}
The timescale related to the apsidal precession due to tides is \citep[e.g.][]{SmeyersWillems2001, Liu2015}:
\begin{equation}
\label{eq:tides_timescale}
    t_{\rm tides} = \left(\frac{M_1}{15 k_{\rm am}\mu_{\rm in}^{1/2}M_2}\right) \left(\frac{a_{\rm in}^{11/2}}{R_1^{5}}\right) \left(\frac{(1 - e_{\rm in}^2)^5}{1 + \frac{3}{2} e_{\rm in}^2 + \frac{1}{8} e_{\rm in}^4}\right),
\end{equation}
where $k_{\rm am}$ the apsidal motion constant, which we assume to be 0.0144 for MS and helium stars, $\mu_{\rm in} = G(M_1+M_2)$, i.e. the standard gravitational parameter for the inner binary and $R_1$ is the radius of the inner star. 
The timescale related to precession due to general relativistic effects is \citep[e.g.][]{MisnerThorneWheeler_Gravitation1_973, Blaes2002, MillerHamiltion2002}:
\begin{equation}
\label{eq:gr_timescale}
   t_{\rm GR} = \frac{c^2}{3\mu_{\rm in}^{3/2}}a_{\rm in}^{5/2} (1 - e_{\rm in}^2).
\end{equation}
If $t_{\rm ZLK}\gg\rm{min}(t_{\rm GR},t_{\rm tides})$, then three-body dynamics are suppressed. If the timescales are comparable, then the maximum eccentricity induced by the ZLK oscillations is diminished. In principle, rotation-induced oblateness in the inner binary also induces apsidal precession \citep[$g_{\rm in, rot}$, see e.g.][]{FabryckyTremaine2007}. However, as long as the rotational period of the inner stars is not shorter than the orbital period (which is true for all systems considered here), $\dot{g}_{\rm tides} \gg \dot{g}_{\rm rot}$ and therefore precession due to stellar rotation does not play a role in suppressing three-body dynamics \citep[][]{Liu2015}.

Secondly, the octupole term in the perturbing function of the Hamiltonian \citep[e.g.][]{Naoz2013} is typically negligible for CHE triples,
as the vast majority of the inner binaries are in contact, which leads to equal mass components in our models (see section \ref{subsec:contact}). 
Furthermore, for the relatively rare detached CHE inner binaries, the mass ratio is always $q_{\rm in} \geq 0.7$ (see section \ref{subsec:init}).

Finally, we estimate the time it takes for the inner binary to merge due to GWs following \citet{Peters64}, if the tertiary is dynamically decoupled from the inner binary. If ZLK oscillations are still relevant during the inspiral phase, we follow the approximation of \citet{MillerHamiltion2002}:
\begin{equation}
\label{eq:modified_peters}
    t_{\rm GW} \approx t_{\rm  GW, Peters}(a_{\rm in},e_{\rm in, max})(1 - e_{\rm in, max})^{-1/2}, 
\end{equation}
where $t_{\rm GW}$ is the time required for the merger, $t_{\rm GW, Peters}$ is the time to merger based on the relation of \citet{Peters64}, $e_{\rm in, max}$ is the maximum eccentricity reached during ZLK oscillations and $a_{\rm in}$ is the initial inner semimajor axis.  The approximation in equation \ref{eq:modified_peters} is  based on \citet{Wen2003} and it neglects the effects of precession due GR. When the latter is taken into account, \citet{Thompson2011} finds that equation \ref{eq:modified_peters} underestimates the actual merger timescale typically by a factor of 2-3.

\subsection{Modelling of chemically homogeneous evolution}

We follow \citet{Riley2021MNRAS.505..663R} in order to incorporate CHE stars in \verb|TRES|.
 That means that we assume a star evolves chemically homogeneously, if the angular frequency of the spin of the star is above a certain critical value, i.e. $\omega_{\rm star} > \omega_{\rm CHE, crit}$. 
\citet{Riley2021MNRAS.505..663R} provides a fit to this critical value based on \verb|MESA| \citep{Paxton_2010} models of \citet[][]{Marchant2016} at different masses and metallicities .
In order to determine whether a star evolves chemically homogeneously, we check whether our simulated star is spinning above $\omega_{\rm CHE, crit}$ at every timestep.
If a star meets this criteria, we determine its radius and luminosity according to the fits of \cite{Hurley2000} for ZAMS stars. We note that the mass of CHE stars are in general affected by stellar winds, therefore, its radius and luminosity do not remain constant during the core-hydrogen burning phase, even with this simplifying assumption \citep[see also][]{Riley2021MNRAS.505..663R}.
We assume that the star by the end of core hydrogen burning forms a helium star with a mass $M_{\rm He, ZAMS} = M_{\rm TAMS}$, where $M_{\rm He, ZAMS}$ is
the initial mass of the helium star and $M_{\rm TAMS}$ is the terminal age main sequence mass of the star. With these assumptions, CHE stars experience an instantaneous drop in radii at the end of their MS phase \citep[compare main sequence stellar evolution with helium star evolution in][]{Hurley2000}. This is a simplification of the results of detailed simulations of CHE stars, where the latter suggests a gradual contraction of the radius during the MS \citep[e.g.][]{Maeder1987}.
If a CHE star loses angular momentum (e.g. due to stellar winds), its rotational frequency decreases. If the frequency reduces to below the critical value, we assume the evolution of the star transitions back to the classical non-CHE case.

For simplicity, we only consider systems in which the stars of the inner binary are CHE from zero-age main sequence (ZAMS). 
Stars that do not evolve chemically homogeneously from ZAMS could, in theory, become CHE stars, if they attained a sufficiently high-spin frequency before a significant chemical gradient is built up in their interior. 
This can be achieved for example, if a star is spun up by accretion during a mass transfer event \citep[e.g.][]{Cantiello2007, Ghodla2022}. We neglect such systems in this study. 

\subsection{Contact binaries}
\label{subsec:contact}
We follow the implementation of \cite{Riley2021MNRAS.505..663R} of modelling contact binaries  for rapid population synthesis codes \citep[which is based on the detailed stellar models of][]{Marchant2016}. 
We assume that contact binaries, i.e. binaries in which both stars fill their Roche-lobes, can maintain co-rotation and consequently survive the contact phase without merging as long as neither of the stars fill the outer Lagrangian points (L2 and L3).
For contact binaries, \cite{Marchant2016} finds that mass is transferred
 between the two stars back and forth until they reach an equal mass ratio. 
If this mass equalisation indeed occurs in nature, the mass ratio distribution of massive contact binaries would exhibit a prominent peak near one, which is not in an agreement with the observations of massive contact binaries residing in the Magellanic Clouds and in the Milky Way
\citep[see ][]{Menon2021, Abdul-Masih2022:constrain_ocbins_period_stability}.
This discrepancy could be due to missing physics in the contact binary models of \cite{Marchant2016},
e.g. related to energy transfer between the two stars \citep[see e.g.][]{Abdul-Masih2021:constrain_ocbins_mixing, Fabry2023}.

We follow \cite{Marchant2016} and approximate the L2 point as
\begin{equation}
\label{eq:outerrlof}
    \frac{R_{\rm{L2,2}} - R_{\rm{RL,2}}}{R_{\rm{RL,2}}} = 0.299 \, \rm{tan}^{-1}(1.84q^{0.397}),
\end{equation}
where $R_{\rm{RL,2}}$ is the Roche-lobe radius of the secondary star, which we approximate following \cite{EggletonApprox}.

If the stars in the inner binary are in contact but without filling their L2 points, 
we assume that the masses of the binary equalise via a fully conservative mass transfer phase.
We follow \cite{{Riley2021MNRAS.505..663R}} and assume this mass equalisation occurs instantaneously and readjust the orbit of the inner binary as \citep[see, e.g.][]{Soberman1997}:
\begin{equation}
\label{eq:orbit_mass_equal}
    \frac{a_{\rm fin}}{a_{\rm init}} = \left( \frac{M_{\rm 1,init} M_{\rm 2,init}}{M_{\rm 1,fin} M_{\rm 2,fin}}\right)^2,
\end{equation}
where $a_{\rm init}$, $a_{\rm fin}$ are the initial and the final orbital separation and $M_{\rm 1,init}$, $M_{\rm 2,init}$ are the initial masses of the primary and the secondary, respectively. The final masses are $M_{\rm 1,fin}$ = $M_{\rm 2,fin}$ = $1/2\cdot(M_{\rm 1,init} + M_{\rm 2,init})$ by definition.
The assumption of mass equalisation for contact binaries results in the prediction of the CHE channel leading to mostly equal-mass binary BH mergers \citep[e.g.][]{Marchant2016}.

\subsection{Stellar winds}
The mass loss rates of stellar winds and their effects on the evolution of the star are determined by \verb|SeBa| \citep{Hurley2000,Toonen_2012}, while the effects on the orbit of the triple are determined by \texttt{TRES} (equation \ref{eq:ODE}). 
In this study, we use the same implementation of stellar winds for massive stars as in \citet{dorozsmai2022} with one difference; the mass loss rates of helium stars and giants are calculated according to the empirical formula of \citet{Hamann1995} instead of \citet{Sander_2020}. 

For reference, we summarise the mass loss rates prescriptions used in this study.
For MS stars, we follow \citet{Vink2001}, if $T_{\rm{eff}} \leq 50$ kK and \citet{Nieuwenhuijzen1990}, if $T_{\rm{eff}} > 50$ kK.
For evolved stars crossing the Hertzsprung gap or core helium burning (CHeB) stars, we follow \citet{Vink2001}, if $T_{\rm{eff}} \geq 8$ kK or the maximum between \citet{Nieuwenhuijzen1990} and \citet{Reimers1975}, if $T_{\rm{eff}} < 8$ kK.
For evolved stars beyond the Humphreys-Davidson limit, we  assume $\dot{M}_{\rm{LBV}} = 1.5\cdot10^{-4}\ M_{\odot}\rm{yr^{-1}}$ \citep{Belczynski2010}.
For Asymptotic Giant Brach stars and double shell burning supergiants, we calculate the maximum between \citet{Nieuwenhuijzen1990}, \citet{Reimers1975} and \citet{Vassiliadis1993}.
Finally, for helium stars we follow the empirical form from \citet{Hamann1995} in the form 
 $\dot{M}_{\rm{WR}} = 0.5\cdot10^{-13}\cdot \left(\frac{L}{L_{\odot}}\right)^{1.5}\left(\frac{Z}{Z_{\odot}}\right)^{0.86}$ with a clumping factor of $\eta = 0.5$ from \cite{Hamann98} and a metallicity scaling of $\dot{M}_{WR}\sim Z^{0.86}$ (\citealt{VinkdeKoter05}).

In order to compute the change in the orbit due to stellar winds, we assume stellar winds are spherically symmetric and fast compared to the orbital velocity. Additionally, we neglect wind accretion by the companions. With these assumptions, the inner and the outer orbit changes due to stellar winds as:
\begin{equation}
\label{eq:stellarwind_inner}
    \dot{a}_{\rm in,wind} = \left( \frac{a_{\rm final}}{a_{\rm init}}\right)_{\rm in} = \frac{M_{\rm 1,init} + M_{\rm 2,init}}{M_{\rm 1,final} + M_{\rm 2,final}}, 
\end{equation}
and
\begin{equation}
\label{eq:stellarwind_outer}
    \dot{a}_{\rm out,wind} = \left( \frac{a_{\rm final}}{a_{\rm init}}\right)_{\rm out} = \frac{M_{\rm 1,init} + M_{\rm 2,init} + M_{\rm 3, init}}{M_{\rm 1,final} + M_{\rm 2,final} + M_{\rm 3,final}},
\end{equation}
where subscripts `init' and `final' refer to properties before and after the stellar winds carried mass away from the stars in a given timestep.
We assume that the eccentricity remains unchanged by stellar winds \citep{Huang1956A,Huang1963}.

We neglect stellar wind accretion by the other stars in the triple system \citep[see e.g.][]{Bondi1944MNRAS.104..273B}.
Neglecting accretion is justified for line-driven winds due to their large terminal velocities \citep[see e.g.][]{Vink2001}. 
We note, however, that the assumptions of fast and spherically symmetric wind might not always be valid for short period binaries \citep[e.g.][]{Brookshaw1993}, furthermore, rapidly rotating stars might not have fully symmetric outflows \citep{Georgy2011}. Particularly, stellar winds in certain binary-configurations might even lead to orbital shrinking \citep{Schroder2021}.

\subsection{Remnant formation}
\label{subsec:remnant_formation}

The mass of the compact object remnant is computed based on the delayed supernova model from \citet{Fryer_2012}.
This prescription determines the mass of the stellar remnant as a function of CO core mass at the onset of the core-collapse. The latter is determined in \texttt{SeBa} based on the fits of \citet{Hurley2000}.
The natal kick velocity for BHs is calculated as
\begin{equation}
\label{eq:kick}
    v_{BH} = (1 - f_{\rm b})\left(\frac{M_{ NS}}{M_{BH}}\right)v_{\rm kick}, 
\end{equation}
where $f_{\rm b}$ is the fallback fraction \citep{Fryer_2012}, $M_{\rm NS}$ is the canonical neutron star mass ($M_{\rm NS} = 1.4 M_{\odot}$) and $v_{\rm kick}$ is a random kick velocity drawn from the distribution inferred by \citet{Verbunt2017} from proper motion measurements of pulsars.
We determine the change in the inner and outer orbit due to the core collapse of any of the stars in the triple system based on the formalism developed in \citet{Pijloo2012}.

Models of \citet{Fryer_2012} predict that the most massive stars collapse directly (typically $M_{\rm ZAMS} \gtrsim 40\,M_{\odot}$), without any ejecta, and the only mass loss during the remnant formation is due to neutrino losses, which is assumed to be 10 per cent of the pre-core-collapse mass of the star.
We note that the actual neutrino mass loss is considered to be uncertain, and other population synthesis codes assume considerably smaller losses, e.g. 1 per cent of pre-collapse mass \citep[][]{Belczynski2020:evolutionary_roads_leading_to}, or a fixed value of 0.1$\,M_{\odot}$ loss \citep[][]{Riley2022:compas_paper}.
Recent observations suggest that mass decrements from neutrino emission could indeed be only about a few percent of the pre-collapse mass \citep[][]{Vigna-Gomez2023}. Additionally, we assume that the neutrino emission is spherically symmetrical and do not impart natal kick onto the BH. In this case, the orbit is only changed due to the instantaneous mass loss \citep[e.g. via Blaauw kick, see][]{Blaauw1961}.
We note that, if the pre-core-collapse orbit is circular, a Blauuw kick due to neutrino losses does not lead to a significant change in the inner orbital elements. However, this is no longer the case for eccentric pre-core-collapse orbits. In particular, if the core collapse occurs near the pericenter, the orbit can become significantly wider \citep[e.g.][]{Hills1983}.

By the onset of core-oxygen burning, the core temperatures of the most massive stars can reach above $T_{\rm core} \sim 3\times 10^9\,K$.
Under these conditions, the emitted gamma-ray photons in the core are energetic enough to form electron-positron pairs.
This leads pair-instability (see e.g. \citealt{Fowler1964}, \citealt{Rakavy1967}, \citealt{Barkat1967}, \citealt{Fraley1968}).
Depending on the mass of the star, this instability can result in a pulsation pair instability supernova, in which the star experiences a series of pulsations leading to severe mass loss \citep[i.e.  or PPISN, see e.g.][]{Yoshida2016, Marchant2019, Woosley2017,  Renzo2020},
or pair instability supernova, in which the star is completely disrupted and no remnant is formed \citep[PISN, see e.g.][]{Yoshida2016,Marchant2019, Woosley2017, Renzo2020}.
For the treatment of pair-instability in massive stars, we follow \citet{Stevenson2019}. If the mass of the helium star pre-core-collapse 
is $M_{\rm HE, pre-SN}\geq 35\,M_{\odot}$, the star is assumed to undergo PPISN, and its remnant mass is determined by the fitting formula of \citet{Stevenson2019}, based on the detailed stellar simulations of \citet{Marchant2019}. 
If $60\leq M_{\rm HE, pre-SN}\leq130\,M_{\odot}$, we assume the star undergoes PISN, and leaves no remnant behind. In principle, if  $M_{\rm HE, pre-SN} \geq 130\,M_{\odot}$, photo-disintegration prevents 
the pair instability supernova  and the star collapses directly into a BH
(\citealt{Bond1982}, \citealt{Woosley1982}, \citealt{Heger2002}, \citealt{duBuisson2020}).  However this does not occur for any of our simulated systems,since in our simulation the maximum stellar mass is limited to 100$\,M_{\odot}$. We note that the above-quoted mass ranges are  sensitively dependent on the the poorly constrained reaction rate of $^{12}\rm{C}(\alpha, \gamma)^{16}\rm{O}$ \citep[see e.g.][]{Takahashi2018A, Farmer2019A, WoosleyHeger:2021:MassGap, Costa2021}. In particular, more recent simulations, with updated $^{12}\rm{C}(\alpha, \gamma)^{16}\rm{O}$ rates \citep[see e.g.][]{deBoer2017}, find that the lower edge of the BH mass gap is located at a considerably higher value ($M\approx60\,M_{\odot}$) than previously determined \citep[see e.g.][]{Mehta2022, Farag2022}.

\subsection{Tertiary mass transfer (TMT) episodes}
\label{subsec:TMT_method}
If the tertiary star fills its Roche-lobe, it will transfer mass to the inner binary. 
There have been some efforts to study and model this process \citep{deVries2014, Leigh2020, Comerford2020, Glanz2021,Soker2021, 2022arXiv220703514M}, but this complex scenario remains to be fully understood.

In order to calculate the Roche-lobe of the tertiary star, we assume the inner binary can be approximated as a point mass and estimate the Roche radius with the fitting formula of \citet{EggletonApprox}.
This assumption is valid in the regime where the orbital separation of the outer star is much larger than that of the inner binary (e.g. $a_{\rm{out}}\gg a_{\rm{in}}$).
\texttt{TRES} determines the stability of TMT based on extrapolating typical methods from binary star evolution, i.e. by using critical mass ratios \citep[see e.g.][]{Toonen2016}. This parameter is defined as $q_{\rm crit} = M_{\rm donor}/M_{\rm accetor}$, i.e. the ratio of the mass of the donor and the mass of the accretor star at the onset of the mass transfer episode. The mass transfer phase is assumed to be dynamically unstable, if the mass ratio of the system is above the critical mass ratio, i.e. $q>q_{\rm crit}$. We obtain $q_{\rm crit}$ for each stellar evolutionary stage from \citet{Hurley2002} and \citet{Claeys2014}. We quote these values for the two most common donor types in our simulations  \citep[a complete description of our assumptions about $q_{\rm crit}$ can be found in][]{Toonen2016}. These are $q_{\rm crit} =  3$ and $q_{\rm crit} = (1.37+2[M_{\rm donor, core}/M_{\rm donor}]^5)/2.13$ for Hertzsprung gap stars 
(i.e. hydrogen shell burning stars which have not regained thermal equilibrium yet) and core helium burning  (CHeB) stars, respectively. The term in the squared bracket is the core mass to total mass ratio of the donor. If this equals to $\sim 0.45$ - $0.65$, which is fairly typical for massive CHeB stars \citep{dorozsmai2022}, then $q_{\rm crit}\approx 0.7$-$0.75$. This reflects the assumption made by \citet{Hurley2000}, that CHeB stars tend to have deep convective envelopes \citep[cf.][]{Klencki2020}, and are therefore more likely to experience unstable mass transfer episodes (see e.g. \citealt{Hjellming1987}, but see \citealt{Woods2011}).

Stable TMT could be accompanied with the formation of a circumbinary disc or it could occur in a ballistic accretion fashion.  These two types of mass transfer phases could lead to significantly different evolution of the inner orbit \citep{deVries2014}.
 We assume that TMT occurs via ballistic accretion, if $a_{\rm in}(1 + e_{\rm in}) \geq R_{\rm cd}$ at the onset of the TMT phase, where $R_{\rm cd}$ is:
\begin{equation}
\label{eq:accretion_disc}
    R_{\rm cd} = 0.0425\,a_{\rm out}(1-e_{\rm out})\left[\frac{1}{q_{\rm out}}\left(1 + \frac{1}{q_{\rm out}}\right) \right]^{1/4},
\end{equation}
where we adopted the fitting formulas for mass transferring binaries of \citealt{LubowShu1975} and \citealt{UlrichBurger1976} to triples.

\subsubsection{TMT: Evolution of the inner orbit}
If the tertiary star fills its Roche-lobe, \texttt{TRES} stops the simulation of the system. However, when discussing potential GW progenitors (Section \ref{sec:GW}), we determine the orbital evolution due to TMT by applying simplified assumptions, if the mass transfer episode is dynamically stable. In this subsection we describe our assumptions about the evolution of the inner orbit during a stable phase of TMT, while in subsection \ref{subsubsec:evol_aouter} we discuss the evolution of the outer orbit.

We distinguish three particular TMT configurations cases, based on the evolutionary stage of the inner binary and on whether or not the transferred mass forms a circumbinary disc around the inner binary: 
\begin{enumerate}
    \item an inner binary with compact objects and with ballistic accretion,
    \item an inner binary with compact objects and with a circumbinary disc,
    \item a non-compact inner binary.
\end{enumerate}

(i) \textit{An inner binary with compact objects and with ballistic accretion}. Hydrodynamical simulations of \cite{deVries2014} showed that in case of a TMT episode with ballistic accretion, the transferred mass
eventually engulfs the inner binary and exerts friction on it. 
This leads to a scenario that could be considered similar to the common-envelope evolution of binaries \citep[e.g.][]{Paczynski1976, Ivanova2013}, since in both cases drag forces exerted by a gaseous medium supplied from the donor star lead to the orbital shrinking of the binary.
Inspired by this similarity, \citet{deVries2014} applied a modified version of $\alpha$-formalism \citep[originally developed for common-envelope evolution, see e.g.][]{Tutukov1979,deKool1987,Dewi2000} to model the inner binary evolution of triples experiencing TMT \citep[see also][]{Hamers2021MNRAS.502.4479H}. For the configuration case (i), we take the same approach.

Below we explain how the post-mass-transfer inner orbit is determined based on this formalism in detail. $\Delta M_{\rm trnsf}$ is the mass that is transferred from the tertiary in a timestep $\Delta t$. When $\Delta M_{\rm trnsf}$ ends up encompassing the inner binary, it has binding energy of $E_{\rm bind}$.
As the inner orbit is shrinking due to the friction during the TMT episode, the orbital energy of the inner binary changes by $\Delta E_{\rm orb}$. We assume that a fraction ($\alpha_{\rm TMT}$) of $\Delta E_{\rm orb}$ is  used to unbind  $\Delta M_{\rm trnsf}$. We can write an equation expressing the energy balance as: 
\begin{equation}
\label{eq:energy_equation}
    \alpha_{\rm TMT}\Delta E_{\rm orb} = E_{\rm bind},
\end{equation}
with
\begin{equation}
\label{eq:delta_orb}
         \Delta E_{\rm orb} = \frac{GM_1M_2}{2a_{\rm in, fin}} - \frac{G\left(M_1 + \Delta M_{\rm trnsf}/2\right) \left(M_2 + \Delta M_{\rm trnsf}/2\right)}{2a_{\rm in, init}},
\end{equation}
and
\begin{equation}
\label{eq:binding_energy}
        E_{\rm bind} = \frac{-G(M_1 + M_2) \Delta M_{\rm trnsf}}{\lambda_{\rm TMT} a_{\rm init}},
\end{equation}
where $\lambda_{\rm TMT}$ is a parameter related to the structure of $\Delta M_{\rm trnsf}$, parameterising its binding energy, 
$a_{\rm in, init}$ is the initial orbital separation before $\Delta M_{\rm trnsf}$ is transferred to the inner binary and $a_{\rm in, fin}$ is the final orbital separation after $\Delta M_{\rm trnsf}$ is expelled from the inner binary. We assume that the total mass transferred to the inner binary throughout the entire TMT episode equals to the mass of the hydrogen envelope of the tertiary $M_{\rm out,env}$ (but see \citealt{Laplace2020}). Then assuming a constant $\alpha_{\rm TMT}$ and $\lambda_{\rm TMT}$, the orbit changes due to the entire TMT episode as: 
 \begin{equation}
 \label{eq:change_in_inner_a_CEE}
     \frac{a_{\rm in, fin}}{a_{\rm in, init}} = \frac{M_1M_2}{  \frac{2(M_1 + M_2) M_{\rm out,env}}{\alpha_{\rm TMT}\lambda_{\rm TMT}} +  \left(M_1 + \frac{M_{\rm out,env}}{2}\right) \left(M_2 + \frac{M_{\rm out,env}}{2}\right)}.
 \end{equation}
 As both $\alpha_{\rm TMT}$ and $\lambda_{\rm TMT}$ are unknown, we combine them and try three different values: $\alpha_{\rm TMT}\lambda_{\rm TMT} = 0.05,\, 0.5,\, 5$. Here $\alpha_{\rm TMT}\lambda_{\rm TMT} = 5$ is the fiducial value used in \citet{Hamers2021MNRAS.502.4479H}, which is in a good agreement with the hydrodynamical simulations of \citet{deVries2014}, in which the inner stars are on the MS during the TMT episode.
 We note that we neglect the possibility of TMT episode with ballistic accretion transitioning to a TMT episode with a circumbinary disc.

Additionally, for configuration type (i), we assume that the inner binaries circularise as a result of the mass transfer phase (as $a_{\rm in, new} = a_{\rm in}(1 - e_{\rm in})$). We note that this assumption might not be correct for highly eccentric inner binaries. For example, \citet{GlanzPerets2021} showed that binaries at the onset of common-envelope events with $e\gtrsim 0.95$ might retain eccentricities as high as $e\sim0.2$.

(ii) \textit{An inner binary with compact objects with  circumbinary disc.} If a circumbinary disc is formed during a mass transfer phase towards an inner BH-BH binary we assume that the orbit of the inner binary remains unchanged. 
The actual physics underlying such a process are very complex \citep[see][for a review on circumbinary accretion from gaseous medium]{LaiMunoz2022}. The circumbinary disc may exert a torque on the inner binary and extract angular momentum from it, while the accreted matter can transfer angular momentum onto the inner binary. Furthermore, the circumbinary disc and the inner binary could be tidally distorted by the tertiary star. 
It is commonly assumed that circumbinary accretion of a BH-BH binary from a  gaseous medium leads to the shrinking of its orbit due to the torques exerted by the circumbinary disc and due to dynamical friction of the gas (e.g. \citealt{ Bartos2017ApJ...835..165B, Stone2017,Antoni2019,Tiede2020, Duffell2020, McKerna2020, Rozner2022}). However, a consesus regarding this physical process is still missing, with some hydrodynamical simulations suggesting that accretion from circumbinary disc could even lead to to orbital widening instead of orbital decay \citep[e.g.][]{Munoz2019,Moody2019}.

(iii) \textit{A non-compact inner binary}. If the mass transfer occurs with a MS-MS accretor, we assume that this results in the merger of the inner binary. We make this assumption because these binaries have very short periods and a sizeable fraction of them are in contact and most likely they would expand due to TMT, overfilling their L2 point, which would lead to merger (see later subsection \ref{sec:GW}). As we discuss in in subsection \ref{sec:GW}, we do not consider GW sources from those triple systems, in which the TMT occurs towards a binary with evolved (i.e. non-MS), non-compact stars.

We do not model unstable phases of TMT (as we will show later, they are very rare among the systems we discuss in this paper) . We note, however, that during this type of mass transfer episode, the outer orbital separation is predicted to rapidly decrease due to the common-envelope-like evolution in triple system; this could result in a regime where the secular approximation from the triple is no longer valid \citep[e.g.][]{Glanz2021,Comerford2020,Soker2021}.

\subsubsection{TMT: Evolution of the outer orbit}
\label{subsubsec:evol_aouter}
When determining the evolution of the outer orbit due to a stable phase of TMT, we apply the same method for all accretor types, irrespective of whether a circumbinary disc is formed. 
We calculate the evolution of the outer orbit during the TMT phase, based on the following relation:
\begin{equation}
\label{eq:change_a_binary}
    \frac{\dot{a}_{\rm out}}{a_{\rm out}} = -2 \frac{\dot{M}_3}{M_3}\left[1 - \beta \frac{M_3}{M_1 + M_2} - (1 - \beta)\left(\gamma + \frac{1}{2}\right)\frac{M_3}{M_{\rm tot}}  \right],
\end{equation}
where $\beta$ is the fraction of mass accreted by the inner binary, $\gamma$ is the specific angular momentum lost from the system as a fraction of the specific angular momentum of the triple and $\dot{M}_3$ is the mass transfer rate from the tertiary star.
Equation \ref{eq:change_a_binary} can be derived from angular momentum arguments.
It is an adaptation of the relation describing the orbital evolution of a circular, mass transferring binary comprised of point particles \citep[see e.g.][]{Soberman1997}, applied to a triple experiencing a TMT episode. This adaptation is valid, if the tertiary star is sufficiently far away from the inner binary, such that the inner binary can be treated as a point particle with a mass of $M_1 + M_2$.
We assume that eventually all the transferred mass is isotropically expelled from the triple ($\beta = 0$), from near the inner binary. This expelled matter thus carries away a specific angular momentum that is equal to that of the inner binary ($\gamma = M_{3}/(M_{1} + M_{2})$, see also \citealt{Hamers2021MNRAS.502.4479H}, for a similar approach). In this case equation \ref{eq:change_a_binary} can be expressed as
\begin{equation}
    \label{eq:iso}
     \frac{a_{\rm out, fin}}{a_{\rm out, init}} = \frac{M_{\rm tot,init}}{M_{\rm tot, fin}}\left(\frac{M_{\rm 3,init}}{M_{\rm 3,fin}}\right)^2\exp\left(2\frac{M_{\rm 3,fin}- M_{\rm 3,init}}{M_{1} + M_{2}}\right).
\end{equation}
In case of BH-BH inner accretors, these assumptions might be valid, as the accretion rate of BHs might be capped by the Eddington-limit, and most of the mass could indeed be expelled from the system, for example in the form of a jet (e.g. \citealt{King2000,vandenheuvel17}). On the other hand, MS stars are likely to accrete more efficiently, and therefore $\beta = 0$ might no longer be a good approximation.

\subsection{Initial conditions}
\label{subsec:init}
We sample $10^5$ triples  at two representative (moderate and low) metallicities: Z = 0.005 and Z = 0.0005. We simulate each hierarchical triples from ZAMS. After drawing the parameters for a given triple system, we further check, if it is dynamically stable \citep[based on the criteria of][]{MardlingAarseth2001} or if the stars in the inner binary are CHE at ZAMS. If any of the two criteria are not met, we do not evolve the triple system and only take it into account for the normalisation of event rate calculations. We terminate the simulation of a triple system when either a Hubble time (assumed to be 13.5 Gyr) has passed, or when the tertiary star fills its Roche lobe, a merger occurs, a dynamical instability occurs or if any of the stars becomes unbound from the triple. 
We also stop the simulation, if any of the stars in the inner binary transitions back from CHE to classical evolution. That is, we only consider triples in which the stars of the inner binary chemically homogeneously evolve throughout their entire MS lifetimes. We refer to this population as \textit{CHE triple population}.

In this study, we motivate the choice of the initial distributions of the parameters of the inner binaries based on recent surveys of massive binaries \citep[e.g.][]{ Sana2012, Kobulnicky2014}. In such surveys, a possible tertiary companion is not always unequivocally identified and therefore it is not clear whether the inferred distributions also hold for triples or only for isolated binaries.

We assume the ZAMS mass of the primary star ($M_{\rm 1,ZAMS}$) follows the power-law mass distribution of \citet{Kroupa}, i.e. $N$$\sim$$M_{ZAMS}^{-2.3}$  for $M_{\rm ZAMS} \geq 0.5\,M_{\odot}$ and $N$$\sim$$M_{ZAMS}^{-1.3}$  for $M_{\rm ZAMS} < 0.5\,M_{\odot}$. We sample $M_{\rm 1,ZAMS}$ from a mass range of 20-100$\,M_{\odot}$. The lower limit approximately coincides with the lowest initial mass at which CHE is still possible in a tidally locked binary \citep[e.g.][]{Riley2021MNRAS.505..663R}, while the upper limit is roughly the maximum mass at which the stellar tracks used in \texttt{TRES} are still reasonably accurate. We assume a flat inner mass-ratio (i.e. $q_{\rm in, ZAMS} = M_{\rm 2,ZAMS}/M_{\rm 1,ZAMS}$) distribution, which is in broad agreement with \cite{Sana2012}. We restrict the range of $q_{\rm in, ZAMS}$ to 0.7-1 given that inner binaries in which both of the stars are chemically homogeneously evolving and have $q_{\rm in}\leq 0.7$ are in contact and merge early during the MS due to outer Langrange overflow (where we found the lower limit of 0.7 from our simulations).
We sample the inner semimajor axis from a log-uniform  distribution (\citealt{OpikLaw1924}; and in broad agreement with \citealt{Sana2012}) in the range of 16 to 40 $R_{\odot}$. 
We assume that the inner binaries are tidally locked at ZAMS. This has three implications: i) the inner binaries have circular orbits, ii) their rotational angular frequency is synchronised with the orbital angular frequency, and iii) the spins of the stars are aligned with the orbital angular momentum vector.

We draw the properties of the outer binary from the same distributions that we assume for the inner binaries, with the exception of outer eccentricities.
Observations of hierarchical multiple systems of galactic solar-type stars support the assumption that the distributions of the initial parameters of the inner and the outer binaries are the same \citep{Tokovinin2014, Tokovinin2006}.
We sample the outer semimajor axis from a loguniform distribution in the range of 100 to $10^5$ $R_{\odot}$.
We assume that the distribution of the outer mass ratio (i.e. $q_{\rm out, ZAMS} = M_{\rm out,ZAMS}/(M_{\rm 1,ZAMS} + M_{\rm 2,ZAMS}$)) is flat on a range of 0.1 to 1, furthermore the mass of the tertiary is restricted to a range of 5-100$\,M_{\odot}$.
We assume non-spinning tertiary stars. The eccentricities of the outer orbit are drawn from a thermal distribution \citep[e.g][]{Heggie1975}.
The mutual inclination between the inner and outer orbit is assumed to be uniform in $\rm{cos}(i_{\rm{ZAMS}})$, where $\rm{i}_{\rm ZAMS}$ is the initial inclination. The initial argument of the pericenter is assumed to be uniformly distributed between $-\pi$ and $\pi$.

In Section \ref{sec:GW}, we compare our CHE triple population to a CHE isolated binary population. To this end, we also perform population synthesis of isolated binaries with CHE stars.  We sample $10^5$ isolated binaries at Z = 0.005 and Z = 0.0005 and evolve them with \texttt{TRES}. We sample from the same initial distributions that we assumed for the inner binaries of our triple population. Similarly to the triple population, we discard systems that are not CHE at ZAMS and stop the simulation, if a Hubble time has passed, or if any of the stars in the binary transitions from CHE to classical evolution. We only analyse binaries, in which the stars remain CHE throughout their entire MS lifetime (hereafter \textit{CHE binaries}).

Throughout the paper, we estimate birth rate and merger rate densities of different evolutionary channels (discussed in detail in appendix, section \ref{subsec:app_norm}). In order to determine each of these quantities, one must know how common single and multiple stellar systems are.
We assume two different stellar populations, with different binary and triple fractions. In the first, we assume that about 73 per cent of massive stars are found in triples \citep[with multiplicity fractions of $f_{\rm single} = 0.06$, $f_{\rm binary} = 0.21$, $f_{\rm triple} = 0.73$, see e.g.][]{MoeDiStefano2017},
 whereas in the second test population, we assume there are no triples and about 70 per cent of massive stars are in binaries  \citep[$f_{\rm single} = 0.3$, $f_{\rm binary} = 0.7$, $f_{\rm triple} = 0$, see e.g. ][]{Sana2012}.

Regarding our first test population, we note that \citet{MoeDiStefano2017} finds that 73 per cent of O stars are either in triples or quadrupoles. Therefore $f_{\rm triple} = 0.73$ should be considered as a rough upper limit. However, \citet{Tokovinin2006} finds that there is a strong correlation between the inner period and the triple multiplicity; among solar type stellar systems, 96 per cent of the spectroscopic binaries with periods less than 3 days has a tertiary companion. Therefore CHE triples, which have also inner binaries with periods of few days, could have exceptionally high triple fractions too. We also note that tertiary companions have been detected for many massive contact binary systems \citep[see e.g.][]{Kennedy2010:ocms, Pavel2013:ocms, Lorenzo2017:ocms, Janssesns2021:ocms}.
Regarding, the second test population, we note that \citet{Sana2012} did not make any statements about triple fractions, but they found that 70 per cent of massive stars have companions that are sufficiently close such that mass exchange will occur some time in their evolution.

\section{Results of population synthesis simulations} 
\label{sec:Results}

\begin{table*}
\caption{An overview of our sampled triples based on the evolutionary type of the inner binary. For the definitions of the different categories, see text in section \ref{sec:mainres}.
}
\resizebox{\textwidth}{!}{\begin{tabular}{@{}lccccc@{}}
\toprule 
& \multicolumn{2}{c}{Z = 0.005}                                       & \multicolumn{2}{c}{ Z = 0.0005}                                      & Combined                                      \\ \midrule
& \multicolumn{1}{l}{\% of simulation} & \multicolumn{1}{l}{\% of CHE at ZAMS} & \multicolumn{1}{l}{\% of simulation} & \multicolumn{1}{l}{\% of CHE at ZAMS} & \multicolumn{1}{l}{Birth rate {[}$\rm{Gpc}^{-3}\rm{yr}^{-1}${]}} \\ \midrule
CHE at ZAMS                                     & 10.3                                 & 100                                   & 9.9                                  & 100                                   & 13.7                                                   \\
\textit{- CHE triple}                           & 7.6                                  & 73.5                                  & 7.5                                  & 75.7                                  & 9.6                                                   \\
\textit{- Transition to classical evolution}  & 1.6                                  & 15.5                                  & 0                                    & 0                                     & 1.9                                                    \\
\textit{- Merges during contact phase} & 0.9                                  & 9                                     & 1.7                                  & 17.5                                  & 1.4                                                    \\
\textit{- Simulation error}                     & 0.2                                  & 2                                     & 0.6                                  & 6.7                                   & 0.2                                                    \\ \bottomrule
\end{tabular}}
\label{tab:mainnumbers}
\end{table*}

In Table \ref{tab:mainnumbers}, we provide an overview of our sampled systems based on the evolutionary type of the inner binary. Out of our sampled population of triples, only about 10 per cent have an inner binary where both stars evolve chemically homogeneously from ZAMS (\textit{CHE at ZAMS triples}, see Table \ref{tab:mainnumbers}). We follow the further evolution only for these systems. 
About 75 per cent of CHE at ZAMS triples qualify as CHE triples and we focus on these systems for the majority of the paper. For the remaining 25 per cent, we distinguish three scenarios:
\begin{itemize}
\item \textbf{The inner stars transition to classical evolution}. As the orbit of the inner binary
widens due to stellar winds, the rotational frequencies of the inner stars decrease, because the stellar tides enforce synchronization between the stellar spins and the (new longer) orbital period. If the inner orbit widens sufficiently, the angular rotational frequencies of the inner stars drop below $\omega_{\rm CHE}$ and therefore these stars transition to classical evolution. This occurs only in our moderate metallicity model (15.5 per cent of all CHE at ZAMS triples at Z = 0.005 and 0 per cent at Z = 0.0005).
\item \textbf{The inner binary does not survive the contact phase during the MS phase of the inner stars}. We assume a merger takes place when both stars overflow their outer Lagrangian point during the contact phase. This occurs during mass equalization in the contact phase or due to GW emission, which lead to shrinkage of the inner orbit. As orbital widening due to stellar winds can counteract both of these processes, inner binary merger occurs more frequently at low metallicities (i.e. about 9 per cent of all CHE at ZAMS triples at Z = 0.005 metallicity and 17.5 per cent at Z=0.0005).
\item \textbf{Computational issue.} Finally, we note that the simulation of about 2 (6.7) per cent of CHE at ZAMS triples fails at Z = 0.005 (Z = 0.0005). This can occur because either no solution is found for the secular orbital evolution of the system, or the computation time exceeds the allowed CPU time (which is 5000 seconds per system). Computational issues arise more often for systems that are close to the dynamical instability (i.e. have low $a_{\rm out}/a_{\rm in}$). Therefore, our estimates for the occurrence rates  of systems, in which stellar merger due to ZLK oscillations or dynamical instability occurs should be considered as a lower limit (see next subsection).
\end{itemize}

\subsection{Main evolutionary outcomes}
\label{sec:mainres}

\begin{table*}
\caption{A summary of the different channels (in bold font) and their sub-channels (in normal font) identified of CHE triples. The rows with bold fonts in the second and third column express the number of systems in each channel as a percentage of all CHE triple systems. The rows with normal fonts in the second and third column express the number of systems in each sub-channel as a percentage of all systems in their respective main channel. See equation \ref{eq:birth_rate} and the accompanying discussion in appendix \ref{subsec:app_norm} for the definition of birth rate density.
}
\begin{tabular}{@{}lccc@{}}
\toprule
\multicolumn{1}{c}{Channel}                  & \% at Z = 0.005 & \% Z = 0.0005 & Birth rate density {[}$\rm{Gpc}^{-3}\rm{yr}^{-1}${]} \\ \midrule
\textbf{No post-MS MT}                                & \textbf{27.2}             & \textbf{10.8}          & \textbf{2.3}                                                      \\
- Inner binary evolution decoupled from tertiary      &    97.4                      &     98.6                   &  2.2                                                     \\
- Driven by three-body dynamics                       &   2.6                       &   1.4                     &  0.1                                                     \\ \midrule
\textbf{Stellar merger of the inner binary due to KL} & \textbf{3.3}             & \textbf{2.5}           &\textbf{0.4}                                                        \\
- double helium star accretor                                  &  74.7                        & 58.5                     &  0.3                                                     \\
- helium star-MS accretor                                    &   7.9                       & 35                   &    0.04                                                   \\
- helium star-BH accretor                            &   17.4                       & 6.5                    &  0.06                                                     \\ \midrule
\textbf{Tertiary mass transfer}                & \textbf{55.4}            & \textbf{52.1}          &\textbf{5.2}                                                       \\
- MS-MS accretor                                      &   58                     & 60.8                   &  3.1                                                     \\
- BH-BH accretor                                      & 30.9                      & 24.3                  &  1.5                                                      \\
- Other types of accretors                            &   11.1                       & 14.9                   &    0.6                                                   \\ \midrule
\textbf{Unbound systems}                              & \textbf{10.7}                & \textbf{34}            & \textbf{1.4}                                                      \\
- Core collapse SN                                    &    100                      & 16                     &   0.9                                                    \\
- (P)PISN                                             &   0                       & 84                     &   0.5                                                    \\ \midrule
\textbf{Dynamical instability}                       &    \textbf{3.5}                      & \textbf{0.6}                    & \textbf{0.3}                                                      \\ \bottomrule
\end{tabular}
\label{tab:secondnumbers}
\end{table*}

In Table \ref{tab:secondnumbers}, we show the most common evolutionary outcomes for CHE triples.
We distinguish 5 different evolutionary channels:
\begin{itemize}
    \item \textbf{No post-MS mass transfer phase:} 
    During the MS, it may be in a contact, but the system does not experience any other form of mass transfer events. The inner binary eventually forms a BH-BH binary in all such systems.  
    \item \textbf{Stellar merger of the inner binary due to ZLK:} Stellar merger occurs in the inner binary due to ZLK oscillations. 
    \item \textbf{Tertiary mass transfer (TMT):} The tertiary star fills its Roche lobe. 
    \item \textbf{Unbound systems:}
    This evolutionary outcome takes place, if any of the stars becomes unbound from the system. 
    This occurs when a stellar remnant is formed in the system, with three major subtypes: (i) natal kick imparted onto the remnant object during the SN explosion, (ii) instantaneous mass loss during pulsational PISN, or (iii) complete disruption of the star due to PISN.
    \item \textbf{Dynamical instability:} These systems eventually become dynamically unstable, where the secular approximation is no longer valid.
\end{itemize}
 We discuss these channels in detail in sections \ref{subsec:tmdt} - \ref{subsec:systems_without_postMS_mt}.

\subsection{Examples for the evolution of a few selected systems}
\label{subsec:example_evol}
In the following, we present the evolution of a few selected systems from some of the channels introduced in section \ref{sec:mainres}. 
In all of these example systems, the initial parameters of the inner binary are the same: $M_{\rm 1,ZAMS} = M_{\rm 2,ZAMS} = 70\,M_{\odot}$, $a_{\rm in, ZAMS} = 22.4\,R_{\odot}$. These have been specifically chosen such that this system would form a GW source via the binary CHE channel within the Hubble time, if it was an isolated binary  
(i.e. in about 8.9 $\rm Gyr$). The inner binary is tidally locked and therefore $e_{\rm in, ZAMS} = 0$. The stars of the inner binary are in contact from ZAMS. 
The initial mutual inclination is $\rm{i}_{\rm ZAMS} = 90^{\circ}$ in all systems discussed below, which allows for ZLK oscillations to develop, unless they are suppressed by short range forces \citep[see e.g.][]{Liu2015}.

In order to understand the evolutionary paths of CHE triples introduced below, we first show which configurations of CHE triples lead to efficient ZLK oscillations (see Fig. \ref{fig:3bd_pm}). We evolve the previously introduced CHE inner binary as an isolated system, and take four snapshots during different evolutionary stages (ZAMS, end of MS, at the onset of core collapse, and at the formation of an inner BH-BH binary). For each snapshot, we show a range of possible tertiary companions to this inner binary with different tertiary masses ($M_{\rm out}$) and outer semi major axes ($a_{\rm out}$) and identify those regions, where three-body dynamics are relevant.

As shown in the leftmost panel, precession due to tides completely 
suppresses three-body dynamics when the inner stars are still on MS for almost the entire parameter space of CHE triples. 
The limited number of triples for which this is not true typically become dynamically unstable later in the evolution (e.g. compare panel 1 with panel 4).
By the time of  hydrogen depletion in the inner stars, the stellar radii of CHE stars shrinks typically by a factor of 3-5 with respect to their ZAMS value.  Therefore, at this stage tides become less efficient (since $t_{\rm tides} \sim R^{-5}$, see equation \ref{eq:tides_timescale}) and precession due to GR becomes the major limitation to three-body dynamics. For the systems shown in Fig. \ref{fig:3bd_pm}, ZLK oscillations occur only, if $a_{\rm out} \lesssim 500\,R_{\odot}$ . During the CHeB phase of the inner stars, the the typical timescale of precession due to GR further increases, as a result of the strong 
Wolf-Rayet winds that significantly widen the inner orbit. 
As long as the inner orbit widens faster than the outer orbit (which is always true for CHE triples, if the tertiary star is the initially least massive star in the system), the timescale related to ZLK oscillations will typically decrease. Therefore, during this stage, the parameter space where three-body dynamics are relevant increases.
This is also shown in the rightmost panel of Fig. \ref{fig:3bd_pm}; by the time the inner binary forms BHs, triples with $a_{\rm out} \lesssim 2000\,R_{\odot}$ will develop ZLK oscillations.


\subsubsection{Example for stellar merger of the inner binary due to ZLK oscillations}
\label{subsubsec:doubleHEmerger}
First, we discuss the evolution of a CHE triple, in which the inner binary merges as a double helium star due to strong ZLK oscillations (shown in Fig.  \ref{fig:he_merger}).
This triple has a tertiary with an initial mass of $M_{\rm out, ZAMS} = 32.1\,M_{\odot}$ and a circular outer orbit with $a_{\rm out, ZAMS}=200\,R_{\odot}$.  
As indicated by Fig. \ref{fig:3bd_pm}, when the stars of the close inner binary are still on the MS, precession associated with strong tides suppresses the effects of the three-body dynamics  \citep[see also e.g.][]{VignaGomez2022}.
At 3.9 Myr, the stars of the inner binary evolve off the MS. By this time, these stars had lost a small amount of mass due to stellar winds and the inner orbit had widened by only 2 per cent as a result. Similarly, the outer orbit also widens only by a negligible amount. Consequently, the timescale of the ZLK oscillations does not change significantly. On the other hand, the tidal effects become much weaker, as the radii of the stars had decreased by a factor of 5 with respect to their ZAMS value. 
As a result, the ZLK oscillations are no longer suppressed (see also second panel of Fig \ref{fig:3bd_pm}). 
At this stage, there are two competing mechanisms that drive the evolution of the pericenter: ZLK oscillations and the strong Wolf-Rayet-like winds, which decrease and increase the pericenter, respectively.
For this triple, the ZLK timescale is extremely short (few years) and a large inner eccentricity of $e_{\rm in}\approx 0.65$ is reached shortly after the onset of CHeB, during which the orbital widening due to stellar winds is negligible. At this stage, the pericenter becomes sufficiently small such that the helium stars fill their Roche-lobes at the point of closest approach. 
We stop the simulation at this point and assume it leads to a merger. In principle it is possible that the system would enter a stable contact phase, however given the eccentricities, the majority of these systems would probably soon experience outer Langrange overflow, before tides would quench ZLKs and before the inner stars would form BHs. More detailed models are necessary to understand the evolution of a double helium star system in contact, in particular when undergoing three-body dynamical effects.

%


\subsubsection{Example for TMT towards an eccentric BH-BH binary}
\label{subsubsec:tmt_bhbh_eccentric}
The next triple we discuss experiences a TMT episode towards an eccentric BH-BH inner binary (shown in Fig \ref{fig:bhbh_eccentric}). This system has the same parameters as the previously discussed triple, but with a slightly larger initial outer semimajor axes: $a_{\rm out,ZAMS} = 421\,R_{\odot}$. 
When the inner stars evolve off MS, ZLK oscillations are quenched by precession due to GR (compare the second panels of Fig. \ref{fig:3bd_pm} and Fig. \ref{fig:bhbh_eccentric}). Three-body dynamics become eventually effective, however, because the orbit of the inner binary widens significantly and faster than the outer orbit due to strong WR winds (compare the third panels of Fig. \ref{fig:3bd_pm} and Fig. \ref{fig:bhbh_eccentric}, although by this stage the parameters of the inner binary differ slightly).
As a result, $t_{\rm GR}$ increases by a factor of 5, while $t_{\rm ZLK}$ barely changes. As ZLK cycles become only effective, once the inner orbit has sufficiently widened, the inner binary does not come into contact despite reaching similarly high inner eccentricities as in the previous system.

As the stars of the inner binary have the same mass, they co-evolve, and they become stellar remnants at the same time. 
This occurs around 4.2 Myr, when the inner eccentricity is $e_{\rm in, max} = 0.75$. Since core-collapse occurs in an eccentric orbit, large range of possbile post-supernova orbits are possible ($a_{\rm in} =$ 42-186 $R_{\odot}$) depending on where exactly the stars are in their orbit. In the particular example shown in Fig \ref{fig:bhbh_eccentric}, the core collapse occurs while both stars are near the pericenter (which is less likely as they spend more time near the apocenter). This leads to an inner semi-major axis of $a_{\rm in} = 171\,R_{\odot}$ after BH-BH formation. As the outer orbit is circular at the onset of the core-collapse, it only widens by a moderate amount. As the inner period to outer period ratio has increased by a factor of 7, the timescale of the ZLK oscillations also further decrease, making the three-body dynamics even more relevant for the further evolution of the system. The evolution of this triple therefore demonstrates, that if the ZLK oscillations are strong enough to induce eccentricities before the formation of an inner BH-BH binary, the importance of three-body dynamics can be significantly increased during the last stages of the evolution of the triple, depending on (i) where the inner stars are in their orbit when the formation of the compact objects occur and (ii) on the eccentricity of the outer orbit.

After the formation of the inner BH-BH binary, the tertiary star evolves off MS, and at 6.1 Myr fills its Roche-lobe and transfers mass to the highly eccentric ($e_{\rm in} = 0.94$) BH-BH binary at a highly inclined orbit ($\rm{i} = 71.5^{\circ}$). At this stage, we stop the simulation (but see later section \ref{sec:GW}, where we predict the further evolution of some of these systems). We note, however, that even if the TMT episode does not affect the inner binary, it still merges due to GWs about a factor of 8 faster than its isolated binary counterpart, just alone due to the high eccentricities induced by the ZLK oscillations.

\subsubsection{Example for TMT towards a circular BH-BH binary}
\label{subsubsec:tmt_bhbh_circular}
\begin{figure*}
\includegraphics[width=\textwidth]{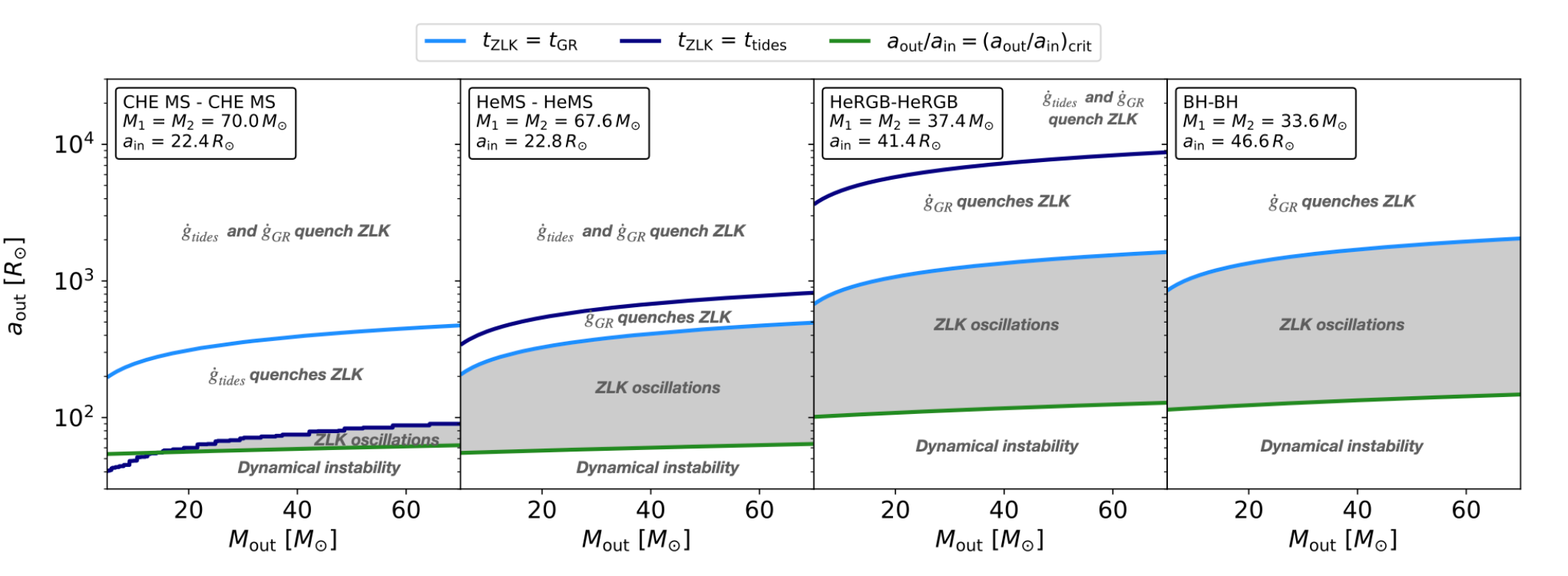}
\caption{We illustrate the parameter space where ZLK oscillations develop for typical CHE triples at different evolutionary stages at Z = 0.005. Each panel represents triples with a specific inner binary (with masses $M_1$, $M_2$ and inner separation $a_{\rm in}$ as indicated in the top left of each panel) but with varying tertiary masses, $M_{\rm out}$ (x-axis) and outer semimajor axes, $a_{\rm out}$ (y-axis, logscale). The parameters of the inner binary  in each panel are the same as that of an isolated CHE binary with $M_{\rm 1, ZAMS}$ = $M_{\rm 2, ZAMS}$ = $70\,M_{\odot}$, $a_{\rm ZAMS} = 22.4\,R_{\odot}$ at different evolutionary stages (see text in section \ref{subsec:example_evol}).  First panel: two CHE main sequence stars at zero-age main sequence, second panel: helium stars at the onset of core-helium burning (HeMS), third panel: two helium stars at the end of core-helium burning (HeRGB), fourth panel: at the formation of a BH-BH inner binary. We assume circular inner and outer orbits and a mutual inclination of $\rm{i} = 90^{\circ}$.
The light blue and the dark blue lines show regions, where the ZLK timescale equals to the timescale of apsidal precession due to tides and GR effects, respectively (see equation  
\ref{eq:zlk_timescale}, \ref{eq:gr_timescale}, and 
\ref{eq:tides_timescale}). Consequently, triples above any of these two lines do not exhibit ZLK oscillations, as they are quenched by these short range forces. The green line represents the boundary of dynamical instability.
The shaded grey region represents triples where ZLK oscillations are effective.}
\label{fig:3bd_pm}
\end{figure*}

\begin{figure*}
\includegraphics[width=\textwidth]{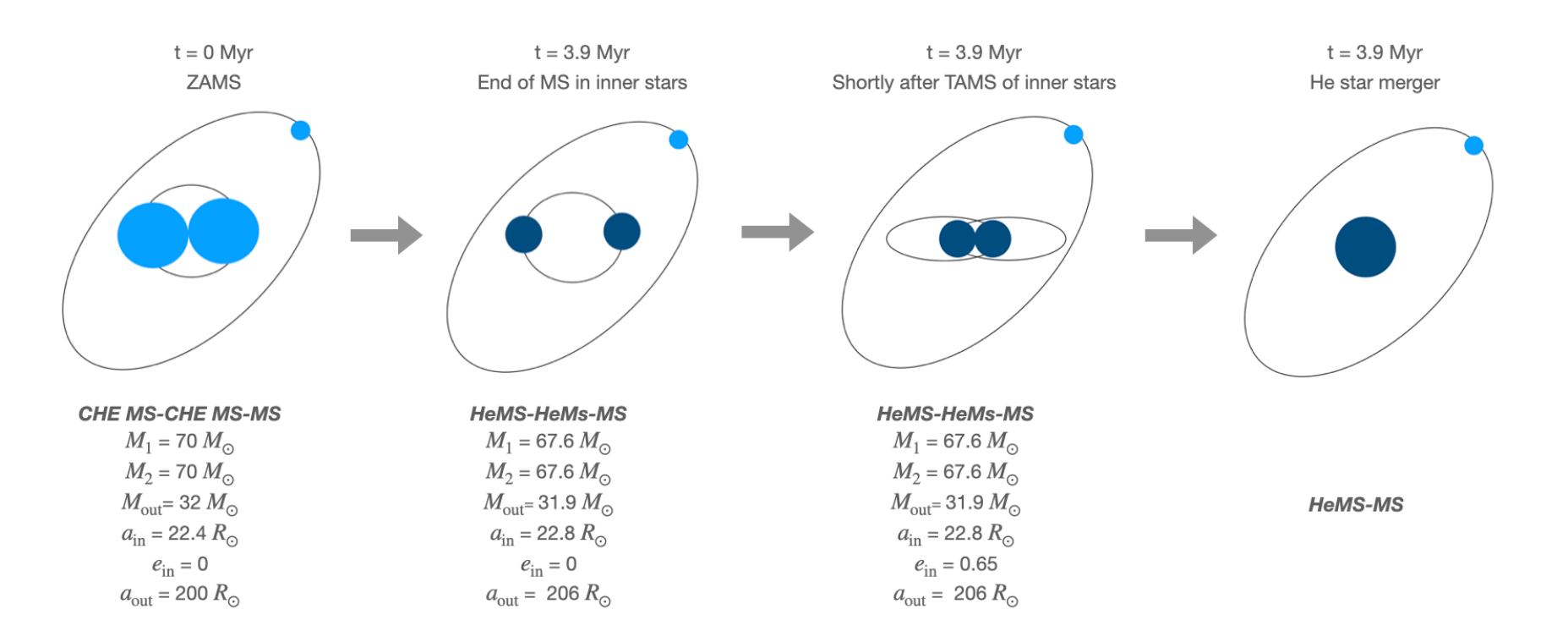}
\caption{A schematic drawing showing the evolution of a triple system in which the stars in the inner binary experience stellar merger due to ZLK oscillations. The first line below each drawing shows the evolutionary stage of each star. The first is for the primary, the second is for the secondary, and the third is for the tertiary star. CHE MS is chemically homogeneously evolving MS star, HeMS is core-helium burning helium star, HeRGB is a helium star that has finished core-helium burning. The parameters of the triple at ZAMS are: $M_{\rm 1,ZAMS} = M_{\rm 2,ZAMS} = 70\,M_{\odot}$, $a_{\rm in, ZAMS} = 22.4\,R_{\odot}$, $\rm{i}_{\rm ZAMS} = 90^{\circ}$, $M_{\rm out, ZAMS} = 32.1\,M_{\odot}$, $a_{\rm out, ZAMS}=200\,R_{\odot}$, $e_{\rm out} = 0$. The outer eccentricity remains $e_{\rm out} \lesssim 0.01$ throughout the evolution.
}
\label{fig:he_merger}
\end{figure*}

\begin{figure*}
\includegraphics[width=\textwidth]{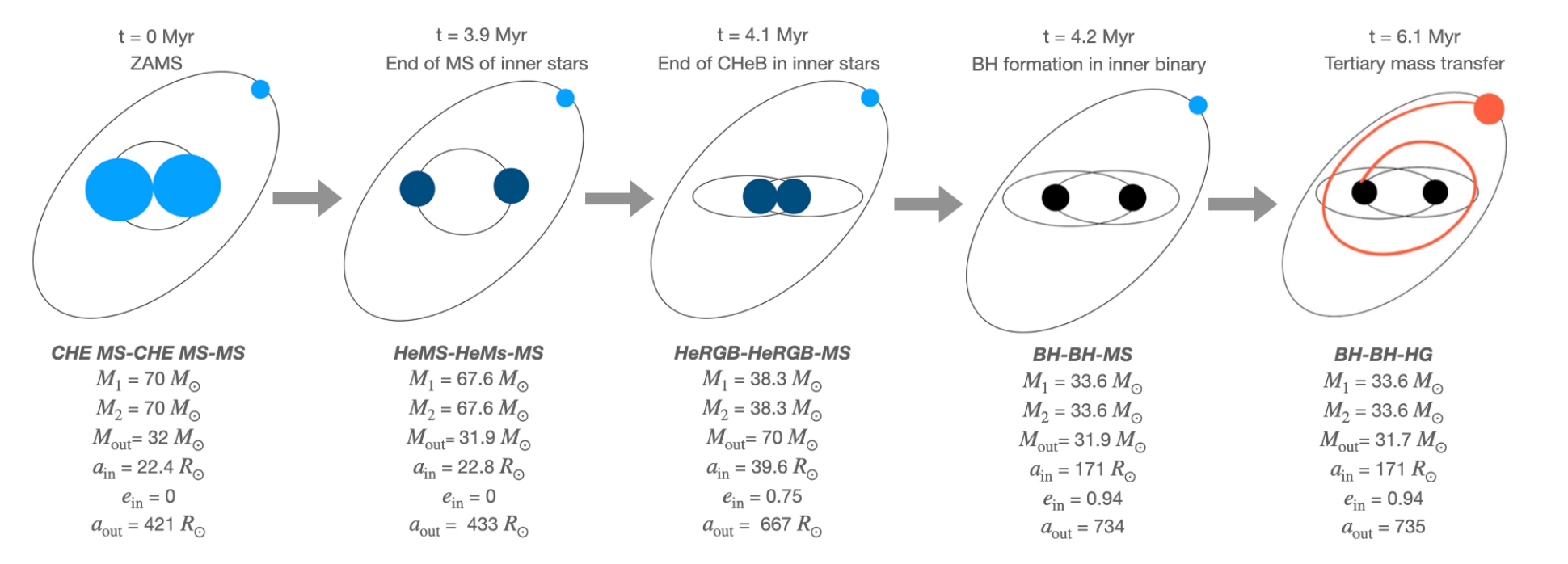}
\caption{A schematic drawing showing the evolution of a triple system with that eventually experiences a TMT episode (i.e. the most common evolutionary outcome, see Table \ref{tab:secondnumbers}).
The triple system with the following initial parameters: $M_{\rm 1,ZAMS} = M_{\rm 2,ZAMS} = 70\,M_{\odot}$, $a_{\rm in, ZAMS} = 22.4\,R_{\odot}$, $\rm{i}_{\rm ZAMS} = 90^{\circ}$, $M_{\rm out, ZAMS} = 32.1\,M_{\odot}$, $a_{\rm out, ZAMS}=200\,R_{\odot}$, $e_{\rm out} = 0$.
The outer eccentricity remains $e_{\rm out} \lesssim 0.01$ throughout the evolution.
}
\label{fig:bhbh_eccentric}
\end{figure*}

Next, we show the evolution of a CHE triple, which also experiences a TMT episode towards a BH-BH binary, but in which three-body dynamics remain suppressed throughout the entire evolution. The initial outer semimajor axis is $a_{\rm out,ZAMS} = 1069\,R_{\odot}$. For this system the timescales of the ZLK oscillations remain too long with respect to the timescale associated with precession due to GR effects throughout the entire post-MS phase. At the onset of the core-collapse, at which the parameter space for ZLK oscillations is the typically the largest for CHE triples with inner binaries composed of non-compact objects, the outer semimajor axis is $a_{\rm out,ZAMS} = 1720\,R_{\odot}$ and the tertiary mass is $M_{\rm out} = 31.9\,M_{\odot}$. Third panel in Fig. \ref{fig:3bd_pm} implies that the three-body dynamics is just quenched by the relativistic precession at this stage. Therefore, the inner orbit remains circular when the BHs are formed, and the inner orbit only widens moderately due to BH formation. The inner and the outer orbit after the formation of a BH-BH binary are $a_{\rm in} = 46.6\,R_{\odot}$ and $a_{\rm out} = 1860\,R_{\odot}$ and therefore the ZLK oscillations remain quenched. At 6 Myr, the tertiary reaches a radius of 547 $R_{\odot}$ and fills its Roche-lobe while crossing the Hertzsprung gap. The last two examples suggest (and we will show in section \ref{subsec:origin} that this is generally true for the vast majority of CHE triples) that three-body dynamics are only relevant for the evolution of CHE triples, if the tertiary star is on a sufficiently short orbit, such that it will eventually fill its Roche-lobe and initiate a TMT episode. Conversely, if the tertiary star remains detached throughout the evolution of the triple, the inner binary evolves effectively as an isolated binary for the vast majority of CHE triples. 

\subsection{No post-MS mass transfer}
\label{subsec:systems_without_postMS_mt}
In these triples, the tertiary star remains bound and detached, while the stars of the inner binary form a BH-BH binary. 
The inner stars are in contact in the majority of the cases (e.g. around 90 per cent  at Z = 0.005). 
There are no any other mass transfer phases during the evolution of these systems (by definition). About 27 per cent of CHE triples evolve this way in our moderate metallicity model (see Table \ref{tab:secondnumbers}). This decreases to 11 per cent at Z = 0.0005. The main reason for this difference is the larger number of PISN that occurs at lower metallicities, which prevent the formation of BHs.

After the formation of the BH-BH binary, the system may merge due to GW emission within a Hubble time.
This occurs for all systems of this type at $Z = 0.0005$. However at $Z = 0.005$,
the stellar winds are strong enough such that 32 per cent of the inner binaries of these triples end up with orbits that are too wide to merge within a Hubble time
due to GW emission. We note that these are not necessarily all of the GW sources from our simulations, as triples in other channels discussed here can also potentially
form merging binary BHs (see discussion in section \ref{sec:GW}).

For the majority of these triples ($>97$ per cent), the inner binary evolves essentially unaffected by the tertiary star (see also section \ref{subsec:origin}). Therefore, the properties of the inner binaries of this channel are nearly indistinguishable from those of isolated CHE binaries. The initial outer pericenters of the triples of this channel are large enough such that the outer star remains detached (i.e. $a_{\rm p, out, ZAMS} \gtrsim 2000$-$3000\,R_{\odot}$ at Z = 0.005, see also section \ref{subsec:origin}). At such large tertiary separations, the three-body dynamics remain suppressed during the entire evolution of the triple.

The properties of the subgroup in which three-body dynamics drive the evolution of the inner binary are very different. Firstly, they have very short initial outer pericenters (i.e, $a_{\rm p, out, ZAMS} \approx 100$-$700\,R_{\odot}$), and secondly, the tertiary has a relatively low mass (typically $M_{\rm out,ZAMS}$ = $10$-$30$ $M_{\odot}$). In these systems, the ZLK oscillations drive the eccentricity of the inner BH-BH binary up to large values (e.g. $e_{\rm in} \gtrsim 0.7$-$0.9$). Above a given eccentricity, the GW emission becomes so efficient that the inner binary decouples from the tertiary and it plunges due to GWs \citep[see e.g.][]{SilsbeeTremaine2017,RodriguezAntonini2018}. These systems typically have a relatively low-mass tertiary star compared to the stars in the inner binary, such that the inner binary merges as a BH-BH binary due to GW emission before the tertiary star would evolve off the MS and fill its Roche-lobe.
Overall, the parameter space for this subgroup is very small, and therefore we predict a negligible GW merger rate (see later discussion in section \ref{sec:GW}).

\subsection{Stellar merger of the inner binary due to ZLK}
In this scenario, the inner binary merges due to three-body dynamics, before it would form a BH-BH binary. At Z = 0.005, about 3.3 per cent of the CHE triple population evolves this way. 
In our low metallicity model, this fraction decreases slightly, to 2.2 per cent. This is because at lower metallicities, the inner period to outer period ratio increases less due to the weaker stellar winds, and therefore ZLK oscillations remain less efficient (see equation \ref{eq:zlk_timescale}).

Mergers in this channel occur in inner binaries, in which one or both of the stars have already evolved off MS, otherwise the strong tidal effects typically quench the ZLK oscillations (see section \ref{subsec:example_evol}).
As shown in Table \ref{tab:secondnumbers},  most of the merger occurs between two helium stars (59-75 per cent). 
The rest occurs between a helium star - MS star or  helium star - BH binaries.
The majority of the double helium star mergers (>90 per cent) originate from triples, in which the stars in the inner binary were in contact during MS and co-evolved. This also implies that the majority of them have equal masses at the time of the merger. The masses of these helium inner stars typically range from 29 to 94 $M_{\odot}$ at Z = 0.005.

The outer orbital period of the triples from this channel has to be sufficiently short, such that the ZLK oscillations are strong enough such that they prompt the inner binary to merge. The outer pericenter at the moment of the merger typically ranges from 100 to 200 $R_{\odot}$ and it does not exceeds 700 $R_{\odot}$.
The eccentricities of the inner binary at the moment of the merger typically have values of $e_{\rm in}\approx$ 0.5-0.9. 
For all of these triples, the tertiary is a MS star at the time of the merger and less massive than the stars of the inner binary, otherwise it would evolve faster than the stars in the inner binary and would fill its Roche-lobe, while the inner stars are still on MS. If the outer orbit does not significantly change after the merger, the tertiary star is expected to transfer mass to the merger product, once it evolved off the MS.

\subsection{Systems with tertiary mass transfer (TMT)}
\label{subsec:tmdt}
Among CHE triples, this is the most common evolutionary pathway. In these systems, the outer star eventually initiates a mass transfer phase while the inner binary is detached or in contact. Approximately 55 (52) per cent at $Z = 0.005$ ($Z = 0.0005$) of all CHE triples 
follow this type of evolutionary path (see Table \ref{tab:secondnumbers}). 
This means that a TMT episode would eventually occur in about 40 per cent of all stellar systems containing a binary with CHE stars (with $f_{\rm binary} = 0.21$, $f_{\rm triple} = 0.73$). 
While systems containing binaries with CHE stars are rare (see e.g. typical birth rates in Table \ref{tab:mainnumbers}), they form GW sources very efficiently \citep[e.g.][]{Mandel2016, de_Mink_2016, Marchant2016}. Therefore, our predictions suggest that the evolution of a non-negligible fraction of potential GW progenitors could experience a TMT episode.
This is an interesting result, as TMT is thought to be very uncommon for classically evolving hierarchical triples, which would have implied that they play a limited role in important astrophysical phenomena \citep[see e.g.][]{deVries2014,Toonen2020, Hamers2022, kummer2023main}. In particular \citet{Toonen2020} found that about 1 per cent of triples with primaries in the intermediate mass range belong to this evolutionary channel. Similarly, \citet{deVries2014} predicts that about only 1 per cent of the observed 725 triples in the catalogue of \citet{Tokovinin_2010} would eventually initiate TMT. 

In the following sections (\ref{subsub:donor_of_tmt}-\ref{subsub:3bd_for_tmt}), we discuss the properties of the triples of this channel at the onset of TMT. While predicting the outcome of a TMT episode is currently extremely challenging, highlighting several important aspects of these systems (e.g. dynamical stability of TMT, timescales of TMT epsiodes, the amount of transferred mass, the type of accretors, etc.) helps to better understand the nature of these systems and the role they potentially play in the evolution of GW progenitors.

\subsubsection{Donors of TMT episodes}
\label{subsub:donor_of_tmt}
Here, we discuss the different stellar evolutionary stages of the donor stars at the onset of the mass transfer phase, as it is highly relevant for determining the stability of the mass transfer episode \citep[which has dramatic effect on the outcome of the TMT, compare e.g.][]{deVries2014,GlanzPerets2021}.

In particular, core-helium-burning or asymptotic giant branch stars tend to develop deep convective envelopes.  Mass transfer episodes initiated by such cool-giant donors with deep convective envelopes are more likely to occur in a dynamically unstable way than mass transfer phases initiated by giant donors with mostly radiative envelopes \citep[e.g.][]{Hjellming1987,Soberman1997,Klencki2020}. 

\begin{figure*}
\includegraphics[trim=0.5cm 0.5cm 2.5cm 0cm, clip, width=\textwidth]{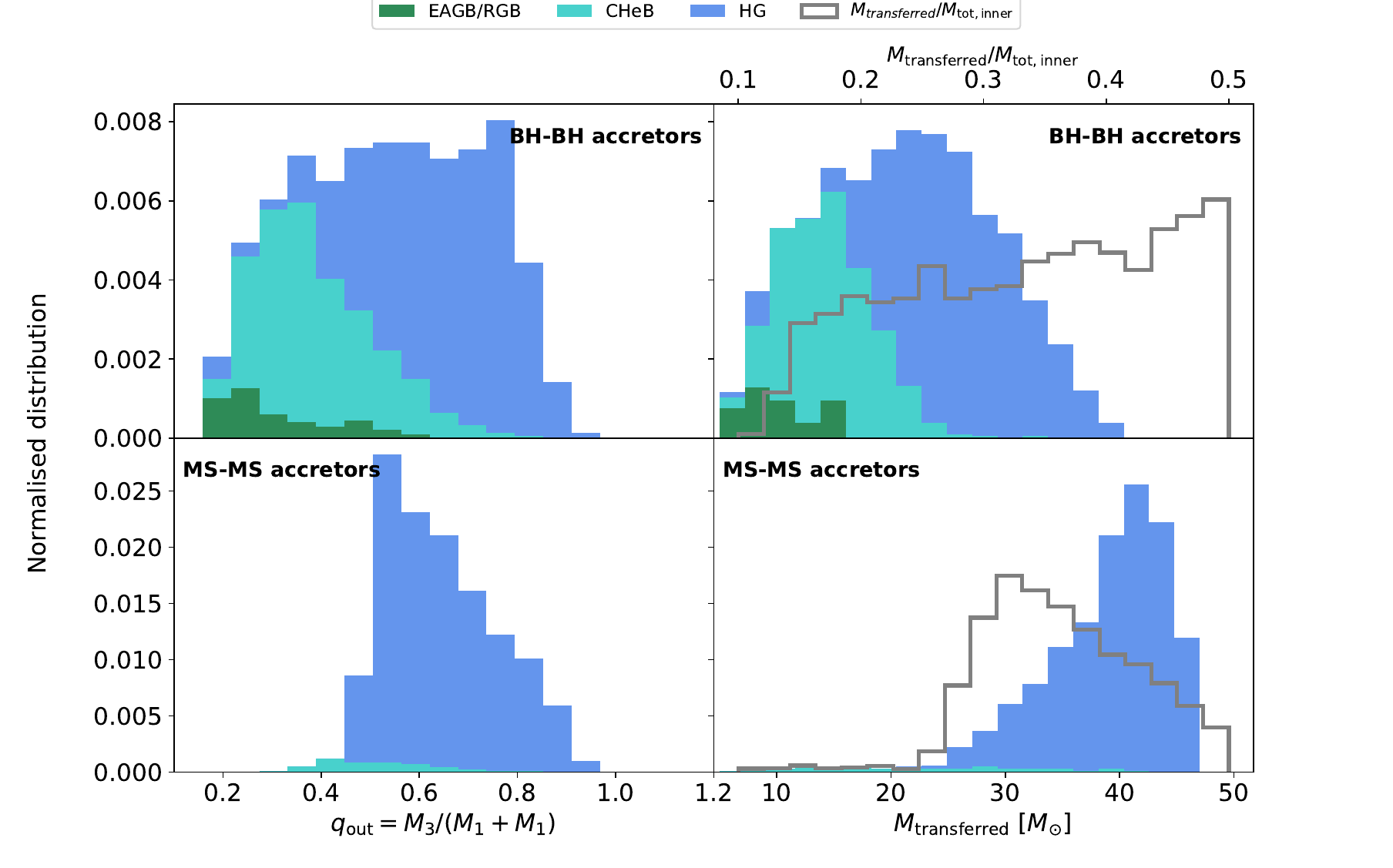}
\caption{We show the properties of the donor star at the onset of tertiary mass transfer episode. The right panels show the outer mass ratio, $q_{\rm out} = M_{\rm out}/(M_1 + M_2)$, while the left panels show the amount of mass transferred from the tertiary to the inner binary for systems undergoing TMT at Z = 0.005, at the onset of the tertiary mass transfer phase. Different colours correspond to different evolutionary stages of the donor as shown by the legend. We exclude cases where the donor is a MS star. Furthermore, we distinguish BH-BH inner binaries (upper panels) and MS-MS inner binaries (lower panels), which are the main types of accreting systems (see e.g. Table \ref{tab:secondnumbers}) . The unfilled histogram in the panels on the left shows the transferred mass as a fraction of the total mass of the inner binary for all donor types. 
All histograms shown have been normalised with respect to the population of CHE evolving triples.} 
\label{fig:m_transferred_ext_z_high}
\end{figure*}

At $Z = 0.005$, around 80 per cent of the donors of TMT systems are stars crossing the Hertzsprung gap.
At this metallicity, the largest expansion in the radius of massive stars occurs during this evolutionary phase, which makes binary interaction during this stage the most probable. The second most common donor type is CHeB star.
with 11.3 per cent, while the rest are either stars on the first giant branch (when the tertiary $M_{\rm out,ZAMS} \lesssim 8\,M_{\odot}$) or stars on the asymptotic giant branch. 

At lower metallicities, CHeB donors are more prevalent. At $Z = 0.0005$, only 58 per cent of the tertiary donors are HG stars while 40 per cent are CHeB stars; this is because the onset of CHeB occurs at a higher effective temperature with respect to systems at Z = 0.005. Consequently, at lower metallicities, the onset of CHeB is followed by a larger increase in radius with respect to their higher metallicity counterparts. This in turn implies that stars are more likely to fill their Roche-lobes at this evolutionary stage.

\subsubsection{Stability of TMT episodes}
\label{subsub:stability_of_tmt}
The vast majority of mass transfer episodes in this channel occur in a dynamically stable way (99.9 per cent at Z = 0.005 and 98.8 per cent at Z = 0.0005). This is due to the relatively low mass ratios at the onset of the mass transfer phase (i.e. typically $q_{\rm out} < q_{\rm crit}$, see right panel of Fig. \ref{fig:m_transferred_ext_z_high} for our moderate metallicity model, and Fig. \ref{fig:m_transferred_ext_z_low} for our low metallicity model). Typical mass ratios for systems with HG donors are $q_{\rm out}$ = 0.4-0.8, while for CHeB donors, they are $q_{\rm out}$ = 0.3-0.5 . The values for CHeB donors are smaller because of the strong LBV winds that CHeB star experience  decrease the mass ratios over time. Unstable mass transfer phases exclusively occur with CHeB donors in our simulations. 

These low mass ratios also imply that the expansion due to stellar evolution drives the TMT episodes \citep[e.g.][]{Soberman1997}. Consequently, we expect TMT episodes with HG donors to last $10^4$ yrs, while TMT epiosdes with CHeB donor could last much longer up to $10^5$-$10^4$ years.

\subsubsection{Accretors of TMT episodes}
\label{subsub:accretors_of_tmt}
In this subsection, we discuss the type of accretors of TMT episodes. The evolutionary stage of the inner binaries has a crucial role in the outcome of TMT episodes. If the inner binary comprises CHE MS stars, a TMT episode probably leads to the merger of the inner binary, as CHE binaries have very short periods and the majority of them are in contact at the onset of the TMT \citep[see also][]{Braudo2022}. 
On the other hand, if the inner binary consists of BHs, TMT episode is is less likely to lead to merger by itself, since the orbit has to shrink by a much larger factor with respect to double MS systems. In fact, in our models, this process never leads directly to coalescence \citep[see later discussion in section \ref{subsubsec:gw_TMT_with_BBH_accretor}, but also see models, in which compact binary mergers are predicted in gaseous environemnts][]{Antoni2019, Rozner2022, LaiMunoz2022, Siwek2023, DorzaioDuffell2021}. Nevertheless, a TMT epsiode with a BH-BH inner binary could be a source of (an observable) X-ray emission \citep[e.g.][]{XrayBook1997}.

As shown in Table \ref{tab:secondnumbers}, the two most common types of accretors are MS-MS and BH-BH binaries.  In only 11-15 per cent of CHE triples experience TMT with different accretors, such as an inner binary consisting of two helium stars or a helium star with a MS or BH companion.

We highlight the relatively large fraction of BH-BH accretors (24-31 per cent of CHE triples experiencing TMT). 
For classically evolving triples, mass transfer towards a BH-BH binary is highly unlikely. Firstly, in systems in which a TMT episode were to occur towards a BH-BH inner binary, the stars of the inner binary need to be more massive than the tertiary, such that they form BHs before the outer star fills its Roche-lobe. Secondly, the outer star has to be sufficiently close, otherwise it would remain detached throughout its evolution.  This, in turn, puts a limit on the largest possible inner orbit, if the system is to remain dynamically stable. The maximum inner orbital separation for such systems is so small that classically evolving inner stars (which eventually expand) would initiate mass transfer and would most likely merge, which would reduce the triple to a binary and a tertiary mass transfer would never occur \citep[see e.g. Fig. 14 in][]{Toonen2020}. On the other hand, if the triple has CHE inner stars, the stars will not expand and  not merge with one another, instead the system will evolve to contain a BH-BH binary by the time the tertiary fills its Roche-lobe.

\subsubsection{Mass transferred towards the inner binary}
We discuss the amount of mass that is transferred during the TMT episode. This is an important aspect, as the relative transferred mass (i.e. $M_{\rm transferred}/M_{\rm tot,inner}$)  determines angular momentum reservoir available to change the orbit of the inner binary.

Assuming that the entire envelope of the donor star is transferred towards the inner binary, the amount of transferred mass ranges between 1-40$\,M_{\odot}$ for BH-BH accretors  and between 10-50$\,M_{\odot}$ for MS-MS accretors (see left panel of Fig. \ref{fig:m_transferred_ext_z_high} for Z = 0.005 and \ref{fig:m_transferred_ext_z_low} in section of \ref{sec:additional_figures} of Appendix for Z = 0.0005). 
Systems with MS-MS accretors typically receive a larger amount of mass than BH-BH accretors, because the tertiary star is typically more massive in the former case. 
This is because for the tertiary to fill its Roche lobe, while the inner stars are still on the MS, the initial tertiary star needs to evolve faster and hence be more massive than the MS stars.
The relative transferred mass expressed as a fraction of the total mass of the inner binary ($M_{\rm transferred}/M_{\rm tot,inner}$) has the same maximum value ($\sim$ 0.5) for both BH-BH and MS-MS accretors (see grey histogram in left panel of Fig. \ref{fig:m_transferred_ext_z_high}).

\subsubsection{Formation of circumbinary disc}
\label{subsub:cbd_or_no_cbd}

As explained in section \ref{subsec:TMT_method}, whether a TMT episode is accompanied by a formation of a circumbinary disc can have important consequences for the evolution of the inner orbit.
In this subsection, we discuss how common it is for TMT systems to develop a circumbinary disc at the onset of the mass transfer episode.

We find that about 63 per cent of all TMT systems develop circumbinary discs in our moderate metallicity model, while in the rest TMT proceeds in a ballistic fashion.
Systems in which a circumbinary disc is formed during the TMT phase typically have larger outer pericenters at the onset of the mass transfer ($a_{\rm p, out} \approx$ 300-6000 $R_{\odot}$)  than in those where TMT proceeds in a ballistic manner ($a_{\rm p, out} \approx$ 100-600 $R_{\odot}$).

 TMT with circumbinary disc is more prevalent at lower metallcities. About 74 per cent of all TMT systems develop circumbinary discs at Z = 0.0005. This occurs because the ratio of the inner and the outer orbital separation decreases less by the onset of the mass transfer phase due to weaker stellar winds (see equation \ref{eq:accretion_disc}).

TMT episodes with inner BH-BH binaries are somewhat more likely to occur in a ballistic fashion than with MS-MS inner binaries. About 45 (23) per cent of TMT systems with BH-BH inner binaries do not develop circumbinary discs at Z = 0.005 (Z = 0.0005), while 32 (27) per cent of TMT episodes with MS-MS inner binaries occur in a ballistic fashion. This is mainly because the inner apocenter to outer pericenter ratios at the onset of TMT are typically higher for inner BH-BHs than for inner MS-MS binaries (see equation \ref{eq:accretion_disc}). This difference is due to Wolf-Rayet winds, supernova kicks and possible ZLK oscillations that BH-BH inner binaries experienced prior to the TMT episode.

\subsubsection{Three-body dynamics prior to TMT}
\label{subsub:3bd_for_tmt}
Three-body dynamics can increase the eccentricities of the inner binary. This can, for example significantly decrease the coalescence time due to GWs \citep[e.g.][]{MillerHamiltion2002,Blaes2002,Wen2003, Thompson2011}.

\begin{figure}
\includegraphics[width=\columnwidth]{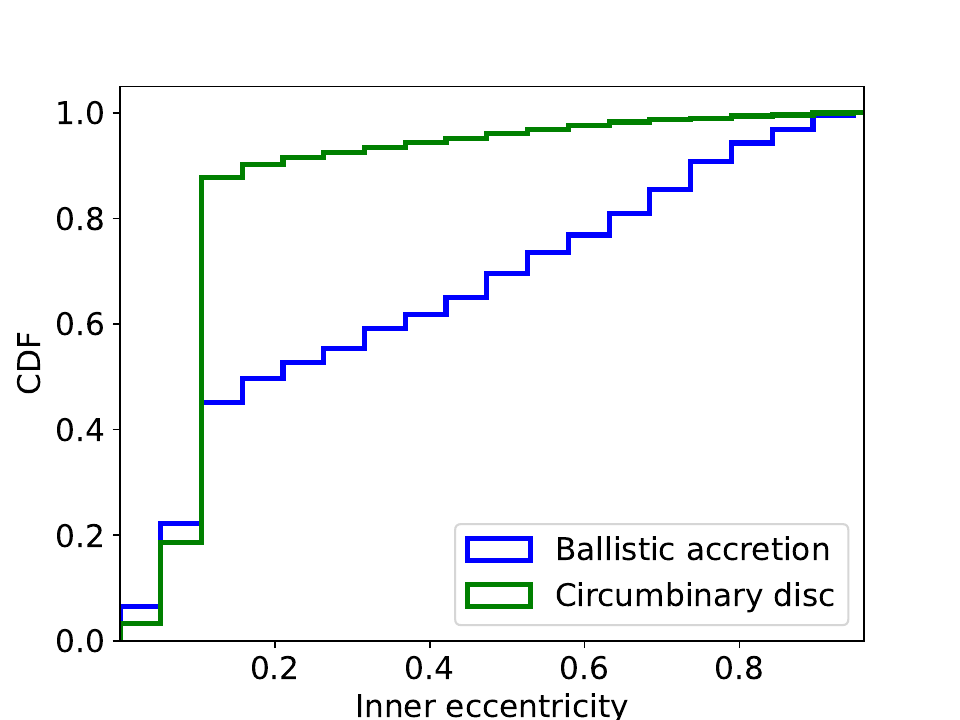}
\caption{The cumulative distribution of the inner binary eccentricities of systems experiencing TMT with BH-BH accretors at the onset of the mass transfer phase at Z = 0.005. The green curve shows the systems with accretions disc formed during the mass transfer phase. The blue curve shows the systems where ballistic accretion occurs.
}
\label{fig:inner_ecc_teritary_rlof}
\end{figure}

Three-body dynamics are almost always suppressed during the MS phase of the inner binaries due to the strong tides (see also section \ref{subsec:example_evol}). Consequently, the inner orbits of TMT systems with MS-MS inner binaries are always circular at the onset of the mass transfer episode. On the other hand, this is no longer the case when the inner stars are in their post-MS. 
In Fig. \ref{fig:inner_ecc_teritary_rlof}, we show the cumulative distribution of the inner binary eccentricities at the onset of the mass transfer phase of TMT systems with BH-BH accretors at Z = 0.005.
We see that systems without circumbinary discs tend to have eccentric inner orbits at the onset of mass transfer. The high eccentricities are caused by ZLK cycles during the post-MS evolution of the inner binary. 
About 40 per cent of such triples have $e_{\rm in}\gtrsim 0.4$ at this stage. 
This is in contrast with the systems with circumbinary discs; about 90 per cent of the systems have eccentricities $e_{\rm in} \lesssim0.1$. The difference is due to the smaller inner period to outer period ratios that systems without circumbinary discs have (see equation \ref{eq:zlk_timescale}).
In our low metallcity model, high eccentricites at the onset of TMT are much less common (see Fig. \ref{fig:inner_ecc_teritary_rlof_z_low} in section of \ref{sec:additional_figures} of Appendix). 
For these systems the inner period to outer period ratio does not increase significantly because of the weak stellar winds. 

\subsubsection{Significance of Humphreys-Davidson limit on the predicted rate of TMT among CHE triples}
\label{subsub:hdlimit}

 There are several poorly understood aspects of stellar physics that make our predictions for the occurrence rate of TMT episodes of CHE triples uncertain, such as mixing processes that trigger CHE, stellar winds, and the radial evolution of classically evolving tertiary stars. The latter is especially uncertain for stars that eventually reach the so-called Humphreys-Davidson (HD) limit (roughly stars with $M_{\rm ZAMS} \gtrsim 40\,M_{\odot}$). This empirically determined limit represents a boundary beyond which no cool supergiants are observed in the Milky Way and the Magellanic Clouds \citep[see e.g.][]{Humphreys1979, Davies2018, DaviesBeasor2020}. 
 While there is still no consensus regarding the origin of the HD limit, its existence could imply that the most massive stars remain relatively compact throughout their lives and never become red supergiants or spend only a very short fraction of their CHeB lifetimes as red supergiants. If the former case is true, the maximum radius of the tertiary stars are significantly overestimated in our models, and therefore so is the occurance rate of TMT episodes among CHE triples. 
 
Several possible explanations for the HD limit have been proposed in the recent years. One of the most common hypothesis is that strong (steady-state or eruptive) stellar winds strip the hydrogen envelopes of stars near the HD limit \citep[e.g.][]{Lamers1988,  Smith2004, VinkSabhahit2023}.
Other studies suggest that different mixing mechanisms could be responsible for preventing the redward evolution of massive stars \citep[see e.g.][]{Langer1995}. More specifically, this could be caused by enhanced energy transport in convective regions near the Eddington limit in evolved stars \citep[][]{Sabhahit2021}, or by efficient semiconvection \citep[][]{Higgins2020}, or by strong convective overshooting \citep[][]{Gilkis2021, Schootemeijer2019}.

Below, we determine the fraction of TMT episodes that occur with donors beyond the HD limit. 
For this, we assume the following metallicity independent HD limit \citep[see e.g.][]{Hurley2000}:
\begin{equation}
\label{eq:HDlimit}
    \left(\frac{L}{L_{\odot}}\right)_{\rm HD} = \rm{max}(3 \cdot 10^{-3} \cdot (T_{\rm eff}/K)^2, 10^{5.5}).
\end{equation}
We note that the metallicity dependence of the HD limit is still debated \citep[see e.g.][]{Davies2018, DaviesBeasor2020}.

64 (67) per cent of the TMT episodes occur with donor stars that have already crossed the HD limit at Z = 0.05 (Z = 0.0005). This implies that, if classically evolving stars indeed never evolve beyond the HD limit, only about 20 (17) per cent of CHE triples would initiate TMT in our moderate (low) metallicity model instead of 55 (52) per cent. 
Furthermore, under such conditions, 32 (31) percent of the TMT episodes would occur with double MS-MS inner binaries, and  62 (60) percent with BH-BH inner binaries at Z = 0.05 (Z = 0.0005).

We do not run additional simulations to predict how CHE triples would evolve, if the radial expansion of the tertiary stars would be restricted by the HD limit.
We note, however, that at the onset of the mass transfer episode, ZLK oscillations are not quenched in 22 per cent of CHE triples experiencing TMT episodes with BH-BH inner binaries and with donor stars beyond the HD limit at Z = 0.005. The typical ZLK timescales of these systems range from few tens to a few hundreds of Myr. This suggests, that in these systems,
the tertiary star could still affect the evolution of the inner binary, even if the radial expansion of the tertiary star was restricted by the HD limit and therefore the TMT episode did not take place.
Specifically, ZLK oscillations could lead to the formation of GW sources with considerably shorter delay times than those formed from isolated CHE binaries.

Whether this would really occur also depends on the exact mechanism responsible for the HD limit. In particular, if strong, steady-state stellar winds are responsible, the mass loss rates have to be $\dot{M}_{\rm LBV}\sim10^{-3}\,M_{\odot}\ \rm{yr}^{-1}$, which is about an order of magnitude larger than assumed in our models \footnote{This mass loss rate follows from the following considerations. Stellar models of \citet{Hurley2000} predict that at metallicities that are typical for the Milky Way and the Magellanic Clouds (i.e. $Z\gtrsim 0.005$), stars typically reach the HD limit, when they are in their very short Hertzsprung gap phase and initiate core-helium burning well beyond the HD limit, as red supergiants (i.e. $T_{\rm eff}\lesssim4800\,K$). Therefore, if steady-state stellar winds are responsible for preventing the formation of red supergiants for stars that cross the HD-limit, the envelope stripping has to occur in a timescale that is similar or shorter than the lifetime of the Hertzsprung gap phase (i.e. $t\lesssim 10^4\rm{yrs}$). Since the typical hydrogen envelope mass for the most massive stars is a few tens of $M_{\odot}$, the mass loss rate has to be $\dot{M}_{\rm LBV}\sim10^{-3}\,M_{\odot}\ \rm{yr}^{-1}$ in this case. We have also confirmed this by varying the LBV mass loss rate of single stars and checking the corresponding stellar tracks in the Hertzsprung-Russell diagram. Similar conclusions can also be found in \citet{Mennekens_2014}.}. As a result of such intense mass loss, the outer orbit would significantly widen soon after the tertiary star evolves off the MS. This, along with the decreased mass of the tertiary star, would lead to a significant increase of the ZLK oscillation timescale, which could potentially quench the three-body dynamics already after few Myr after ZAMS. On the other hand, if the radial expansion is prevented by various mixing processes instead of stellar winds, then the mass loss rates are comparable with that of our models. This could lead a non-negligible systems in which the tertiary remains detached but ZLK oscillations still dominate the evolution of the inner binary.

\subsection{Unbound systems} 
\label{subsec:unbound_systems}
In this channel, one of the stars in the triple becomes unbound as a result of core-collapse.
We distinguish systems based on whether this occurs via PISN or via classical core collapse \citep[e.g. ][]{Fryer_2012}. As shown in Table \ref{tab:secondnumbers}, PISN does not occur in our moderate metallicity model, whereas at Z = 0.0005, it becomes quite prevalent; about 84 per cent of the unbound systems occur due to PISN.

If the triples becomes unbound as a result of a classical core-collapse, we further distinguish whether it is due to the core-collapse occuring in the inner binary (97 per cent of all classical core-collapse systems at Z = 0.005 and 99 at Z = 0.0005) or of the tertiary star (3 per cent at Z = 0.005 or 1 per cent at Z = 0.0005). As the inner binary consists of CHE stars, they have
large initial masses (i.e. $M_{\rm ZAMS}\gtrsim 30\,M_{\odot}$) and furthermore they develop more massive CO cores than their classically evolving counterparts. Therefore, they get weak (if any) natal kicks when they form BHs according to our implemented natalk kick prescription 
Yet weak natal kicks, or even completely symmetrical instantaneous mass losses due to neutrino losses \citep[which we assume to be 10 per cent of the pre-collapse mass according to][]{Fryer_2012} can unbind the tertiary star, if the outer star has high eccentricities. We find that in systems in which one of the stars becomes unbound due to the core-collapse in the inner binary, the outer eccentricities are large, about 70 per cent of them $e_{\rm out} \geq 0.8$. In the vast majority of the cases (about 99 per cent of such unbound systems), only the tertiary is ejected, while the inner binary remains bound.

If the triple becomes unbound due to the core-collapse of the tertiary star and with low outer eccentricity, it almost always occurs as a result of a strong natal kick. Consequently, most of such unbound systems have initial tertiary masses of $M_{\rm out,ZAMS}\approx 8$-$25\,M_{\odot}$ (see also discussion in section \ref{subsec:origin}), as these systems are expected to receive the largest kicks according the supernova prescription of \citet{Fryer_2012} .

\subsection{Systems which become dynamically unstable}
\label{subsec:dynin}
These triples typically have very short initial outer pericenters ($a_{\rm p, out, ZAMS} \approx 70$-$400\,R_{\odot}$) and therefore are very close to the stability limit at ZAMS.  Such systems can transition to non-secular or non-hierarchical evolution, if $a_{\rm in}/a_{\rm out}$, $e_{\rm out}$ or $q_{\rm out}$, significantly increases during evolution \citep[see][]{MardlingAarseth2001}. Among CHE triples, there are primarily two processes that can trigger this change: stellar winds and core collapse. 

 If the relative wind mass loss rate (e.g. $\dot{M}/M$)  in the inner binary is higher than that of the tertiary star, $a_{\rm in}/a_{\rm out}$ and $q_{\rm out}$ will increase, which can prompt the triple to experience a dynamical instability \citep[see also][]{Kiseleva1994,Iben1999,PeretsKratter2012, Toonen2022}. 30 per cent of the systems of this channel destabilise due to stellar winds and the destabilisation occurs when the stars of the inner binary are in their post-MS phase. At this stage, the inner stars experience strong Wolf-Rayet winds, while the tertiary star is still on the MS with significantly lower mass loss rates. 

 In the remaining 70 per cent, 
the instability sets in due to core-collapse in one of the inner stars. 
As noted in section \ref{subsec:remnant_formation}, CHE stars typically form BHs via direct collapse, such that  $q_{\rm out}$ only increases slightly. Furthermore, the direct collapse 
is expected to be accompanied by a weak Blauw-kick due to neutrino losses such that $a_{\rm in}/a_{\rm out}$ and $e_{\rm out}$  only increase significantly, if the inner or the outer pre-core-collapse orbits are eccentric, respectively.
The pre-core-collapse inner orbit is eccentric in 72 per cent of the systems of this channel. This high inner eccentricity is caused by ZLK oscillations.  In the remaining 28 per cent, three body-dynamics is not efficient in driving up the eccentricity because the mutual inclination is outside of the critical Kozai range \citep[see e.g.][]{Naoz2016}. Therefore, the core collapse occurs in circular inner orbits. These systems still become unstable during the BH formation, because 1) either $a_{\rm in}/a_{\rm out}$ already increased strongly  due to stellar wind mass losses before the BH formation or 2) the outer orbit is eccentric and the core collapse occurs, while the tertiary star is near the outer pericenter (leading to a significant increase in $e_{\rm out}$).

The occurrence rate of this channel is strongly dependent on metallicity (3.5 per cent of all CHE triples at Z = 0.005 and 0.7 per cent at Z = 0.0005, see Table \ref{tab:secondnumbers}). This dependence is due to the reduced strength of stellar winds and ZLK oscillations (which are responsible for any eccentricity in CHE inner binaries)  at lower metallicities. 

\section{The origin of each evolutionary channel}
\label{subsec:origin}
In this section, we discuss the initial parameters of the triples from each evolutionary channel introduced in section \ref{sec:mainres}. We find that initial parameters can be used as a proxy to determine the final evolutionary outcome of CHE triples. 
In particular, the evolutionary outcome can be parameterised by the initial mass and orbital separation of the tertiary star. The parameters of the inner binary play a less important role in this regard, as the parameter space for CHE inner binaries is already quite reduced.
We illustrate this in the left panel of Fig. \ref{fig:grid_z0d005} by showing an ensemble of CHE triples at $Z = 0.005$, in which the parameters of the inner binary are the same, but the mass and the orbital separation of the tertiary star are varied (therefore this grid represents only a small subset of the entire CHE population discussed in section \ref{sec:mainres}).
The inner binary consists of two 70 $M_{\odot}$ stars and with a circular initial orbit with $a_{\rm in, ZAMS} = 22.4\,R_{\odot}$ (similarly to the example systems discussed in section \ref{subsec:example_evol}). The initial tertiary mass ranges from 5 to 100$\,M_{\odot}$, while $a_{\rm out,ZAMS}$
ranges from 200 to $10^4\,R_{\odot}$.

\subsection{Initial parameters of systems of different evolutionary channels}

 The majority of the triples shown in the left panel of Fig. \ref{fig:grid_z0d005} experience TMT episodes. Their initial outer orbital separations are relatively short and range roughly from 100 to 3300 $R_{\odot}$.  The evolutionary phase of the inner stars at the onset of the TMT episode depends on the initial mass of the tertiary star. For the systems shown in the left panel of Fig. \ref{fig:grid_z0d005}, the inner binary at the onset of TMT comprise of BHs, if $M_{\rm out, ZAMS}\lesssim59\,M_{\odot}$, helium stars, if $59\,M_{\odot} \lesssim M_{\rm out, ZAMS}\leq 70\,M_{\odot}$, and MS stars, if $M_{\rm out, ZAMS} \geq 70\,M_{\odot}$. The majority (53 per cent) of the TMT systems in the left panel of Fig. \ref{fig:grid_z0d005} have a BH-BH inner binaries. For the entire population of CHE triples presented in section \ref{sec:mainres}, the same percentage is smaller (i.e 31 per cent) at the same metallicity (see Table \ref{tab:mainnumbers}). As shown in Fig. \ref{fig:m2_bh_frac_z_high}, this quantity (i.e. the ratio of the number of TMT systems with BH-BH inner binaries and the number of all TMT system) scales proportionally to the initial mass of the secondary star in the inner binary. This means that TMT episodes occur more frequently with BH-BH accretors among CHE triples with more massive inner stars. This is due to our assumptions about the initial distribution of the triples (section \ref{subsec:init}). If the TMT occurs towards a BH-BH inner binary, the tertiary has to be initially the least massive in the triple. With increasing $M_{\rm 2,ZAMS}$, the fraction of triples for which $M_{\rm 2,ZAMS} > M_{\rm out,ZAMS}$ increases because of our assumptions of a maximum initial stellar mass of $M_{\rm ZAMS, max} = 100\,M_{\odot}$ and a flat outer mass ratio distribution.

In 15 per cent of the 
triples shown in the left panel of Fig. \ref{fig:grid_z0d005}, the inner binary merges before BH formation or before a TMT episode occurs. All such mergers in the grid occur between two helium stars, and are due to ZLK oscillations that arise when the stars of the inner binary evolve off the MS. The initial outer orbital separations in this channel are very short, i.e. 200 to 241$\,R_{\odot}$, while the tertiary masses range between $32 \leq M_{\rm out, ZAMS}/M_{\odot}\leq 68$. For lower tertiary masses ($M_{\rm out, ZAMS}<32\ M_{\odot}$), the ZLK oscillations are not strong enough to boost the inner eccentricity and cause a mass transfer episode in the inner binary. 
For larger tertiary masses ($M_{\rm out, ZAMS}>70\ M_{\odot}$), the tertiary typically fills its Roche-lobe  before the stars of the inner binary evolve off the MS. However, during the main sequence phase of the inner stars, the effects of ZLK cycles are quenched and consequently no mergers are prompted by three-body dynamics before the tertiary initiates a TMT episode.

Unbound systems shown in Fig. \ref{fig:grid_z0d005} have a specific initial tertiary mass range of $8\lesssim M_{\rm out, ZAMS}/M_{\odot} \lesssim 25$. Tertiary stars in this mass range receive strong natal kicks during remnant formation, which can lead to unbinding the triple system. Unbound systems from the entire population of CHE triples, however, are not confined to this specific range of initial tertiary masses. In fact, as it was mentioned in section \ref{subsec:unbound_systems}, the majority of CHE triples becomes unbound due to the core-collapse occurring in the inner binary. The main reason why this does not occur in any of the triples shown in Fig. \ref{fig:grid_z0d005} is because these systems have circular outer orbits and therefore a weak Blauuw kick does not change the outer orbit significantly.

Triples of the no post-MS MT channel in the left panel of Fig. \ref{fig:grid_z0d005} have initial outer orbits $a_{\rm{out}}\gtrsim 2000$-$3000\ R_{\odot}$. 
Their initial tertiary mass is also typically outside of the range of $\sim$8-25 $M_{\odot}$, such that the system does not dissociate due to SN kicks.
As we shown in the next subsection, three-body dynamics are not important for the evolution of these systems.

In left panel of Fig. \ref{fig:uter_perti_types_z_high}, we show the inital pericenter ($a_{\rm outer,ZAMS}$) distribution of the entire CHE triple population for each evolutionary channel at Z = 0.005. As it can be seen, the range of initial pericenters are in agreement with those shown in Fig. \ref{fig:grid_z0d005} for all channels except for the unbound systems 
This again confirms that the parameters of the tertiary star play the most important role in determining the evolutionary path of a CHE triple. As shown in left panel of 
Fig. \ref{fig:uter_perti_types_z_high}, the range of $a_{\rm outer,ZAMS}$ of systems with TMT episodes increases with decreasing metallicity. At lower metallicity, the stellar winds are weaker and consequently, the outer orbit widens less. Therefore, the maximum $a_{\rm outer,ZAMS}$ at which the tertiary stars can still fill their Roche-lobes also increases with decreasing metallicity.

\subsection{Initial parameters of triples with three-body dynamics}
\begin{figure*}
  \includegraphics[trim=0 1.5cm 1.5cm 1.5cm, clip,width=\textwidth]{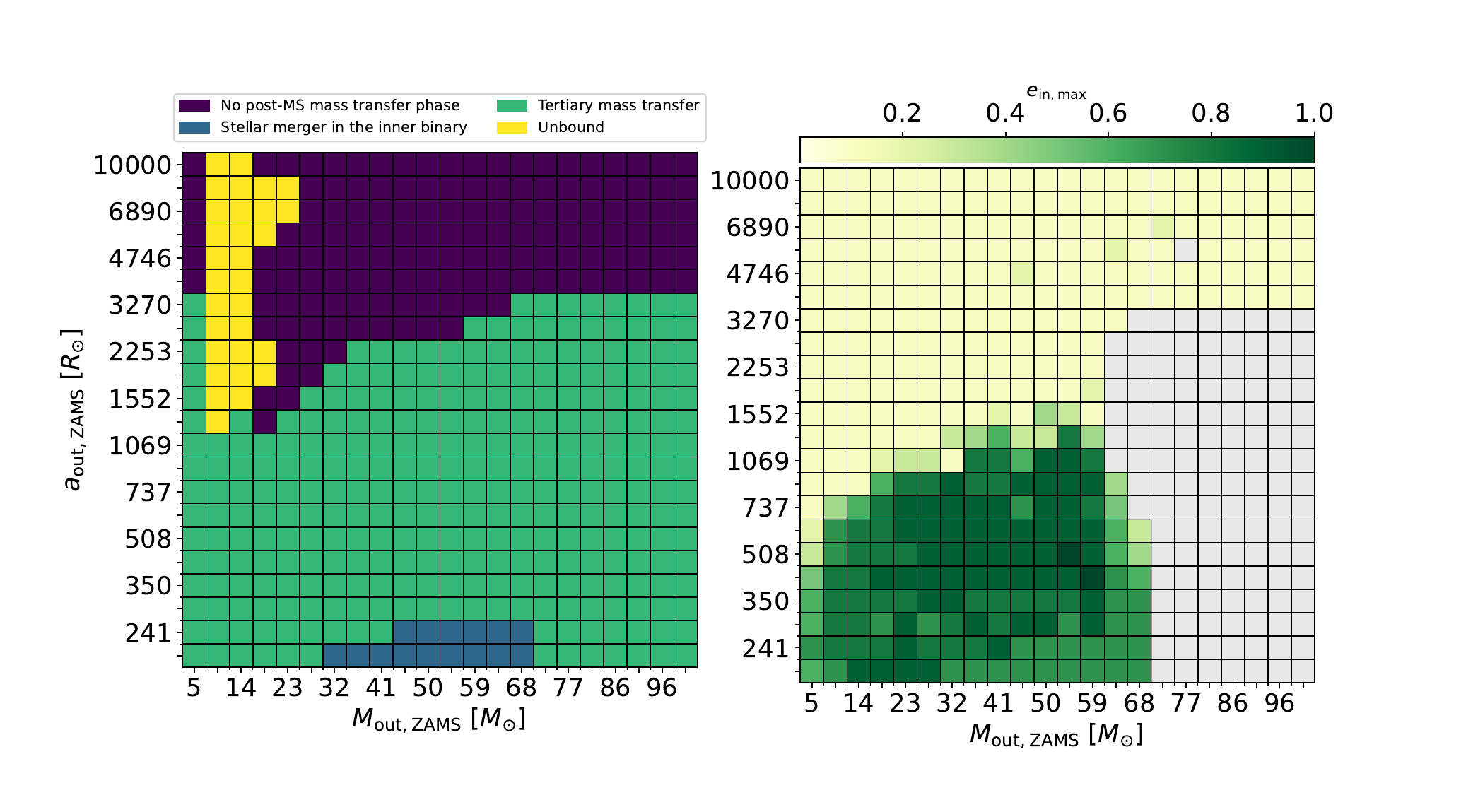} 
  \caption{Left panel: we show the evolutionary outcome of collection of triples with fixed inner binary parameters and different parameters for the tertiary at a metallicity $Z = 0.005$. The parameters of the inner binary are the same for each system shown in the grid, i.e. $M_{\rm 1, ZAMS} = M_{\rm 2, ZAMS} = 70,M_{\odot}$, $a_{\rm in, ZAMS} = 22.4\,R_{\odot}$, $e_{\rm in} = 0$. 
  The initial tertiary mass $M_{\rm out}$ ranges from 5 to 100$\,M_{\odot}$ on a linear scale, while $a_{\rm out, ZAMS}$  ranges from 200 to 1000$\,R_{\odot}$ on a logarithmic scale. 
  We assume zero outer eccentricity and an initial inclination of $90^{\circ}$. Right panel: we show the maximum eccentricity of the inner binary reached during the evolution. The parameters of the inner binary are the same as in left panel}
  \label{fig:grid_z0d005}
\end{figure*}

\begin{figure}
\includegraphics[trim=0 0 0.0cm 0, clip,width=\columnwidth]{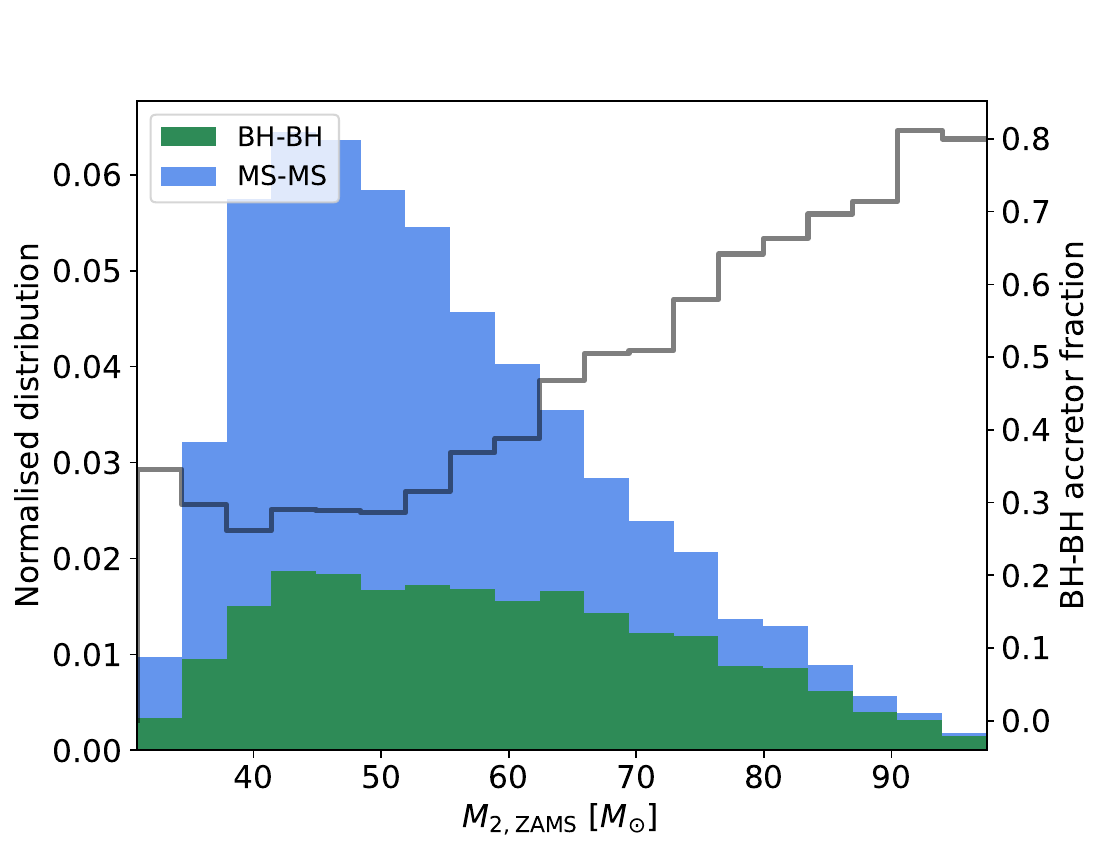}
\caption{The distribution of the initial secondary masses of the entire CHE triple population that undergo tertiary mass transfer at Z = 0.005. We only show those systems which have either BH-BH or MS-MS accretors. The grey unfilled histogram, with the corresponding secondary x-axis on the right hand side, shows the number of systems with BH-BH accretors as a fraction of all systems undergoing tertiary mass transfer phases. }
\label{fig:m2_bh_frac_z_high}
\end{figure}

In the right panel of Fig. \ref{fig:grid_z0d005}, 
we show the maximum eccentricities  that the inner binaries reach during their evolution ($e_{\rm in, max}$).
About 29 per cent of the triples shown in the right panel of Fig. \ref{fig:grid_z0d005}  reach $e_{\rm in, max}\geq0.4$ due to ZLK cycles.
In all of these triples, the tertiary star eventually fills its Roche-lobe (although in some cases, the inner binary merges first).

For the systems shown in Fig. \ref{fig:grid_z0d005}, ZLK cycles are efficient when $a_{\rm out, ZAMS}\lesssim1200\,R_{\odot}$ and $M_{\rm out, ZAMS}\lesssim70\,M_{\odot}$. When the outer orbit is $a_{\rm out, ZAMS}\gtrsim1200\,R_{\odot}$, the ZLK cycles are quenched by various short range forces (e.g. precession caused by tides or general relativistic effects). If $a_{\rm out, ZAMS}\lesssim1200\,R_{\odot}$ but $M_{\rm out, ZAMS}\gtrsim70\,M_{\odot}$, the tertiary star fills its Roche-lobe while the stars in the inner binary are still on the MS. The inner binaries of these triples do not develop high eccentricities, as ZLK cycles are quenched during MS due to strong tides (see section \ref{subsec:example_evol}), and TMT episode with MS-MS accretors are expected to result in the merger of the inner binary (see section \ref{sec:GW}).

The right panel of Fig.  \ref{fig:grid_z0d005} also shows that $e_{\rm in, max}$ does not decrease smoothly with decreasing outer orbital separations, instead it drops rather abruptly across $a_{\rm out, ZAMS}\approx1200\,R_{\odot}$. 
Triples with $a_{\rm out, ZAMS}\approx1200\,R_{\odot}$ reach very large inner eccentricties ($e_{\rm in, max} \approx 0.9$), while at slightly larger orbital separations  (i.e. $a_{\rm out, ZAMS}\approx1500\,R_{\odot}$) the ZLK cycles are completely quenched.

\begin{figure*}
\includegraphics[trim=0 0 1cm 0, clip,width=\columnwidth]
{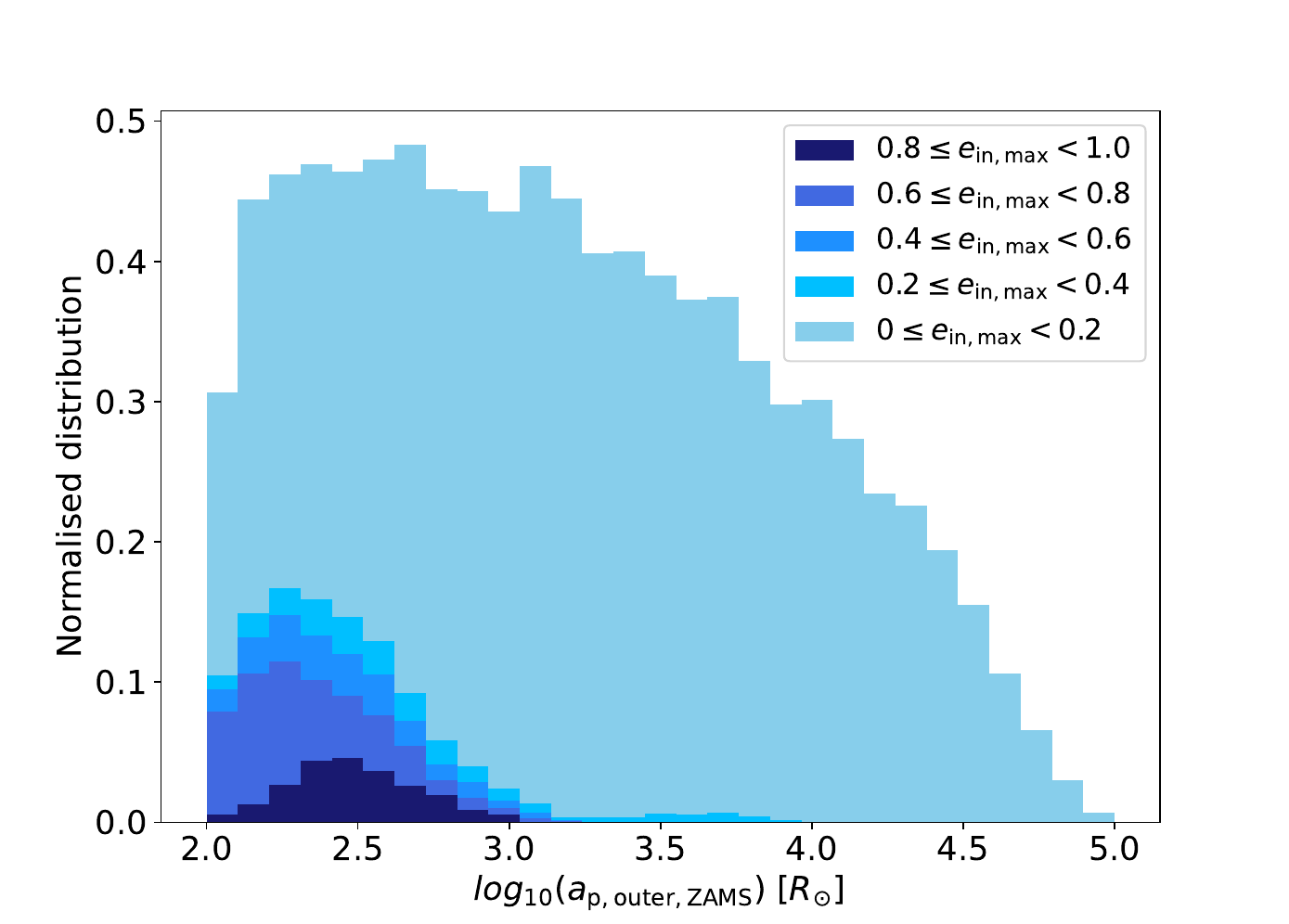} \hfill
\includegraphics[trim=0 0 1cm 0, clip,width=\columnwidth]
{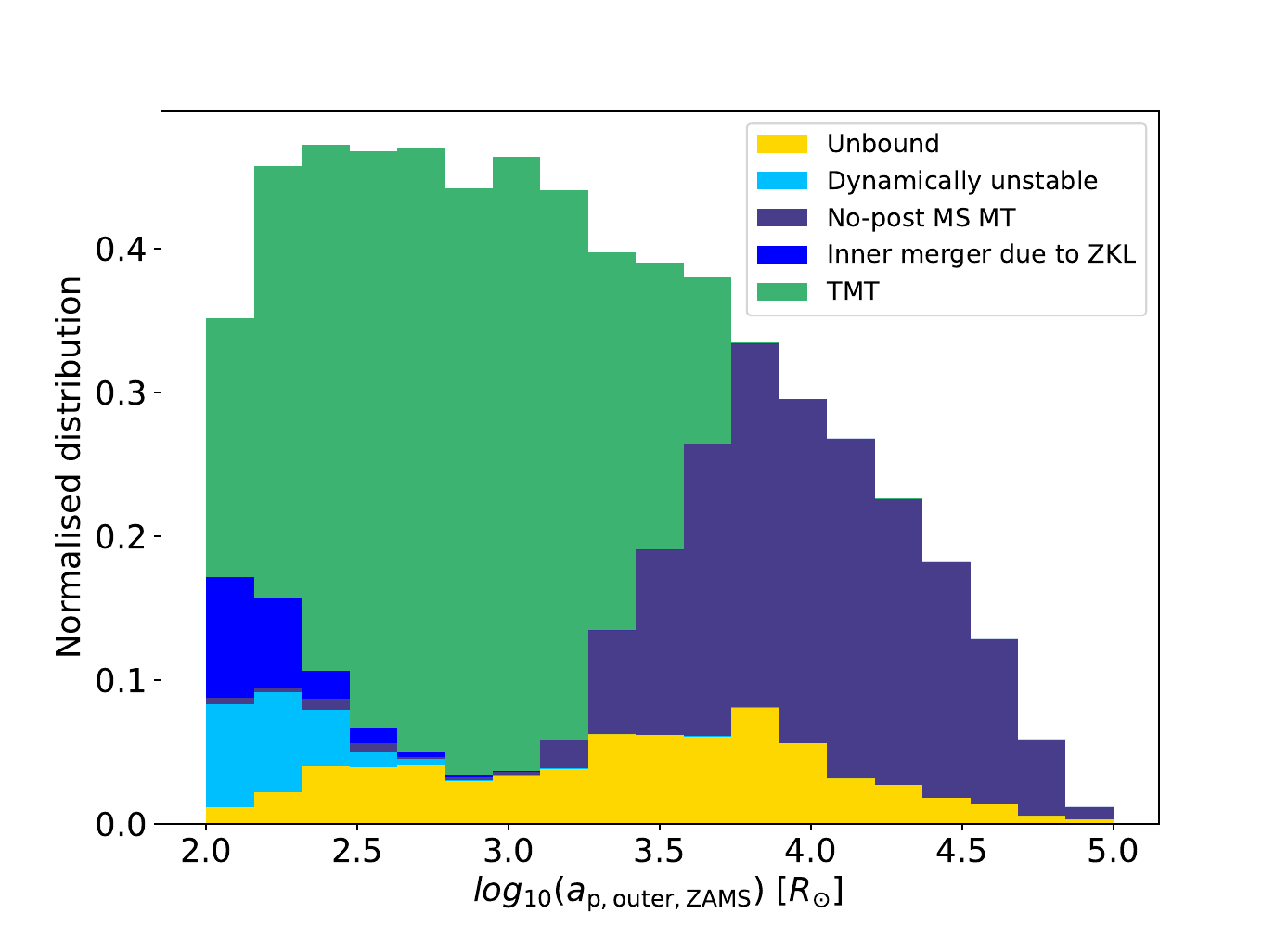}
\caption{Left panel: The distribution of the initial pericenter of a few selected evolutionary types of the entire CHE triple population at Z = 0.005. The histograms have been normalised to the full population of the CHE evolving triples. The histograms shown are stacked. Each colour represents a different evolutionary type. For clarity we do not show all the types introduced in section \ref{sec:mainres}, see text for discussion. Right panel: The distribution of the initial outer pericenter of the entire CHE triple population at Z =  0.005. We distinguish systems based on the maximum eccentricity of the inner binary reached during their evolution. The histograms are normalised to one and stacked. } 
\label{fig:uter_perti_types_z_high}
\end{figure*}

These above mentioned effects are qualitatively also true for the entire CHE triple population presented in section \ref{sec:mainres} (see right panel of  Fig. \ref{fig:uter_perti_types_z_high}).
At Z = 0.005, the ZLK oscillations are only efficient, if $a_{\rm p, out, ZAMS} \lesssim 1200\,R_{\odot}$.
This implies that three-body dynamics are only relevant for those triples, in which the tertiary star would eventually fill its Roche-lobe (compare right and left panel of Fig. \ref{fig:uter_perti_types_z_high}). Consequently, if the tertiary in a CHE triple remains detached throughout its evolution, the evolution of the inner binary will almost always be kinetically decoupled from the tertiary star. If $a_{\rm p, out, ZAMS} \lesssim 1200\,R_{\odot}$, a wide range of inner eccentricites are possible ($e_{\rm in, max}
=$ 0-0.9) for all $a_{\rm p, out, ZAMS}$. In this case, the value of $e_{\rm in, max}$ is primarily determined by the mutual inclination of the triple \citep[see also e.g.][]{Kassandra2017}. 


In our low metallicity model ($Z = 0.0005$) the maximum initial outer pericenter at which three-body dynamics are still relevant is lower compared to our moderate metallicity model (right panel in Fig. \ref{fig:uter_perti_types_z_low} in section \ref{sec:additional_figures} of Appendix). At such low metallicities, stellar winds do not widen the orbit of the inner binary significantly and thus the timescales of the ZLK cycles do not decrease as much as at Z = 0.005.



%

\section{Gravitational waves sources}
\label{sec:GW}
We now discuss the possible formation channels of GW sources that originate from CHE triples and their properties. 
In section \ref{subsec:rates_of_GW} we predict the merger rate densities and compare them to that of GW sources from isolated CHE binaries. For this, we assume two test populations with different stellar multiplicity fractions.
One population is composed of only single and binary stellar systems (i.e. with stellar multiplicity fractions at ZAMS of $f_{\rm single} = 0.3$, $f_{\rm binary} = 0.7$, $f_{\rm triple} = 0$), in the other triples dominate  ($f_{\rm single} = 0.06$, $f_{\rm binary} = 0.21$, $f_{\rm triple} = 0.73$). 

In sections \ref{subsec:gw_isolated_binaries} - \ref{subsubsec:gw_TMT_with_MSMS_accretors} we discuss the properties of each GW formation channel from CHE triples and binaries. These predictions are based on the synthetic populations discussed previously, and in cases where the simulations are stopped 
before the formation of a BH-BH binary, we predict the further evolution of CHE triples beyond the stopping conditions (Section\,\ref{subsec:init}) by applying simple assumptions (as detailed below).

\begin{figure*}
  \includegraphics[trim=0 0 0 0, clip, width=\textwidth]{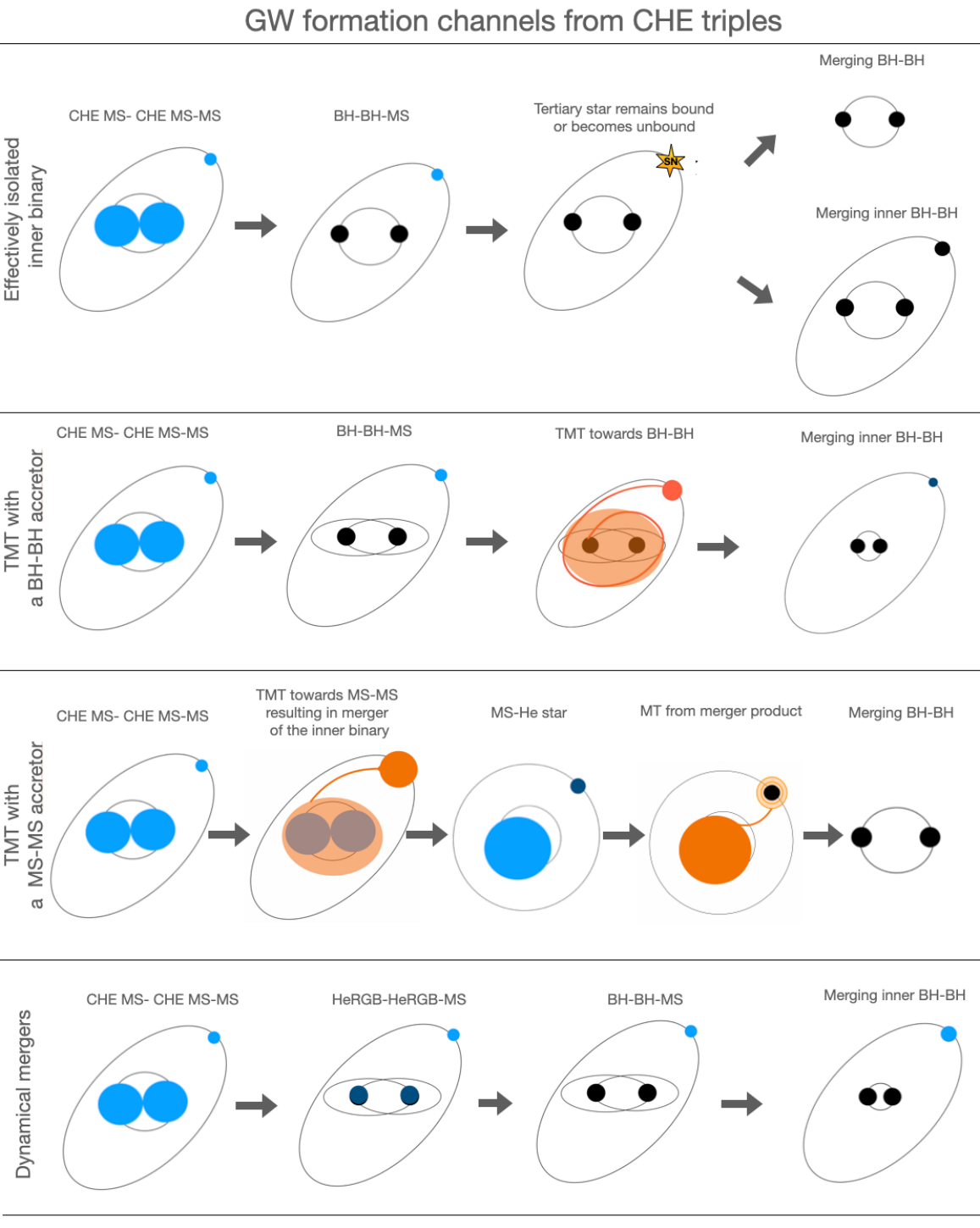}
  \caption{The possible formation channels of merging binary BHs from our CHE triples population. }
  \label{fig:formation_channels_GW}
\end{figure*}

The four main identified formation channels of GW sources within our CHE triple population are (see also Fig. \ref{fig:formation_channels_GW}): 
\begin{itemize}
  \item \textbf{Effectively isolated inner binary:}
    For such triples, three-body dynamics is suppressed by various short-range forces 
    and the tertiary star remains detached throughout the entire evolution. The inner binary therefore evolves effectively as an isolated binary and the properties of these GW sources are indistinguishable from those of the CHE binary channel. There are two ways these systems can form: i) with the tertiary star bound to the triple (systems from the no post-MS MT channel, see section \ref{subsec:systems_without_postMS_mt}) and ii) systems in which the tertiary star becomes unbound from the triple (from the unbound channel discussed in section \ref{subsec:unbound_systems}). For the latter, we assume that the orbit of the inner binary is not affected by the tertiary unbinding from the triple system.


    \item \textbf{TMT with a BH-BH accretor:}
    This channel comprises systems in which the tertiary star fills its Roche-lobe when the inner binary is a BH-BH binary.
     The inner binary components do not coalesce during the TMT phase, but will merge afterwards due to GW emission. 
     In these systems, the tertiary star can affect the evolution of the inner binary in two major ways, via TMT episode and via three-body dynamics (see section \ref{subsec:tmdt}). In section \ref{subsec:TMT_method}, we introduced our assumptions regarding the evolution of the inner orbiy during a TMT episode.

    \item \textbf{TMT with a MS-MS accretor:}
    In this scenario, there are two sequential mergers taking place in the system 
    \citep[see also e.g.][]{Stegmann2022}.
    First, the inner binary merges when the stars are still on the MS as a result of  mass transfer from the tertiary to the inner binary. This reduces the triple to a binary.
    We assume that the merger product of the inner binary evolves further in a classical way (as opposed to CHE). Consequently, the merger product expands and eventually fills its Roche-lobe and transfers mass to the initial tertiary star. 
    The orbit shrinks due to this second phase of mass transfer and as a result, a merging double compact object is formed.
    The second phase of mass transfer is essential. Systems in which no mass transfer takes place after the inner binary merger might form detached BH-BH binaries but are too wide to merge due to GWs within the Hubble time.
   We note that double MS mergers among CHE triples typically occur due to TMT episodes as three-body dynamics are suppressed during the MS phase. 

    

    \item{\textbf{Dynamical mergers:}} 
    In the triples of this channel, ZLK oscillations are very efficient and drive up the inner eccentricities to $e_{\rm in} \approx 0.6$-$0.9$  after the stars of the inner binary have become BHs. Such systems merge due to GW emission within a few Myr. The tertiary remains detached until the inner binary merges and therefore these triples belong to the no post-MS MT channel. As discussed in section \ref{subsec:systems_without_postMS_mt}, these systems are rare.

\end{itemize}
    
\begin{table*}
\caption{Summary of the statistics of GW sources from the population with triples (with $f_{\rm single} = 0.06 $, $f_{\rm binary} = 0.21 $, $f_{\rm triple} = 0.73$), and from the population without triples (with $f_{\rm single} = 0.3 $, $f_{\rm binary} = 0.7 $, $f_{\rm triple} = 0$). The 'of all CHE systems' is the number of systems, expressed as a fraction of all systems containing a binary that contains two, tidally-locked CHE stars. Formation efficiency gives the number of systems expressed as a faction of all stellar systems (see equation \ref{eq:formation_efficiency}). Merger rate density is the merger rate density in the local universe (see equation \ref{eq:event_rate}).}
\label{tab:gwnumbers}
\resizebox{\textwidth}{!}{\begin{tabular}{@{}lccccc@{}}
\toprule
                                   & \multicolumn{1}{l}{\begin{tabular}[c]{@{}l@{}}of all CHE systems\\ at Z = 0.005 {[}\%{]}\end{tabular}} & \multicolumn{1}{l}{\begin{tabular}[c]{@{}l@{}}of all CHE systems\\ at Z = 0.0005 {[}\%{]}\end{tabular}} & \multicolumn{1}{l}{\begin{tabular}[c]{@{}l@{}}Formation efficiency\\  at Z = 0.005\end{tabular}} & \multicolumn{1}{l}{\begin{tabular}[c]{@{}l@{}}Formation efficiency\\  at  Z = 0.0005\end{tabular}} & \multicolumn{1}{l}{\begin{tabular}[c]{@{}l@{}}Merger rate density \\ {[}$\rm{Gpc^{-3}yr^{-1}}${]}\end{tabular}} \\ \midrule
\multicolumn{6}{c}{\textbf{Population with triples}}                                                                                                                                                                                                                                                                                                                                                                                                                                                                                                                                                                                                                                                                                                                                                                                            \\
\textbf{CHE triple channels:}      & \textbf{29.1/31.9}                                                                                     & \textbf{23.7/23.7}                                                                                          & \textbf{$\boldsymbol{6.9\cdot10^{-7}$/$7.2\cdot10^{-7}}$}                                        & \textbf{9.2}                                                                                       & \textbf{12.7/11.8}                                                                                                   \\
- Effectively isolated inner binary       & 19.3                                                                                                   & 12                                                                                                    & $4.6\cdot10^{-7}$                                                                                & $3.0\cdot10^{-7}$                                                                                    & 8.8                                                                                                             \\
- TMT with BBH \& BA (sc. 1/sc. 2) & 3.8/6.6                                                                                                & 2.3/2.3                                                                                                     & $9\cdot10^{-8}$/$1.2\cdot10^{-7}$                                                                & $5.8\cdot10^{-8}$/  $5.8\cdot10^{-8}$                                                                               & 1.4/0.5                                                                                                         \\
- TMT with BBH  \& CBD             & 5.9                                                                                                    & 7.8                                                                                                     & $1.4\cdot10^{-7}$                                                                                & $1.9\cdot10^{-7}$                                                                                  & 2.4                                                                                                             \\
- TMT with MS-MS channel           & 0.3                                                                                                   & 5                                                                                                     & $6\cdot10^{-9}$                                                                                & $1.2\cdot10^{-7}$                                                                                  & 0.2                                                                                                             \\
- Dynamical mergers           & 0.2                                                                                                   & 0.1                                                                                                     & $4\cdot10^{-9}$                                                                                & $2.0\cdot10^{-9}$                                                                                  & 0.05                                                                                                             \\
\textbf{CHE binaries}              & \textbf{8.7}                                                                                           & \textbf{12.9}                                                                                           & \textbf{$\boldsymbol{7.2\cdot10^{-7}}$}                                                          & \textbf{$\boldsymbol{5.2\cdot10^{-7}}$}                                                            & \textbf{11}                                                                                                     \\ \midrule
\multicolumn{6}{c}{\textbf{Population without triples}}                                                                                                                                                                                                                                                                                                                                                                                                                                                                                                                                                                                                                                                                                                                                                                                                                                                                                                                                                                                      \\
\textbf{CHE binaries}              & \textbf{65}                                                                                            & \textbf{88.6}                                                                                           & \textbf{$\boldsymbol{1.2\cdot10^{-6}}$}                                                          & \textbf{$\boldsymbol{1.2\cdot10^{-6}}$}                                                            & \textbf{44.2}                                                                                                   \\ \bottomrule
\end{tabular}}
\end{table*}

\begin{figure}
  \includegraphics[width=\columnwidth]{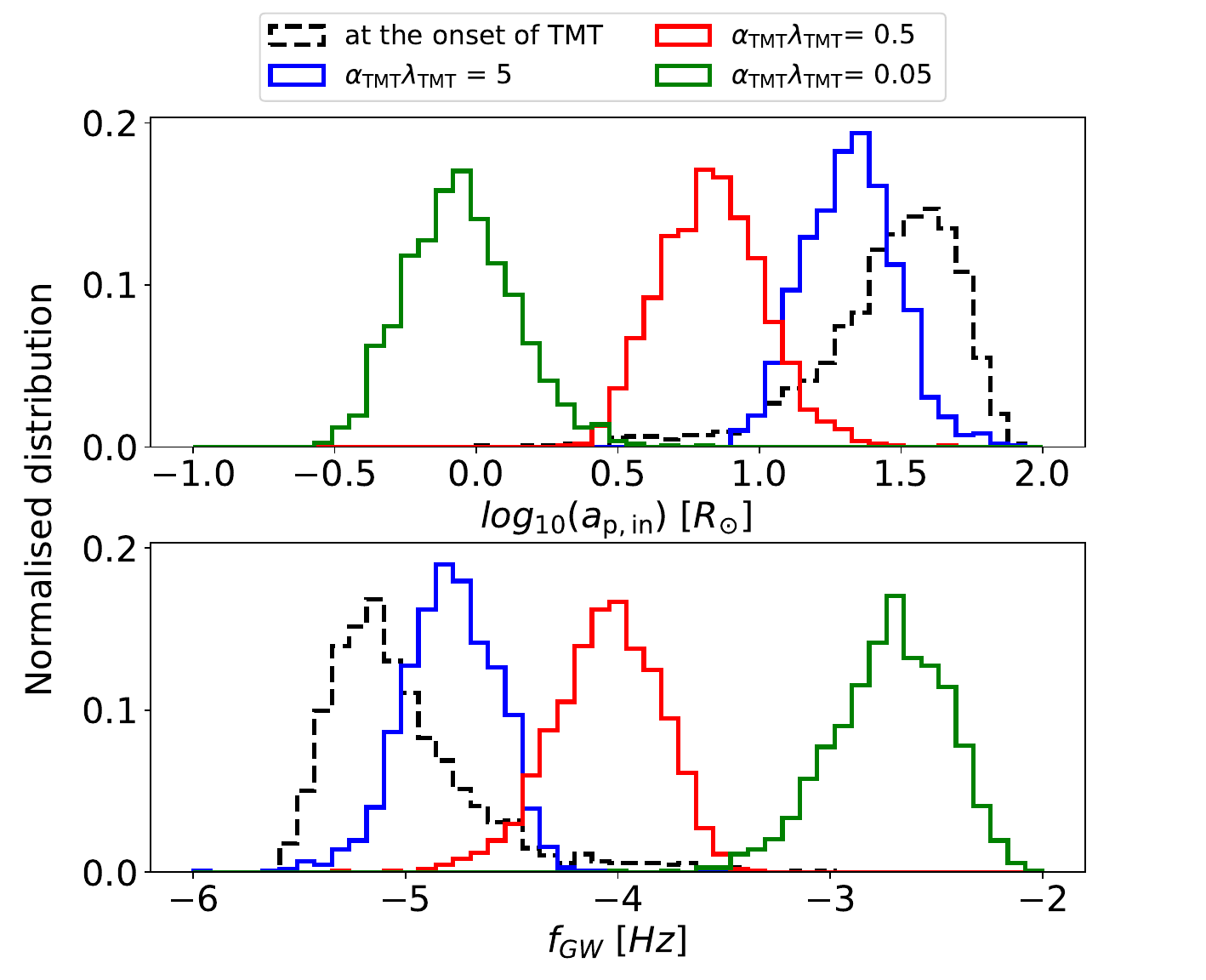}
  \caption{Orbital parameters of GW sources that evolve through the TMT channel with BH-BH accretors and ballistic accretion at Z=0.005. In the upper panel we show the inner pericenter of the inner BH-BH binary ($a_{\rm p, in}$). The black line represents $a_{\rm p, in}$ at the onset of TMT. The coloured lines show the orbital characteristics at the end of the mass transfer for \textit{scenario 2}, where the change in the orbit has been estimated using an energy formalism with three different values of $\alpha_{\rm TMT}\lambda_{\rm TMT}$ (equation \ref{eq:delta_orb} and \ref{eq:binding_energy}).
    Lower panel: we show the frequency of the GW radiated by the inner BH-BH binary before and after the tertiary mass transfer phase.
  }
  \label{fig:a_inner_evol}
\end{figure}

\begin{figure}
    \includegraphics[trim=0 0 1cm 0, clip,width=\columnwidth]{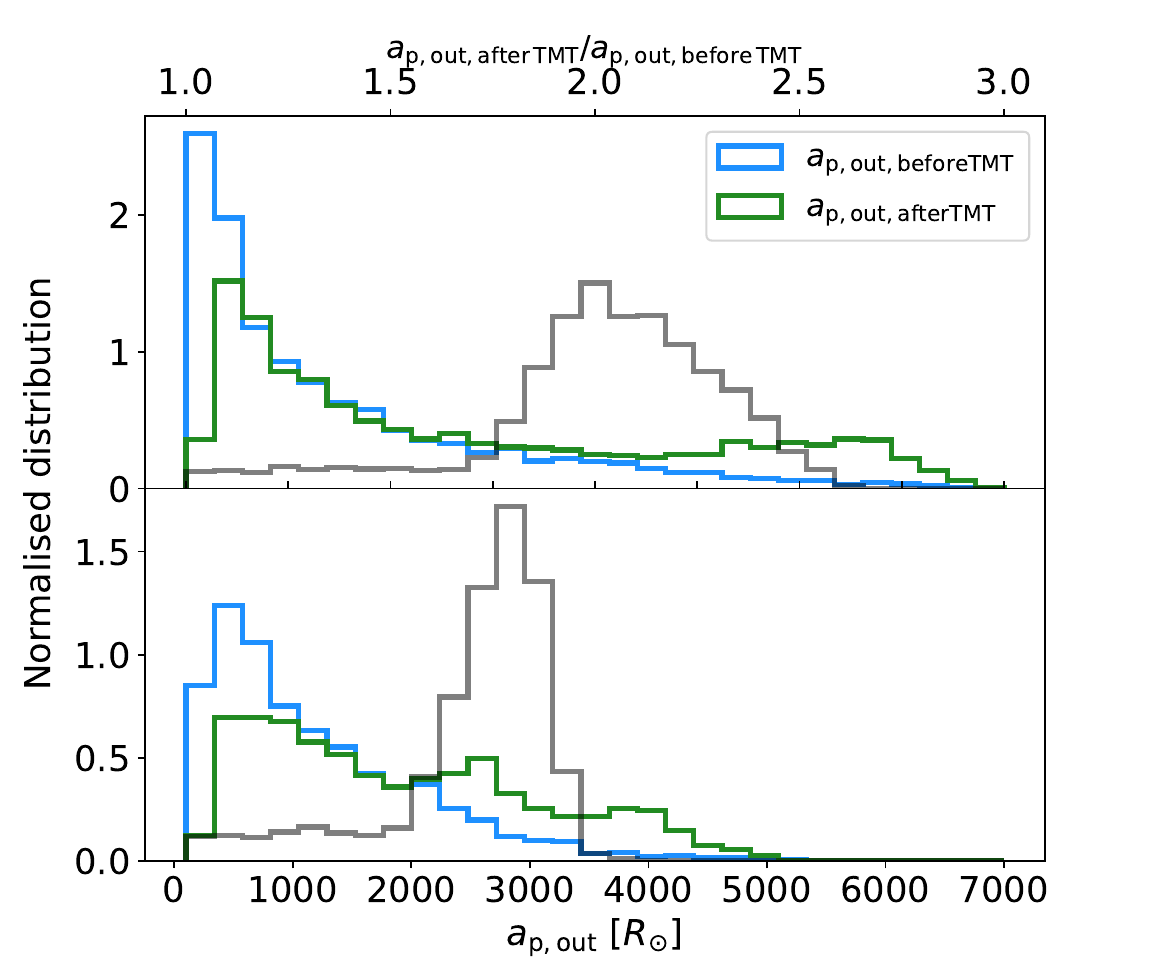}
  \caption{The pericenter of the outer orbit before and at the onset of TMT for MS-MS inner  (upper panel) and BH-BH inner binary accretors (lower panel) as calculated with equation \ref{eq:iso} at Z = 0.005. The grey line shows the ratio of the pericentre after the TMT and at the before the TMT episode (i.e. $a_{\rm p, out, after TMT}/a_{\rm p, out, before TMT}$ with the corresponding values shown in the upper x-axis). We note that for we have normalised these distributions to $10^4$ (such that the numbers on the y-axis are in the order of unity). }  
  \label{fig:a_outer_evol}
\end{figure}

We ignore the possibility of GW source forming in a CHE triple through a stellar merger that do not occur between two MS stars. Such mergers can occur due to TMT or three-body dynamics with (i) helium star-MS binary or (ii) double helium star binaries. We justify the omission of the first type, as they are relatively rare. This type of merger occurs in 0.2-2 per cent of all CHE triples depending on metallicity.  For the second type, the merger product is a helium star, it is not expected to significantly expand and it is unlikely to ever fill its Roche-lobe. Without a phase of mass transfer that leads to orbital shrinkage, the binary remains too wide to merge within a Hubble time. However, if the merger remnant can accrete matter during the TMT phase it could regain a hydrogen-rich envelope,  and expand later in its evolution. For simplicity, we neglect this scenario.

\subsection{Rates of GW mergers}
\label{subsec:rates_of_GW}

In the population without triples, the predicted merger rate density is $R_{\rm merger} = 44.2\,\rm{Gpc^{-3}yr^{-1}}$ (see Table \ref{tab:gwnumbers}). This is about a factor of two higher than predicted by \citet{Riley2021MNRAS.505..663R}, giving a rough agreement given the simplicity of our rate calculation (see discussion in Appendix \ref{subsec:app_norm}). The total merger rate density of the population containing triples is $R_{\rm merger}$ = 23 $\rm{Gpc^{-3}yr^{-1}}$. This is about a factor of two lower than that of the population without triples. There are two reasons for this difference. Firstly, stellar mergers frequently occur in CHE triples, preventing the formation of compact BH-BH binaries. While all CHE binaries form BH-BH binaries, only about 60 (45) per cent of CHE triples form (inner) BH-BH binaries at Z = 0.005 (Z = 0.0005).

Secondly, the number of systems formed in the population with triples 
is always lower per unit stellar mass formed than in the population without triples, as triple systems, on average, have larger total masses than binaries and/or single stars.

In the population with triples, about half of the GW mergers originate from formation channels involving CHE originate from triples. The role of the tertiary is negligible for 69 per cent of GW progenitors from CHE triples. In the remaining 31 per cent, the evolution of the inner binary is affected by the tertiary star via TMT and/or three-body dynamics. 

\subsection{Isolated binaries}
\label{subsec:gw_isolated_binaries}

At $Z = 0.005$, about 68 per cent of the CHE binary population forms a BH-BH binary that merges within the Hubble time, while at $Z = 0.0005$, all CHE binaries merge due to GWs within the age of the universe. In our moderate metallicity model, the delay times of these BH-BH binaries from this population ranges from 3 to 50 Gyr (and therefore the delay time of GW sources ranges from 3 to 13.5 Gyr).
In our low metallicity model, the delay times are considerably shorter, ranging roughly from 100 to 600 Myr. At Z = 0.005, only those binaries merge which were in contact during their MS phase. At Z = 0.0005, about 97 per cent of all GW progenitors were in contact during their MS phase. Since we assume such binaries equalise in mass, we predict that the vast majority of GW sources consist of equal mass black home binaries from this population (in broad agreement with \citealt{Marchant2016}).
The masses of the merging binary black
holes from this channel range from 20 to 42 $M_{\odot}$ at Z = 0.005 and 33 to
54 $M_{\odot}$  at Z = 0.0005.

\subsection{Effectively isolated inner binaries}
\label{subsubsec:gw_effectively_isolated_binaries}

This is the dominant channel among CHE triples with a predicted merger rate density of 8.8 $\rm{Gpc^{-3}yr^{-1}}$. At Z = 0.005 (Z = 0.0005), about 19 (12) per cent of all CHE systems (e.g CHE binaries and CHE triples, see section \ref{subsec:init}) are expected to form GW sources via this channel. In 53 per cent of the GW progenitors of this channel, the tertiary  star becomes unbound by the time both stars in the inner binaries form BHs. This percentage drops to 38 per cent at Z = 0.0005.


The demographics of this channel are nearly indistinguishable from the isolated binary population. 
The merger efficiency of this channel, which we define as the GW sources as a fraction of 
BH-BH inner binaries formed via a certain channel, is 68 per cent. 
Unsurprisingly, 
 this is the same as the merger efficiency of the isolated CHE binary channel.
Similarly to the CHE binary case, the majority of the inner binaries of these triples were also in contact during their MS phase and therefore this channel also produces overwhelmingly equal mass mergers.


\subsection{TMT with a BH-BH accretor}
\label{subsubsec:gw_TMT_with_BBH_accretor}

This is the dominant formation channel in which the evolution of the inner binary is affected by the tertiary star.
The predicted merger rate density is $R_{\rm merger}$ = 3.8 $\rm{Gpc^{-3}yr^{-1}}$, which accounts for about 16 per cent of all GW mergers from CHE systems. About 10 per cent of all CHE systems form merging binary BHs via this channel.


With our simplistic models of TMT (see subsection \ref{subsec:TMT_method}), we predict that the outer orbit widens as a result of the TMT episode for all triples considered in this study.  
In the lower panel of Fig. \ref{fig:a_outer_evol}, we show how the outer pericenter changes after the mass transfer phase for 
 triples experiencing TMT  with a BH-BH inner binary accretor for our moderate metallicity model (and in the lower panel of Fig. \ref{fig:a_outer_evol_zlow} for our low metallicity model).
 The orbital separations widen typically by a factor 1.5-2.
 
 Even, if the inner orbit remains unchanged due to TMT, the outer orbit widens so much, such that three body-dynamics become typically negligible after the TMT episode for the majoirty of these triples.
For example, at Z = 0.005, in those TMT systems, in which ZKL oscillation are effective prior to the mass transfer event, 70 per cent of the inner binary becomes decoupled from the tertiary star after the TMT episode. If the evolution of the inner BH-BH inner binary is decoupled from the tertiary, its orbital evolution is solely determined by the emission of GWs (and therefore the coalescence time can be determined according \citealt{Peters64}, otherwise, we use equation \ref{eq:modified_peters}). 
 
 


As noted in section \ref{subsec:TMT_method}, we make different assumptions about the evolution of the inner orbit based on whether a circumbinary disc is formed during TMT. We therefore discuss the properties of GW sources from these two subtypes separately.

 \subsubsection{Accretion through a circumbinary disc}
\label{subsubsec:gw_TMT_with_BBH_accretor_cbd}
The predicted merger rate of this channel is 2.4 $\rm{Gpc^{-3}yr^{-1}}$.
The merger rate efficiency is just 6 per cent higher than the merger rate efficiency from isolated binaries. The slight increase is due to the small number of eccentric inner binaries at the onset of the mass transfer ($\sim$10 per cent of systems undergoing TMT with BH-BH accretors and circumbinary discs have $e_{\rm in}$>0.4, see Fig. \ref{fig:inner_ecc_teritary_rlof}). 
The small difference is not surprising as we have assumed here that the orbit of the inner binary does not change due to circumbinary disc accretion. However, if circumbinary disc accretion leads to a significant increase (decrease) in the inner period, the compact object merger fraction decreases (increases) significantly as well. Clearly, better models are required to understand circumbinary accretion of a BH binary from a mass transferring tertiary star. 

 \subsubsection{Ballistic accretion}
\label{subsubsec:gw_TMT_with_BBH_accretor_ba_sc1}

\begin{figure*}
\includegraphics[width = \textwidth]{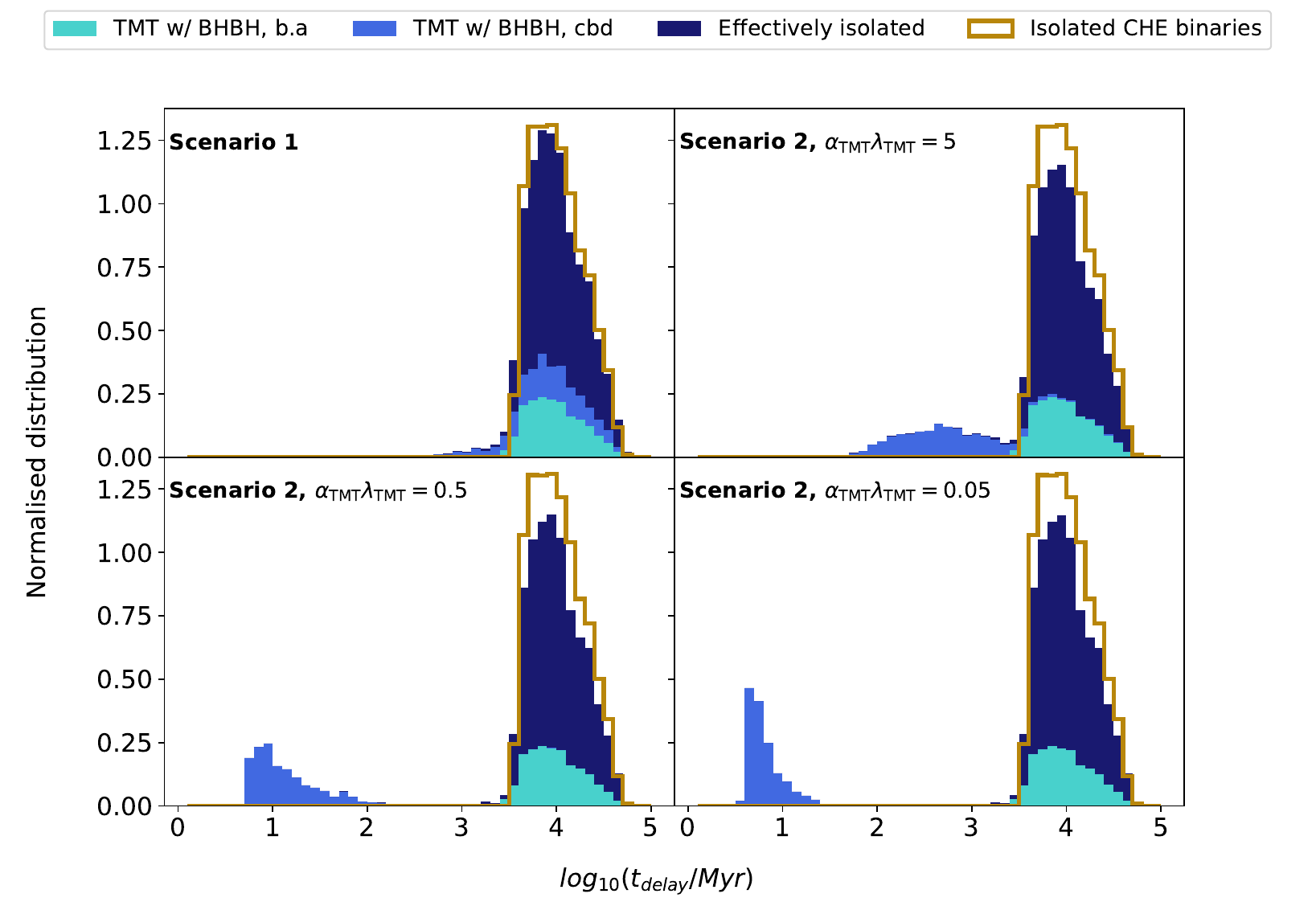}
\caption{The time delay distribution of the GW sources of CHE triple population shown by a stacked histogram. As a comparison, we also show the time delay distribution of the GW sources from CHE isolated binaries at the same metallicity (gold). The different colours of the stacked histograms refer to different evolutionary paths of the triples (shown by the legend in the top of the panel)} 
\label{fig:time_delays_GW}
\end{figure*}

The properties of these GW sources depend on how the inner binary evolves due to TMT. If we simplistically assume that that the inner orbit does not change (i.e. \textit{Scenario 1}, see section \ref{subsec:TMT_method}), then
the merger rate density of this channel in the local universe is $R_{\rm merger}$ = 1.4 $\rm{Gpc^{3}yr^{-1}}$. In this case about 3.8 (2.3) per cent of all stellar systems containing a CHE binary form GW sources via this channel at Z = 0.005 (Z = 0.0005). The merger efficiency of this channel is 75 per cent at Z = 0.005, which is slightly higher than that of the CHE binary population (68 per cent). %
 As discussed in section \ref{subsec:tmdt}, a considerable fraction  of these sources have high eccentricities, namely, 48 per cent with $e_{\rm in}\gtrsim 0.4$ at Z = 0.005 and 10 per cent at Z = 0.0005. 
This results in shorter delay times and more mergers with respect to the isolated CHE binary channel (top left panel of Fig. \ref{fig:time_delays_GW}).

 If the orbital evolution can be described by equation \ref{eq:change_in_inner_a_CEE} (i.e. \textit{Scenario 2}, see section \ref{subsec:TMT_method}), then the inner pericenters of BH-BH binaries decrease by 1-3 orders of magnitude due to the TMT episode, depending on the efficiency parameter $\alpha_{\rm TMT}$.  In this case, all inner bineries become dynamically decoupled from the tertiary star after the TMT episode. As shown in the left panel of Fig. \ref{fig:a_inner_evol}, the peak of the orbital separation distribution shifts from $32\,R_{\odot}$ to 25, 5 and 1 $R_{\odot}$ with $\alpha_{\rm TMT}\lambda_{\rm TMT}=0.05,0.5,5$.
 With such short periods, nearly all (i.e. typically $\gtrsim$ 99 per cent) of the inner binaries eventually emerge. However, none of the inner binaries merge during the mass transfer, in fact they merge due to GW emission afterwards. In Fig. \ref{fig:time_delays_GW}, we show that the typical delay times in \textit{Scenario 2} are also orders of magnitude shorter with respect to that of isolated CHE binaries.
With $\alpha_{\rm TMT} = 0.05$, the delay times of these GW sources is dominated by the stellar evolution. Such timescales could make TMT episodes relevant in young clusters in which star-formation is still active. Even when assuming a weaker 
friction exerted by the transferred mass (i.e. $\alpha_{\rm{TMT}}\lambda_{\rm TMT} =5$) resulting in the smallest orbital shrinkage in our models, most of the BHs merge within a few hundred Myr at Z = 0.005.

Despite the higher merger efficiency, 
the predicted merger rate density for \textit{Scenario 2} is considerably lower (i.e. $R_{\rm merger} = 0.5\,\rm{Gpc^{-3}yr^{-1}}$) than in \textit{Scenario 1}. 
This is due to the extremely short delay times, implying the progenitor stars must have formed recently, when the cosmic star formation rate is low \citep[e.g.][]{Madau_2017}. 
As the cosmic star formation rate is expected to increase strongly from $z=0$ to $z=2$ , we expect the merger rate density of this channel to be significantly higher at $z\approx2$ than at $z = 0$. This would make these sources  more relevant for third-generation GW detectors.

We mention two interesting aspects of this channel. Firstly, depending on the efficiency parameter of the TMT episode, these systems could be in the LISA  frequency band \citep{LISAwhitepaper} during the mass transfer phase. In the right panel panel of Fig. \ref{fig:a_inner_evol}, we show the frequency at which the BH-BH binaries emit GWs after the mass transfer episode. With $\alpha_{\rm{TMT}} = 0.5$, about half, and with $\alpha_{\rm{TMT}} = 0.05$, all of our systems enter the mHZ regime during the mass transfer phase. The evolution through the LISA frequency range would be primarily driven by gas dynamics  instead of GW emission \citep[see also][]{Renzo_2021}. Such sources would be detectable by LISA, if the corresponding luminosity distances are not larger than $\sim10\,\rm{kpc}$ and $\sim10^4\,\rm{kpc}$ in case of $\alpha_{\rm{TMT}} = 0.5$ and $\alpha_{\rm{TMT}} = 0.05$, respectively \citep[see e.g. Fig. 1 in ][]{LISAwhitepaper}. 

Secondly, a TMT episode could be accompanied by a detectable electromagnetic signal, as the transferred mass is expected to heat up when it reaches the inner BH binary. If the delay time between this signal and the GW merger is within the lifetimes of typical observing missions, then the GW merger could be associated with this electromagnetic counterpart \citep[see also e.g. ][]{deMinkKing2017}.
We find that the time between the end of the TMT episode and the GW merger in case of $\alpha_{\rm TMT}\lambda_{\rm TMT} = 0.05$ is shorter than a year for 6 per cent of these sources at Z = 0.0005. This implies that in this case a electromagnetic counterpart could be detected, shortly before the GW merger. This is in contrast with the possible electromagnetic signatures associated with BH mergers in AGN discs, where the electromagnetic counterpart would occur after the GW merger \citep[see e.g.][]{McKernan2019}

\subsection{TMT with a MS-MS accretor}
\label{subsubsec:gw_TMT_with_MSMS_accretors}

This channel has a low merger rate density of $R_{\rm merger}$ = 0.2 $\rm{Gpc^{-3}yr^{-1}}$. Even though 25 per cent of all systems containing a CHE binary experience a double MS merger in the inner binary at Z = 0.005, only 1.1 per cent of them form merging binary BHs. This low merging efficiency is due to two reasons. Firstly, if the mass transfer episode between the merger product and the tertiary star proceeds in a dynamically unstable way, the process mostly ends in stellar merger and no double compact binary is formed. Secondly, if the same mass transfer proceeds instead in a stable way, the binary BH typically has too wide orbit to merge within the Hubble time. We note, however, that these predictions are sensitively dependent on uncertain stellar physics (such as the efficiency of CEE phase, mass-loss radius exponent and binding energy of stars with $M_{\rm ZAMS} \gtrsim 100\,M_{\odot}$). We also note that the merger efficiency is significantly higher in our low metallicity model, 12.3 per cent of triples with double MS merger forms merging binary BHs. As the merger efficiency seems to increase with decreasing metallicity, and we only calculate the merger rate density based on two metallicities, it is likely that we underestimate the merger rate density for this channel (see more detailed explanation in Appendix section \ref{subsec:app_norm}).

    In case of a TMT episode with a MS-MS accretor, we always assume that the inner binary merges due the mass transfer phase.  We justify this assumption by the fact that that CHE MS-MS binaries tend to be on very close orbits ($\sim$20-30 $R_{\odot}$) compared to their stellar radii ($\sim$5-10$\,R_{\odot}$). A significant fraction of them are already in contact.  
   Furthermore, these stars may swell up as a result of accretion, this type mass transfer event is likely to end in merger \citep[e.g.][]{Braudo2022,Leigh2020}.

The merger product is a rejuvenated MS star with a mass of $M_{1+2} = M_1 + M_2$. This means that we neglect any accretion during TMT and we assume a fully conservative merger without mass outflows. At Z = 0.005, the mass of the inner binary merger remnant $M_{\rm 1+2}$ ranges from 65 to 188 $M_{\odot}$. The distribution has a peak around $\sim 100\,M_{\odot}$. At Z = 0.0005, the mass of the merger product ranges from 70 $M_{\odot}$ to 190 $M_{\odot}$.

 The orbital separations after the TMT episode are shown in the upper panel of Fig. \ref{fig:a_outer_evol} (and Fig. \ref{fig:a_outer_evol_zlow} for our low metallicity model). We can see that the outer orbit widens typically by a factor of 1.7-2.5 and the orbital separations range from 150 to 6800 $R_{\odot}$. While the ranges are similar at both metallicities, at Z = 0.0005, the typical orbital separations are significantly shorter.

 Most of the systems experience a second phase of mass transfer after the TMT episode (62 per cent at Z = 0.005 and 96 per cent at Z = 0.0005) and typically the donor star is on the Hertzsprung gap during this second phase of mass transfer ( about 99 per cent at Z = 0.005, and about 86 per cent at Z = 0.0005). More evolved donor stars are not expected to occur frequently, as the onset of CHeB occurs at a cooler effective temperature with increasing mass with and followed by a less significant subsequent radial expansion \citep[][]{Hurley2000}. 
In particular for $M_{\rm ZAMS} \gtrsim 100\,M_{\odot}$, stars are predicted to expand negligibly after the CHeB, even at low metallicities. 

Regarding the stability of the mass transfer between the merger remnant and the initial tertiary, we find that it occurs in an dynamically unstable manner in 66 (30) per cent of cases at Z = 0.005 (Z = 0.0005). 
We assume that CE phases with a donor star on the Hertzsprung gap result in a merger, following \citet{Dominik2012} \citep[but see also][]{Klencki2020, Marchant2021}. 
 At both metallicities, binary BHs are only produced when the second phase of mass transfer proceeds in a stable manner. Furthermore, in order to form a GW source, the orbit needs to be compact enough ($a_{\rm out} \lesssim 1000\,R_{\odot}$) at the onset of the second mass transfer event. This only occurs in about 5 per cent (30 per cent) of systems with stable mass transfer at Z = 0.005 (Z = 0.0005).

This is the only GW formation channel of CHE triples that yields a significantly different mass and mass ratio distributions than the CHE binary channel. The masses of the merging binary BHs range from 16 to 27 $M_{\odot}$ at Z = 0.005 and 17 to 54 $M_{\odot}$ at Z = 0.0005. The mass ratios range from 0.7 to 0.8 at Z = 0.005 and 0.5 to 1.0 at Z = 0.0005. All other channels produce merging binary BHs with masses that range from 20 to 42 at Z = 0.005 and 33 to 54 at Z = 0.0005. The vast majority ($\gtrsim$ 90 per cent) of these systems have equal masses, as the inner binaries had been in a contact during their MS phase.

\subsection{Dynamical mergers}
\label{subsubsec:fast_merger}

The merger rate density of these channel is very low,  $R_{\rm merger}$ = 0.05 $\rm{Gpc^{-3}yr^{-1}}$. 
The delay times of these systems are very short and range from 4 to 20 Myr. Similarly to the GW progenitors that have experienced TMT episodes with ballistic accretion, the short delay times imply that the merger rate density could be about an order of magnitude larger at z $\approx$ 2. About 25 per cent of these systems have eccentricities $e_{\rm in} \gtrsim 10^{-4}$ when the characteristic GW frequency reaches 10 Hz, making eccentricities detectable by third-generation detectors \citep{Lower2018}.

For all systems, the tertiary star is still on the MS when the inner binary merges due to GWs with outer pericenters of $a_{\rm{p,out}}\approx$ 120-790$\,R_{\odot}$. It is therefore expected that the initial tertiary star will eventaully fill its Roche-lobe, once it evolves of the MS.

\section{Conclusion}
We studied the evolution of hierarchical triples with CHE stars in the inner binary with a rapid populations synthesis approach. We performed simulations with the triple population synthesis code \texttt{TRES} at two representative metallicities: Z = 0.005 and Z = 0.0005. 
We showed that the evolution of CHE stars can be altered by the presence of a tertiary star in several ways. 
This can potentially lead to a formation of a number of diverse and unique astrophysical phenomena, e.g. TMT phases with BH-BH accretors, highly eccentric mergers of helium stars, and mergers of binary BHs with very short (few Myr) delay times. 


To summarise our main findings:
\begin{enumerate}
    \item \textbf{Tertiary mass transfer (TMT) episodes are common among CHE triples:} 
    Unlike in classically evolving hierarchical triples, we predict that TMT phase is very common among CHE triples. The tertiary star fills its Roche-lobe in about 50 per cent of all triples with CHE inner binaries. 
    The same fraction for classically evolving systems is predicted to be a few percent at best \citep[see e.g.][]{deVries2014, Toonen2020,Hamers2022,kummer2023main}.
    We find that the mass transfer episodes initiated by the tertiary star typically occurs in a dynamically stable way. 
    \item \textbf{BH-BH inner binaries that accrete from tertiary star are also common:}  About 31 (24) per cent of the tertiary-driven mass-transfer episodes occur with BH-BH accretors at Z = 0.005 (Z = 0.0005). 
     Previous population synthesis studies suggest that such scenario is probably not possible for triples with classically evolving stars  \citep[see e.g.][]{ Toonen2020,Hamers2022}.
    Therefore, mass transfer towards a BH-BH inner binary represents a unique scenario for triples (or higher-order multiples) with CHE stars in the inner binaries. An exciting prospects would be a possible EM counterpart from such an event \citep[e.g.][]{deMinkKing2017}.
    \item \textbf{Importance of three-body dynamics:} ZLK oscillations can be effective for CHE triples, if the stars in the inner binary have evolved off MS (otherwise precession due to strong tides quench ZLK cycles) and if the initial outer pericenter is $a_{\rm p, outer,ZAMS} \lesssim 2000\,R_{\odot}$ (otherwise ZLK cycles are quenched by various short range forces throughout the entire evolution of the inner binary). ZLK oscillations are only present in those CHE triples, in which the outer pericenter is so short, such that the tertiary star would eventually fill its Roche-lobe. The inner eccentricities of these systems can reach values up to $e_{\rm in, max}\sim0.9$
    (left panel of Fig. \ref{fig:uter_perti_types_z_high}). 
   The effects of three-body dynamics are negligible for those CHE triples in which the triple remains detached. In this case, the inner binary evolves effectively as an isolated binary.

    \item \textbf{Three-body dynamics can drive the inner binary to a stellar merger:}
     In about 3 per cent of CHE triples, the inner binary merges before BH-BH formation. The most common type is a merger of a double helium star binary, that comes into contact in a highly eccentric orbit (Table \ref{tab:secondnumbers}).   
    \item \textbf{CHE triples form GW sources efficiently:} About 30 (24) per cent of the CHE triple population forms BH binaries that merge due to GWs within Hubble time at Z = 0.005 (Z = 0.0005). We predict a merger rate density of GW sources from CHE triples of $R_{\rm merger} \approx 12\,\rm{Gpc^{-3}yr^{-1}}$ (Table \ref{tab:gwnumbers}). We also predict that about half of the GW sources from CHE systems originate from triples. In 69 per cent of all GW sources from CHE triples, the inner binary evolves effectively as an isolated binary and therefore its properties are indistinguishable from those of CHE binaries. In the remaining 31 per cent, the evolution of the GW progenitor is affected by three-body dynamics and/or TMT episodes. 
    
    \item \textbf{Tertiary mass transfer and three-body dynamics could lead to the formation of BH-BH binaries that merge within Myr}

    The vast majority of those GW progenitors of CHE triples, in which the evolution of the inner binary is not decoupled from the tertiary object, experience a TMT episode with a BH-BH inner binary. In this case, we model the evolution of the  inner binary during the TMT phase with energy arguments \citep[following][see also subsection \ref{subsec:TMT_method}]{deVries2014} and with different assumptions on how efficiently the transferred mass shrinks the orbit of the inner binary. 
    We find typical values for the delay time of these GW sources of few hundred Myr and few Myr in our model variation with the least and the most orbital shrinkage, respectively. 
\end{enumerate}

\section*{Acknowledgements}
SdM acknowledges Fabio Antonini, Adrian Hamers and Lieke van Son for insightful discussions.
AD acknowledges travel grant from the HPC3 Europa programme for providing computational resources at the Snelius supercomputer in the Netherlands and acknowledges support fro API for allowing an extended visit.
 Computational work was performed by the Snelius supercomputer in the Netherlands and by the University of Birmingham's BlueBEAR HPC service. 
ST acknowledges support from the Netherlands Research Council NWO (VENI 639.041.645 and VIDI 203.061 grants).
SdM acknowledges funding  by the Netherlands Organization for Scientific Research (NWO) as part of
the Vidi research program BinWaves with project number 639.042.728.

\section*{Data Availability}

The data underlying this article will be shared on reasonable request to the corresponding author.



\bibliographystyle{mnras}
\bibliography{example} 

\begin{thebibliography}{}
\makeatletter
\relax
\def\mn@urlcharsother{\let\do\@makeother \do\$\do\&\do\#\do\^\do\_\do\%\do\~}
\def\mn@doi{\begingroup\mn@urlcharsother \@ifnextchar [ {\mn@doi@} {\mn@doi@[]}}
\def\mn@doi@[#1]#2{\def\@tempa{#1}\ifx\@tempa\@empty \href {http://dx.doi.org/#2} {doi:#2}\else \href {http://dx.doi.org/#2} {#1}\fi \endgroup}
\def\mn@eprint#1#2{\mn@eprint@#1:#2::\@nil}
\def\mn@eprint@arXiv#1{\href {http://arxiv.org/abs/#1} {{\tt arXiv:#1}}}
\def\mn@eprint@dblp#1{\href {http://dblp.uni-trier.de/rec/bibtex/#1.xml} {dblp:#1}}
\def\mn@eprint@#1:#2:#3:#4\@nil{\def\@tempa {#1}\def\@tempb {#2}\def\@tempc {#3}\ifx \@tempc \@empty \let \@tempc \@tempb \let \@tempb \@tempa \fi \ifx \@tempb \@empty \def\@tempb {arXiv}\fi \@ifundefined {mn@eprint@\@tempb}{\@tempb:\@tempc}{\expandafter \expandafter \csname mn@eprint@\@tempb\endcsname \expandafter{\@tempc}}}

\bibitem[\protect\citeauthoryear{{Abbott} et~al.,}{{Abbott} et~al.}{2019a}]{Abbott2019_GWTC-1}
{Abbott} B.~P.,  et~al., 2019a, \mn@doi [Physical Review X] {10.1103/PhysRevX.9.031040}, \href {https://ui.adsabs.harvard.edu/abs/2019PhRvX...9c1040A} {9, 031040}

\bibitem[\protect\citeauthoryear{Abbott et~al.,}{Abbott et~al.}{2019b}]{Abbott_2019}
Abbott B.~P.,  et~al., 2019b, \mn@doi [The Astrophysical Journal] {10.3847/2041-8213/ab3800}, 882, L24

\bibitem[\protect\citeauthoryear{{Abbott} et~al.,}{{Abbott} et~al.}{2021}]{Abbott2021_GWTC2}
{Abbott} R.,  et~al., 2021, \mn@doi [Physical Review X] {10.1103/PhysRevX.11.021053}, \href {https://ui.adsabs.harvard.edu/abs/2021PhRvX..11b1053A} {11, 021053}

\bibitem[\protect\citeauthoryear{{Abdul-Masih} et~al.,}{{Abdul-Masih} et~al.}{2021}]{Abdul-Masih2021:constrain_ocbins_mixing}
{Abdul-Masih} M.,  et~al., 2021, \mn@doi [\aap] {10.1051/0004-6361/202040195}, \href {https://ui.adsabs.harvard.edu/abs/2021A&A...651A..96A} {651, A96}

\bibitem[\protect\citeauthoryear{{Abdul-Masih}, {Escorza}, {Menon}, {Mahy}  \& {Marchant}}{{Abdul-Masih} et~al.}{2022}]{Abdul-Masih2022:constrain_ocbins_period_stability}
{Abdul-Masih} M.,  {Escorza} A.,  {Menon} A.,  {Mahy} L.,   {Marchant} P.,  2022, \mn@doi [\aap] {10.1051/0004-6361/202244148}, \href {https://ui.adsabs.harvard.edu/abs/2022A&A...666A..18A} {666, A18}

\bibitem[\protect\citeauthoryear{Amaro-Seoane et~al.,}{Amaro-Seoane et~al.}{2022}]{LISAwhitepaper}
Amaro-Seoane P.,  et~al., 2022, Astrophysics with the Laser Interferometer Space Antenna, \mn@doi{10.48550/ARXIV.2203.06016}, \url {https://arxiv.org/abs/2203.06016}

\bibitem[\protect\citeauthoryear{{Anderson}, {Lai}  \& {Storch}}{{Anderson} et~al.}{2017}]{Kassandra2017}
{Anderson} K.~R.,  {Lai} D.,   {Storch} N.~I.,  2017, \mn@doi [\mnras] {10.1093/mnras/stx293}, \href {https://ui.adsabs.harvard.edu/abs/2017MNRAS.467.3066A} {467, 3066}

\bibitem[\protect\citeauthoryear{{Antognini}, {Shappee}, {Thompson}  \& {Amaro-Seoane}}{{Antognini} et~al.}{2014}]{Antognini2014}
{Antognini} J.~M.,  {Shappee} B.~J.,  {Thompson} T.~A.,   {Amaro-Seoane} P.,  2014, \mn@doi [\mnras] {10.1093/mnras/stu039}, \href {https://ui.adsabs.harvard.edu/abs/2014MNRAS.439.1079A} {439, 1079}

\bibitem[\protect\citeauthoryear{{Antoni}, {MacLeod}  \& {Ramirez-Ruiz}}{{Antoni} et~al.}{2019}]{Antoni2019}
{Antoni} A.,  {MacLeod} M.,   {Ramirez-Ruiz} E.,  2019, \mn@doi [\apj] {10.3847/1538-4357/ab3466}, \href {https://ui.adsabs.harvard.edu/abs/2019ApJ...884...22A} {884, 22}

\bibitem[\protect\citeauthoryear{{Antonini}, {Toonen}  \& {Hamers}}{{Antonini} et~al.}{2017}]{Antonini2017}
{Antonini} F.,  {Toonen} S.,   {Hamers} A.~S.,  2017, \mn@doi [\apj] {10.3847/1538-4357/aa6f5e}, \href {https://ui.adsabs.harvard.edu/abs/2017ApJ...841...77A} {841, 77}

\bibitem[\protect\citeauthoryear{{Barkat}, {Rakavy}  \& {Sack}}{{Barkat} et~al.}{1967}]{Barkat1967}
{Barkat} Z.,  {Rakavy} G.,   {Sack} N.,  1967, \mn@doi [\prl] {10.1103/PhysRevLett.18.379}, \href {https://ui.adsabs.harvard.edu/abs/1967PhRvL..18..379B} {18, 379}

\bibitem[\protect\citeauthoryear{{Bartos}, {Kocsis}, {Haiman}  \& {M{\'a}rka}}{{Bartos} et~al.}{2017}]{Bartos2017ApJ...835..165B}
{Bartos} I.,  {Kocsis} B.,  {Haiman} Z.,   {M{\'a}rka} S.,  2017, \mn@doi [\apj] {10.3847/1538-4357/835/2/165}, \href {https://ui.adsabs.harvard.edu/abs/2017ApJ...835..165B} {835, 165}

\bibitem[\protect\citeauthoryear{{Belczynski}, {Bulik}, {Fryer}, {Ruiter}, {Valsecchi}, {Vink}  \& {Hurley}}{{Belczynski} et~al.}{2010}]{Belczynski2010}
{Belczynski} K.,  {Bulik} T.,  {Fryer} C.~L.,  {Ruiter} A.,  {Valsecchi} F.,  {Vink} J.~S.,   {Hurley} J.~R.,  2010, \mn@doi [\apj] {10.1088/0004-637X/714/2/1217}, \href {https://ui.adsabs.harvard.edu/abs/2010ApJ...714.1217B} {714, 1217}

\bibitem[\protect\citeauthoryear{{Belczynski} et~al.,}{{Belczynski} et~al.}{2020}]{Belczynski2020:evolutionary_roads_leading_to}
{Belczynski} K.,  et~al., 2020, \mn@doi [\aap] {10.1051/0004-6361/201936528}, \href {https://ui.adsabs.harvard.edu/abs/2020A&A...636A.104B} {636, A104}

\bibitem[\protect\citeauthoryear{{Blaauw}}{{Blaauw}}{1961}]{Blaauw1961}
{Blaauw} A.,  1961, \bain, \href {https://ui.adsabs.harvard.edu/abs/1961BAN....15..265B} {15, 265}

\bibitem[\protect\citeauthoryear{{Blaes}, {Lee}  \& {Socrates}}{{Blaes} et~al.}{2002}]{Blaes2002}
{Blaes} O.,  {Lee} M.~H.,   {Socrates} A.,  2002, \mn@doi [\apj] {10.1086/342655}, \href {https://ui.adsabs.harvard.edu/abs/2002ApJ...578..775B} {578, 775}

\bibitem[\protect\citeauthoryear{{Bond}, {Arnett}  \& {Carr}}{{Bond} et~al.}{1982}]{Bond1982}
{Bond} J.~R.,  {Arnett} W.~D.,   {Carr} B.~J.,  1982, in {Rees} M.~J.,  {Stoneham} R.~J.,  eds,  NATO Advanced Study Institute (ASI) Series C Vol. 90, Supernovae: A Survey of Current Research. pp 303--311

\bibitem[\protect\citeauthoryear{{Bondi} \& {Hoyle}}{{Bondi} \& {Hoyle}}{1944}]{Bondi1944MNRAS.104..273B}
{Bondi} H.,  {Hoyle} F.,  1944, \mn@doi [\mnras] {10.1093/mnras/104.5.273}, \href {https://ui.adsabs.harvard.edu/abs/1944MNRAS.104..273B} {104, 273}

\bibitem[\protect\citeauthoryear{{Braudo}, {Bear}  \& {Soker}}{{Braudo} et~al.}{2022}]{Braudo2022}
{Braudo} J.,  {Bear} E.,   {Soker} N.,  2022, \mn@doi [\mnras] {10.1093/mnras/stab3758}, \href {https://ui.adsabs.harvard.edu/abs/2022MNRAS.510.4242B} {510, 4242}

\bibitem[\protect\citeauthoryear{{Brookshaw} \& {Tavani}}{{Brookshaw} \& {Tavani}}{1993}]{Brookshaw1993}
{Brookshaw} L.,  {Tavani} M.,  1993, \mn@doi [\apj] {10.1086/172789}, \href {https://ui.adsabs.harvard.edu/abs/1993ApJ...410..719B} {410, 719}

\bibitem[\protect\citeauthoryear{{Brott} et~al.,}{{Brott} et~al.}{2011}]{Brott2011}
{Brott} I.,  et~al., 2011, \mn@doi [\aap] {10.1051/0004-6361/201016113}, \href {https://ui.adsabs.harvard.edu/abs/2011A&A...530A.115B} {530, A115}

\bibitem[\protect\citeauthoryear{{Cantiello}, {Yoon}, {Langer}  \& {Livio}}{{Cantiello} et~al.}{2007}]{Cantiello2007}
{Cantiello} M.,  {Yoon} S.~C.,  {Langer} N.,   {Livio} M.,  2007, \mn@doi [\aap] {10.1051/0004-6361:20077115}, \href {https://ui.adsabs.harvard.edu/abs/2007A&A...465L..29C} {465, L29}

\bibitem[\protect\citeauthoryear{{Claeys}, {Pols}, {Izzard}, {Vink}  \& {Verbunt}}{{Claeys} et~al.}{2014}]{Claeys2014}
{Claeys} J.~S.~W.,  {Pols} O.~R.,  {Izzard} R.~G.,  {Vink} J.,   {Verbunt} F.~W.~M.,  2014, \mn@doi [\aap] {10.1051/0004-6361/201322714}, \href {https://ui.adsabs.harvard.edu/abs/2014A&A...563A..83C} {563, A83}

\bibitem[\protect\citeauthoryear{{Comerford} \& {Izzard}}{{Comerford} \& {Izzard}}{2020}]{Comerford2020}
{Comerford} T.~A.~F.,  {Izzard} R.~G.,  2020, \mn@doi [\mnras] {10.1093/mnras/staa2539}, \href {https://ui.adsabs.harvard.edu/abs/2020MNRAS.498.2957C} {498, 2957}

\bibitem[\protect\citeauthoryear{{Costa}, {Bressan}, {Mapelli}, {Marigo}, {Iorio}  \& {Spera}}{{Costa} et~al.}{2021}]{Costa2021}
{Costa} G.,  {Bressan} A.,  {Mapelli} M.,  {Marigo} P.,  {Iorio} G.,   {Spera} M.,  2021, \mn@doi [\mnras] {10.1093/mnras/staa3916}, \href {https://ui.adsabs.harvard.edu/abs/2021MNRAS.501.4514C} {501, 4514}

\bibitem[\protect\citeauthoryear{{D'Orazio} \& {Duffell}}{{D'Orazio} \& {Duffell}}{2021}]{DorzaioDuffell2021}
{D'Orazio} D.~J.,  {Duffell} P.~C.,  2021, \mn@doi [\apjl] {10.3847/2041-8213/ac0621}, \href {https://ui.adsabs.harvard.edu/abs/2021ApJ...914L..21D} {914, L21}

\bibitem[\protect\citeauthoryear{{Davies} \& {Beasor}}{{Davies} \& {Beasor}}{2020}]{DaviesBeasor2020}
{Davies} B.,  {Beasor} E.~R.,  2020, \mn@doi [\mnras] {10.1093/mnras/staa174}, \href {https://ui.adsabs.harvard.edu/abs/2020MNRAS.493..468D} {493, 468}

\bibitem[\protect\citeauthoryear{{Davies}, {Crowther}  \& {Beasor}}{{Davies} et~al.}{2018}]{Davies2018}
{Davies} B.,  {Crowther} P.~A.,   {Beasor} E.~R.,  2018, \mn@doi [\mnras] {10.1093/mnras/sty1302}, \href {https://ui.adsabs.harvard.edu/abs/2018MNRAS.478.3138D} {478, 3138}

\bibitem[\protect\citeauthoryear{{Debes} \& {Sigurdsson}}{{Debes} \& {Sigurdsson}}{2002}]{DebesSigurdsson2002}
{Debes} J.~H.,  {Sigurdsson} S.,  2002, \mn@doi [\apj] {10.1086/340291}, \href {https://ui.adsabs.harvard.edu/abs/2002ApJ...572..556D} {572, 556}

\bibitem[\protect\citeauthoryear{{Dewi} \& {Tauris}}{{Dewi} \& {Tauris}}{2000}]{Dewi2000}
{Dewi} J.~D.~M.,  {Tauris} T.~M.,  2000, \mn@doi [\aap] {10.48550/arXiv.astro-ph/0007034}, \href {https://ui.adsabs.harvard.edu/abs/2000A&A...360.1043D} {360, 1043}

\bibitem[\protect\citeauthoryear{{Dominik}, {Belczynski}, {Fryer}, {Holz}, {Berti}, {Bulik}, {Mandel}  \& {O'Shaughnessy}}{{Dominik} et~al.}{2012}]{Dominik2012}
{Dominik} M.,  {Belczynski} K.,  {Fryer} C.,  {Holz} D.~E.,  {Berti} E.,  {Bulik} T.,  {Mandel} I.,   {O'Shaughnessy} R.,  2012, \mn@doi [\apj] {10.1088/0004-637X/759/1/52}, \href {https://ui.adsabs.harvard.edu/abs/2012ApJ...759...52D} {759, 52}

\bibitem[\protect\citeauthoryear{{Dong}, {Katz}  \& {Socrates}}{{Dong} et~al.}{2014}]{Dong2014}
{Dong} S.,  {Katz} B.,   {Socrates} A.,  2014, \mn@doi [\apjl] {10.1088/2041-8205/781/1/L5}, \href {https://ui.adsabs.harvard.edu/abs/2014ApJ...781L...5D} {781, L5}

\bibitem[\protect\citeauthoryear{{Dorozsmai} \& {Toonen}}{{Dorozsmai} \& {Toonen}}{2022}]{dorozsmai2022}
{Dorozsmai} A.,  {Toonen} S.,  2022, arXiv e-prints, \href {https://ui.adsabs.harvard.edu/abs/2022arXiv220708837D} {p. arXiv:2207.08837}

\bibitem[\protect\citeauthoryear{{Duffell}, {D'Orazio}, {Derdzinski}, {Haiman}, {MacFadyen}, {Rosen}  \& {Zrake}}{{Duffell} et~al.}{2020}]{Duffell2020}
{Duffell} P.~C.,  {D'Orazio} D.,  {Derdzinski} A.,  {Haiman} Z.,  {MacFadyen} A.,  {Rosen} A.~L.,   {Zrake} J.,  2020, \mn@doi [\apj] {10.3847/1538-4357/abab95}, \href {https://ui.adsabs.harvard.edu/abs/2020ApJ...901...25D} {901, 25}

\bibitem[\protect\citeauthoryear{{Eggleton}}{{Eggleton}}{1983}]{EggletonApprox}
{Eggleton} P.~P.,  1983, \mn@doi [\apj] {10.1086/160960}, \href {https://ui.adsabs.harvard.edu/abs/1983ApJ...268..368E} {268, 368}

\bibitem[\protect\citeauthoryear{{Eggleton} \& {Kiseleva-Eggleton}}{{Eggleton} \& {Kiseleva-Eggleton}}{2001}]{Eggleton2001}
{Eggleton} P.~P.,  {Kiseleva-Eggleton} L.,  2001, \mn@doi [\apj] {10.1086/323843}, \href {https://ui.adsabs.harvard.edu/abs/2001ApJ...562.1012E} {562, 1012}

\bibitem[\protect\citeauthoryear{{Evans}}{{Evans}}{2011}]{EvansRemage2011}
{Evans} N.~R.,  2011, Bulletin de la Societe Royale des Sciences de Liege, \href {https://ui.adsabs.harvard.edu/abs/2011BSRSL..80..663E} {80, 663}

\bibitem[\protect\citeauthoryear{{Fabry}, {Marchant}, {Langer}  \& {Sana}}{{Fabry} et~al.}{2023}]{Fabry2023}
{Fabry} M.,  {Marchant} P.,  {Langer} N.,   {Sana} H.,  2023, \mn@doi [\aap] {10.1051/0004-6361/202346277}, \href {https://ui.adsabs.harvard.edu/abs/2023A&A...672A.175F} {672, A175}

\bibitem[\protect\citeauthoryear{{Fabrycky} \& {Tremaine}}{{Fabrycky} \& {Tremaine}}{2007}]{FabryckyTremaine2007}
{Fabrycky} D.,  {Tremaine} S.,  2007, \mn@doi [\apj] {10.1086/521702}, \href {https://ui.adsabs.harvard.edu/abs/2007ApJ...669.1298F} {669, 1298}

\bibitem[\protect\citeauthoryear{{Farag}, {Renzo}, {Farmer}, {Chidester}  \& {Timmes}}{{Farag} et~al.}{2022}]{Farag2022}
{Farag} E.,  {Renzo} M.,  {Farmer} R.,  {Chidester} M.~T.,   {Timmes} F.~X.,  2022, \mn@doi [\apj] {10.3847/1538-4357/ac8b83}, \href {https://ui.adsabs.harvard.edu/abs/2022ApJ...937..112F} {937, 112}

\bibitem[\protect\citeauthoryear{{Farmer}, {Renzo}, {de Mink}, {Marchant}  \& {Justham}}{{Farmer} et~al.}{2019}]{Farmer2019A}
{Farmer} R.,  {Renzo} M.,  {de Mink} S.~E.,  {Marchant} P.,   {Justham} S.,  2019, \mn@doi [\apj] {10.3847/1538-4357/ab518b}, \href {https://ui.adsabs.harvard.edu/abs/2019ApJ...887...53F} {887, 53}

\bibitem[\protect\citeauthoryear{{Ford}, {Kozinsky}  \& {Rasio}}{{Ford} et~al.}{2004}]{Ford_2004}
{Ford} E.~B.,  {Kozinsky} B.,   {Rasio} F.~A.,  2004, \mn@doi [\apj] {10.1086/382349}, \href {https://ui.adsabs.harvard.edu/abs/2004ApJ...605..966F} {605, 966}

\bibitem[\protect\citeauthoryear{{Fowler} \& {Hoyle}}{{Fowler} \& {Hoyle}}{1964}]{Fowler1964}
{Fowler} W.~A.,  {Hoyle} F.,  1964, \mn@doi [\apjs] {10.1086/190103}, \href {https://ui.adsabs.harvard.edu/abs/1964ApJS....9..201F} {9, 201}

\bibitem[\protect\citeauthoryear{{Fragione} \& {Loeb}}{{Fragione} \& {Loeb}}{2019}]{Fragione2019}
{Fragione} G.,  {Loeb} A.,  2019, \mn@doi [\mnras] {10.1093/mnras/stz1131}, \href {https://ui.adsabs.harvard.edu/abs/2019MNRAS.486.4443F} {486, 4443}

\bibitem[\protect\citeauthoryear{{Fraley}}{{Fraley}}{1968}]{Fraley1968}
{Fraley} G.~S.,  1968, \mn@doi [\apss] {10.1007/BF00651498}, \href {https://ui.adsabs.harvard.edu/abs/1968Ap&SS...2...96F} {2, 96}

\bibitem[\protect\citeauthoryear{Fryer, Belczynski, Wiktorowicz, Dominik, Kalogera  \& Holz}{Fryer et~al.}{2012}]{Fryer_2012}
Fryer C.~L.,  Belczynski K.,  Wiktorowicz G.,  Dominik M.,  Kalogera V.,   Holz D.~E.,  2012, \mn@doi [The Astrophysical Journal] {10.1088/0004-637x/749/1/91}, 749, 91

\bibitem[\protect\citeauthoryear{{Georgy}, {Meynet}  \& {Maeder}}{{Georgy} et~al.}{2011}]{Georgy2011}
{Georgy} C.,  {Meynet} G.,   {Maeder} A.,  2011, \mn@doi [\aap] {10.1051/0004-6361/200913797}, \href {https://ui.adsabs.harvard.edu/abs/2011A&A...527A..52G} {527, A52}

\bibitem[\protect\citeauthoryear{Ghodla, Eldridge, Stanway  \& Stevance}{Ghodla et~al.}{2022}]{Ghodla2022}
Ghodla S.,  Eldridge J.~J.,  Stanway E.~R.,   Stevance H.~F.,  2022, \mn@doi [\mnras] {10.1093/mnras/stac3177}, 518, 860

\bibitem[\protect\citeauthoryear{{Gilkis}, {Shenar}, {Ramachandran}, {Jermyn}, {Mahy}, {Oskinova}, {Arcavi}  \& {Sana}}{{Gilkis} et~al.}{2021}]{Gilkis2021}
{Gilkis} A.,  {Shenar} T.,  {Ramachandran} V.,  {Jermyn} A.~S.,  {Mahy} L.,  {Oskinova} L.~M.,  {Arcavi} I.,   {Sana} H.,  2021, \mn@doi [\mnras] {10.1093/mnras/stab383}, \href {https://ui.adsabs.harvard.edu/abs/2021MNRAS.503.1884G} {503, 1884}

\bibitem[\protect\citeauthoryear{{Glanz} \& {Perets}}{{Glanz} \& {Perets}}{2021a}]{Glanz2021}
{Glanz} H.,  {Perets} H.~B.,  2021a, \mn@doi [\mnras] {10.1093/mnras/staa3242}, \href {https://ui.adsabs.harvard.edu/abs/2021MNRAS.500.1921G} {500, 1921}

\bibitem[\protect\citeauthoryear{{Glanz} \& {Perets}}{{Glanz} \& {Perets}}{2021b}]{GlanzPerets2021}
{Glanz} H.,  {Perets} H.~B.,  2021b, \mn@doi [\mnras] {10.1093/mnras/stab2291}, \href {https://ui.adsabs.harvard.edu/abs/2021MNRAS.507.2659G} {507, 2659}

\bibitem[\protect\citeauthoryear{{Hamann} \& {Koesterke}}{{Hamann} \& {Koesterke}}{1998}]{Hamann98}
{Hamann} W.~R.,  {Koesterke} L.,  1998, \aap, \href {https://ui.adsabs.harvard.edu/abs/1998A&A...335.1003H} {335, 1003}

\bibitem[\protect\citeauthoryear{{Hamann}, {Koesterke}  \& {Wessolowski}}{{Hamann} et~al.}{1995}]{Hamann1995}
{Hamann} W.~R.,  {Koesterke} L.,   {Wessolowski} U.,  1995, \aap, \href {https://ui.adsabs.harvard.edu/abs/1995A&A...299..151H} {299, 151}

\bibitem[\protect\citeauthoryear{{Hamers} \& {Thompson}}{{Hamers} \& {Thompson}}{2019}]{HamersThompson2019}
{Hamers} A.~S.,  {Thompson} T.~A.,  2019, \mn@doi [\apj] {10.3847/1538-4357/ab3b06}, \href {https://ui.adsabs.harvard.edu/abs/2019ApJ...883...23H} {883, 23}

\bibitem[\protect\citeauthoryear{{Hamers}, {Pols}, {Claeys}  \& {Nelemans}}{{Hamers} et~al.}{2013}]{hamers2013}
{Hamers} A.~S.,  {Pols} O.~R.,  {Claeys} J.~S.~W.,   {Nelemans} G.,  2013, \mn@doi [\mnras] {10.1093/mnras/stt046}, \href {https://ui.adsabs.harvard.edu/abs/2013MNRAS.430.2262H} {430, 2262}

\bibitem[\protect\citeauthoryear{{Hamers}, {Rantala}, {Neunteufel}, {Preece}  \& {Vynatheya}}{{Hamers} et~al.}{2021}]{Hamers2021MNRAS.502.4479H}
{Hamers} A.~S.,  {Rantala} A.,  {Neunteufel} P.,  {Preece} H.,   {Vynatheya} P.,  2021, \mn@doi [\mnras] {10.1093/mnras/stab287}, \href {https://ui.adsabs.harvard.edu/abs/2021MNRAS.502.4479H} {502, 4479}

\bibitem[\protect\citeauthoryear{{Hamers}, {Glanz}  \& {Neunteufel}}{{Hamers} et~al.}{2022a}]{Hamers2022}
{Hamers} A.~S.,  {Glanz} H.,   {Neunteufel} P.,  2022a, \mn@doi [\apjs] {10.3847/1538-4365/ac49e7}, \href {https://ui.adsabs.harvard.edu/abs/2022ApJS..259...25H} {259, 25}

\bibitem[\protect\citeauthoryear{{Hamers}, {Perets}, {Thompson}  \& {Neunteufel}}{{Hamers} et~al.}{2022b}]{HamersTEDI2022}
{Hamers} A.~S.,  {Perets} H.~B.,  {Thompson} T.~A.,   {Neunteufel} P.,  2022b, \mn@doi [\apj] {10.3847/1538-4357/ac400b}, \href {https://ui.adsabs.harvard.edu/abs/2022ApJ...925..178H} {925, 178}

\bibitem[\protect\citeauthoryear{{Harrington}}{{Harrington}}{1968}]{Harrington1968}
{Harrington} R.~S.,  1968, \mn@doi [\aj] {10.1086/110614}, \href {https://ui.adsabs.harvard.edu/abs/1968AJ.....73..190H} {73, 190}

\bibitem[\protect\citeauthoryear{{Hastings}, {Langer}  \& {Koenigsberger}}{{Hastings} et~al.}{2020}]{Hastings2020}
{Hastings} B.,  {Langer} N.,   {Koenigsberger} G.,  2020, \mn@doi [\aap] {10.1051/0004-6361/202038499}, \href {https://ui.adsabs.harvard.edu/abs/2020A&A...641A..86H} {641, A86}

\bibitem[\protect\citeauthoryear{{Heger} \& {Woosley}}{{Heger} \& {Woosley}}{2002}]{Heger2002}
{Heger} A.,  {Woosley} S.~E.,  2002, \mn@doi [\apj] {10.1086/338487}, \href {https://ui.adsabs.harvard.edu/abs/2002ApJ...567..532H} {567, 532}

\bibitem[\protect\citeauthoryear{{Heggie}}{{Heggie}}{1975}]{Heggie1975}
{Heggie} D.~C.,  1975, \mn@doi [\mnras] {10.1093/mnras/173.3.729}, \href {https://ui.adsabs.harvard.edu/abs/1975MNRAS.173..729H} {173, 729}

\bibitem[\protect\citeauthoryear{{Higgins} \& {Vink}}{{Higgins} \& {Vink}}{2020}]{Higgins2020}
{Higgins} E.~R.,  {Vink} J.~S.,  2020, \mn@doi [\aap] {10.1051/0004-6361/201937374}, \href {https://ui.adsabs.harvard.edu/abs/2020A&A...635A.175H} {635, A175}

\bibitem[\protect\citeauthoryear{{Hills}}{{Hills}}{1983}]{Hills1983}
{Hills} J.~G.,  1983, \mn@doi [\apj] {10.1086/160871}, \href {https://ui.adsabs.harvard.edu/abs/1983ApJ...267..322H} {267, 322}

\bibitem[\protect\citeauthoryear{{Hjellming} \& {Webbink}}{{Hjellming} \& {Webbink}}{1987}]{Hjellming1987}
{Hjellming} M.~S.,  {Webbink} R.~F.,  1987, \mn@doi [\apj] {10.1086/165412}, \href {https://ui.adsabs.harvard.edu/abs/1987ApJ...318..794H} {318, 794}

\bibitem[\protect\citeauthoryear{{Holman}, {Touma}  \& {Tremaine}}{{Holman} et~al.}{1997}]{Holman1997}
{Holman} M.,  {Touma} J.,   {Tremaine} S.,  1997, \mn@doi [\nat] {10.1038/386254a0}, \href {https://ui.adsabs.harvard.edu/abs/1997Natur.386..254H} {386, 254}

\bibitem[\protect\citeauthoryear{{Huang}}{{Huang}}{1956}]{Huang1956A}
{Huang} S.~S.,  1956, \mn@doi [\aj] {10.1086/107290}, \href {https://ui.adsabs.harvard.edu/abs/1956AJ.....61...49H} {61, 49}

\bibitem[\protect\citeauthoryear{{Huang}}{{Huang}}{1963}]{Huang1963}
{Huang} S.-S.,  1963, \mn@doi [\apj] {10.1086/147659}, \href {https://ui.adsabs.harvard.edu/abs/1963ApJ...138..471H} {138, 471}

\bibitem[\protect\citeauthoryear{{Humphreys} \& {Davidson}}{{Humphreys} \& {Davidson}}{1979}]{Humphreys1979}
{Humphreys} R.~M.,  {Davidson} K.,  1979, \mn@doi [\apj] {10.1086/157301}, \href {https://ui.adsabs.harvard.edu/abs/1979ApJ...232..409H} {232, 409}

\bibitem[\protect\citeauthoryear{{Hurley}, {Pols}  \& {Tout}}{{Hurley} et~al.}{2000}]{Hurley2000}
{Hurley} J.~R.,  {Pols} O.~R.,   {Tout} C.~A.,  2000, \mn@doi [\mnras] {10.1046/j.1365-8711.2000.03426.x}, \href {https://ui.adsabs.harvard.edu/abs/2000MNRAS.315..543H} {315, 543}

\bibitem[\protect\citeauthoryear{{Hurley}, {Tout}  \& {Pols}}{{Hurley} et~al.}{2002}]{Hurley2002}
{Hurley} J.~R.,  {Tout} C.~A.,   {Pols} O.~R.,  2002, \mn@doi [\mnras] {10.1046/j.1365-8711.2002.05038.x}, \href {https://ui.adsabs.harvard.edu/abs/2002MNRAS.329..897H} {329, 897}

\bibitem[\protect\citeauthoryear{{Iben} \& {Tutukov}}{{Iben} \& {Tutukov}}{1999}]{Iben1999}
{Iben} Icko J.,  {Tutukov} A.~V.,  1999, \mn@doi [\apj] {10.1086/306672}, \href {https://ui.adsabs.harvard.edu/abs/1999ApJ...511..324I} {511, 324}

\bibitem[\protect\citeauthoryear{{Innanen}, {Zheng}, {Mikkola}  \& {Valtonen}}{{Innanen} et~al.}{1997}]{Innanen1997}
{Innanen} K.~A.,  {Zheng} J.~Q.,  {Mikkola} S.,   {Valtonen} M.~J.,  1997, \mn@doi [\aj] {10.1086/118405}, \href {https://ui.adsabs.harvard.edu/abs/1997AJ....113.1915I} {113, 1915}

\bibitem[\protect\citeauthoryear{{Ivanova} et~al.,}{{Ivanova} et~al.}{2013}]{Ivanova2013}
{Ivanova} N.,  et~al., 2013, \mn@doi [\aapr] {10.1007/s00159-013-0059-2}, \href {https://ui.adsabs.harvard.edu/abs/2013A&ARv..21...59I} {21, 59}

\bibitem[\protect\citeauthoryear{{Janssens}, {Shenar}, {Mahy}, {Marchant}, {Sana}  \& {Bodensteiner}}{{Janssens} et~al.}{2021}]{Janssesns2021:ocms}
{Janssens} S.,  {Shenar} T.,  {Mahy} L.,  {Marchant} P.,  {Sana} H.,   {Bodensteiner} J.,  2021, \mn@doi [\aap] {10.1051/0004-6361/202039305}, \href {https://ui.adsabs.harvard.edu/abs/2021A&A...646A..33J} {646, A33}

\bibitem[\protect\citeauthoryear{{Katz} \& {Dong}}{{Katz} \& {Dong}}{2012}]{KatzDong2012}
{Katz} B.,  {Dong} S.,  2012, arXiv e-prints, \href {https://ui.adsabs.harvard.edu/abs/2012arXiv1211.4584K} {p. arXiv:1211.4584}

\bibitem[\protect\citeauthoryear{{Kennedy}, {Dougherty}, {Fink}  \& {Williams}}{{Kennedy} et~al.}{2010}]{Kennedy2010:ocms}
{Kennedy} M.,  {Dougherty} S.~M.,  {Fink} A.,   {Williams} P.~M.,  2010, \mn@doi [\apj] {10.1088/0004-637X/709/2/632}, \href {https://ui.adsabs.harvard.edu/abs/2010ApJ...709..632K} {709, 632}

\bibitem[\protect\citeauthoryear{{King}, {Taam}  \& {Begelman}}{{King} et~al.}{2000}]{King2000}
{King} A.~R.,  {Taam} R.~E.,   {Begelman} M.~C.,  2000, \mn@doi [\apjl] {10.1086/312475}, \href {https://ui.adsabs.harvard.edu/abs/2000ApJ...530L..25K} {530, L25}

\bibitem[\protect\citeauthoryear{{Kinoshita} \& {Nakai}}{{Kinoshita} \& {Nakai}}{1999}]{KinoshitaNakai1999}
{Kinoshita} H.,  {Nakai} H.,  1999, \mn@doi [Celestial Mechanics and Dynamical Astronomy] {10.1023/A:1008321310187}, \href {https://ui.adsabs.harvard.edu/abs/1999CeMDA..75..125K} {75, 125}

\bibitem[\protect\citeauthoryear{{Kiseleva}, {Eggleton}  \& {Orlov}}{{Kiseleva} et~al.}{1994}]{Kiseleva1994}
{Kiseleva} L.~G.,  {Eggleton} P.~P.,   {Orlov} V.~V.,  1994, \mn@doi [\mnras] {10.1093/mnras/270.4.936}, \href {https://ui.adsabs.harvard.edu/abs/1994MNRAS.270..936K} {270, 936}

\bibitem[\protect\citeauthoryear{Klencki, Nelemans, Istrate  \& Pols}{Klencki et~al.}{2020}]{Klencki2020}
Klencki J.,  Nelemans G.,  Istrate A.~G.,   Pols O.,  2020, \mn@doi [Astronomy & Astrophysics] {10.1051/0004-6361/202037694}, 638, A55

\bibitem[\protect\citeauthoryear{{Kobulnicky} et~al.,}{{Kobulnicky} et~al.}{2014}]{Kobulnicky2014}
{Kobulnicky} H.~A.,  et~al., 2014, \mn@doi [\apjs] {10.1088/0067-0049/213/2/34}, \href {https://ui.adsabs.harvard.edu/abs/2014ApJS..213...34K} {213, 34}

\bibitem[\protect\citeauthoryear{{K{\"o}hler} et~al.,}{{K{\"o}hler} et~al.}{2015}]{Kohler2015}
{K{\"o}hler} K.,  et~al., 2015, \mn@doi [\aap] {10.1051/0004-6361/201424356}, \href {https://ui.adsabs.harvard.edu/abs/2015A&A...573A..71K} {573, A71}

\bibitem[\protect\citeauthoryear{{Kozai}}{{Kozai}}{1962}]{Kozai1962}
{Kozai} Y.,  1962, \mn@doi [\aj] {10.1086/108790}, \href {https://ui.adsabs.harvard.edu/abs/1962AJ.....67..591K} {67, 591}

\bibitem[\protect\citeauthoryear{{Kroupa}}{{Kroupa}}{2001}]{Kroupa}
{Kroupa} P.,  2001, \mn@doi [\mnras] {10.1046/j.1365-8711.2001.04022.x}, \href {https://ui.adsabs.harvard.edu/abs/2001MNRAS.322..231K} {322, 231}

\bibitem[\protect\citeauthoryear{Kummer, Toonen  \& de Koter}{Kummer et~al.}{2023}]{kummer2023main}
Kummer F.,  Toonen S.,   de Koter A.,  2023, The Main Evolutionary Pathways of Massive Hierarchical Triple Stars (\mn@eprint {arXiv} {2306.09400})

\bibitem[\protect\citeauthoryear{{Lai} \& {Mu{\~n}oz}}{{Lai} \& {Mu{\~n}oz}}{2022}]{LaiMunoz2022}
{Lai} D.,  {Mu{\~n}oz} D.~J.,  2022, arXiv e-prints, \href {https://ui.adsabs.harvard.edu/abs/2022arXiv221100028L} {p. arXiv:2211.00028}

\bibitem[\protect\citeauthoryear{{Lamers} \& {Fitzpatrick}}{{Lamers} \& {Fitzpatrick}}{1988}]{Lamers1988}
{Lamers} H. J.~G.~L.~M.,  {Fitzpatrick} E.~L.,  1988, \mn@doi [\apj] {10.1086/165894}, \href {https://ui.adsabs.harvard.edu/abs/1988ApJ...324..279L} {324, 279}

\bibitem[\protect\citeauthoryear{Langer}{Langer}{2012}]{Langer_2012}
Langer N.,  2012, \mn@doi [Annual Review of Astronomy and Astrophysics] {10.1146/annurev-astro-081811-125534}, 50, 107–164

\bibitem[\protect\citeauthoryear{{Langer} \& {Maeder}}{{Langer} \& {Maeder}}{1995}]{Langer1995}
{Langer} N.,  {Maeder} A.,  1995, \aap, \href {https://ui.adsabs.harvard.edu/abs/1995A&A...295..685L} {295, 685}

\bibitem[\protect\citeauthoryear{{Laplace}, {G{\"o}tberg}, {de Mink}, {Justham}  \& {Farmer}}{{Laplace} et~al.}{2020}]{Laplace2020}
{Laplace} E.,  {G{\"o}tberg} Y.,  {de Mink} S.~E.,  {Justham} S.,   {Farmer} R.,  2020, \mn@doi [\aap] {10.1051/0004-6361/201937300}, \href {https://ui.adsabs.harvard.edu/abs/2020A&A...637A...6L} {637, A6}

\bibitem[\protect\citeauthoryear{{Leigh}, {Toonen}, {Portegies Zwart}  \& {Perna}}{{Leigh} et~al.}{2020}]{Leigh2020}
{Leigh} N. W.~C.,  {Toonen} S.,  {Portegies Zwart} S.~F.,   {Perna} R.,  2020, \mn@doi [\mnras] {10.1093/mnras/staa1670}, \href {https://ui.adsabs.harvard.edu/abs/2020MNRAS.496.1819L} {496, 1819}

\bibitem[\protect\citeauthoryear{{Lewin}, {van Paradijs}  \& {van den Heuvel}}{{Lewin} et~al.}{1997}]{XrayBook1997}
{Lewin} W. H.~G.,  {van Paradijs} J.,   {van den Heuvel} E. P.~J.,  1997, {X-ray Binaries}.
Cambridge University Press

\bibitem[\protect\citeauthoryear{{Lidov}}{{Lidov}}{1962}]{Lidov1962}
{Lidov} M.~L.,  1962, \mn@doi [\planss] {10.1016/0032-0633(62)90129-0}, \href {https://ui.adsabs.harvard.edu/abs/1962P&SS....9..719L} {9, 719}

\bibitem[\protect\citeauthoryear{{Liu}, {Mu{\~n}oz}  \& {Lai}}{{Liu} et~al.}{2015}]{Liu2015}
{Liu} B.,  {Mu{\~n}oz} D.~J.,   {Lai} D.,  2015, \mn@doi [\mnras] {10.1093/mnras/stu2396}, \href {https://ui.adsabs.harvard.edu/abs/2015MNRAS.447..747L} {447, 747}

\bibitem[\protect\citeauthoryear{{Lorenzo}, {Sim{\'o}n-D{\'\i}az}, {Negueruela}, {Vilardell}, {Garcia}, {Evans}  \& {Montes}}{{Lorenzo} et~al.}{2017}]{Lorenzo2017:ocms}
{Lorenzo} J.,  {Sim{\'o}n-D{\'\i}az} S.,  {Negueruela} I.,  {Vilardell} F.,  {Garcia} M.,  {Evans} C.~J.,   {Montes} D.,  2017, \mn@doi [\aap] {10.1051/0004-6361/201731352}, \href {https://ui.adsabs.harvard.edu/abs/2017A&A...606A..54L} {606, A54}

\bibitem[\protect\citeauthoryear{{Lower}, {Thrane}, {Lasky}  \& {Smith}}{{Lower} et~al.}{2018}]{Lower2018}
{Lower} M.~E.,  {Thrane} E.,  {Lasky} P.~D.,   {Smith} R.,  2018, \mn@doi [\prd] {10.1103/PhysRevD.98.083028}, \href {https://ui.adsabs.harvard.edu/abs/2018PhRvD..98h3028L} {98, 083028}

\bibitem[\protect\citeauthoryear{{Lubow} \& {Shu}}{{Lubow} \& {Shu}}{1975}]{LubowShu1975}
{Lubow} S.~H.,  {Shu} F.~H.,  1975, \mn@doi [\apj] {10.1086/153614}, \href {https://ui.adsabs.harvard.edu/abs/1975ApJ...198..383L} {198, 383}

\bibitem[\protect\citeauthoryear{{Madau} \& {Dickinson}}{{Madau} \& {Dickinson}}{2014}]{MadauDickinson2014}
{Madau} P.,  {Dickinson} M.,  2014, \mn@doi [\araa] {10.1146/annurev-astro-081811-125615}, \href {https://ui.adsabs.harvard.edu/abs/2014ARA&A..52..415M} {52, 415}

\bibitem[\protect\citeauthoryear{Madau \& Fragos}{Madau \& Fragos}{2017}]{Madau_2017}
Madau P.,  Fragos T.,  2017, \mn@doi [The Astrophysical Journal] {10.3847/1538-4357/aa6af9}, 840, 39

\bibitem[\protect\citeauthoryear{{Maeder}}{{Maeder}}{1987}]{Maeder1987}
{Maeder} A.,  1987, \aap, \href {https://ui.adsabs.harvard.edu/abs/1987A&A...178..159M} {178, 159}

\bibitem[\protect\citeauthoryear{{Mandel} \& {de Mink}}{{Mandel} \& {de Mink}}{2016}]{Mandel2016}
{Mandel} I.,  {de Mink} S.~E.,  2016, \mn@doi [\mnras] {10.1093/mnras/stw379}, \href {https://ui.adsabs.harvard.edu/abs/2016MNRAS.458.2634M} {458, 2634}

\bibitem[\protect\citeauthoryear{{Marchant}, {Langer}, {Podsiadlowski}, {Tauris}  \& {Moriya}}{{Marchant} et~al.}{2016}]{Marchant2016}
{Marchant} P.,  {Langer} N.,  {Podsiadlowski} P.,  {Tauris} T.~M.,   {Moriya} T.~J.,  2016, \mn@doi [\aap] {10.1051/0004-6361/201628133}, \href {https://ui.adsabs.harvard.edu/abs/2016A&A...588A..50M} {588, A50}

\bibitem[\protect\citeauthoryear{{Marchant}, {Renzo}, {Farmer}, {Pappas}, {Taam}, {de Mink}  \& {Kalogera}}{{Marchant} et~al.}{2019}]{Marchant2019}
{Marchant} P.,  {Renzo} M.,  {Farmer} R.,  {Pappas} K. M.~W.,  {Taam} R.~E.,  {de Mink} S.~E.,   {Kalogera} V.,  2019, \mn@doi [\apj] {10.3847/1538-4357/ab3426}, \href {https://ui.adsabs.harvard.edu/abs/2019ApJ...882...36M} {882, 36}

\bibitem[\protect\citeauthoryear{{Marchant}, {Pappas}, {Gallegos-Garcia}, {Berry}, {Taam}, {Kalogera}  \& {Podsiadlowski}}{{Marchant} et~al.}{2021}]{Marchant2021}
{Marchant} P.,  {Pappas} K. M.~W.,  {Gallegos-Garcia} M.,  {Berry} C. P.~L.,  {Taam} R.~E.,  {Kalogera} V.,   {Podsiadlowski} P.,  2021, \mn@doi [\aap] {10.1051/0004-6361/202039992}, \href {https://ui.adsabs.harvard.edu/abs/2021A&A...650A.107M} {650, A107}

\bibitem[\protect\citeauthoryear{{Mardling} \& {Aarseth}}{{Mardling} \& {Aarseth}}{2001}]{MardlingAarseth2001}
{Mardling} R.~A.,  {Aarseth} S.~J.,  2001, \mn@doi [\mnras] {10.1046/j.1365-8711.2001.03974.x}, \href {https://ui.adsabs.harvard.edu/abs/2001MNRAS.321..398M} {321, 398}

\bibitem[\protect\citeauthoryear{{Martinez}, {Rodriguez}  \& {Fragione}}{{Martinez} et~al.}{2022}]{Martinez2022}
{Martinez} M. A.~S.,  {Rodriguez} C.~L.,   {Fragione} G.,  2022, \mn@doi [\apj] {10.3847/1538-4357/ac8d55}, \href {https://ui.adsabs.harvard.edu/abs/2022ApJ...937...78M} {937, 78}

\bibitem[\protect\citeauthoryear{{Mayer}, {Drechsel}, {Harmanec}, {Yang}  \& {{\v{S}}lechta}}{{Mayer} et~al.}{2013}]{Pavel2013:ocms}
{Mayer} P.,  {Drechsel} H.,  {Harmanec} P.,  {Yang} S.,   {{\v{S}}lechta} M.,  2013, \mn@doi [\aap] {10.1051/0004-6361/201322153}, \href {https://ui.adsabs.harvard.edu/abs/2013A&A...559A..22M} {559, A22}

\bibitem[\protect\citeauthoryear{{Mazeh} \& {Shaham}}{{Mazeh} \& {Shaham}}{1979}]{Mazeh+Shaham1979}
{Mazeh} T.,  {Shaham} J.,  1979, \aap, \href {https://ui.adsabs.harvard.edu/abs/1979A&A....77..145M} {77, 145}

\bibitem[\protect\citeauthoryear{{McKernan} et~al.,}{{McKernan} et~al.}{2019}]{McKernan2019}
{McKernan} B.,  et~al., 2019, \mn@doi [\apjl] {10.3847/2041-8213/ab4886}, \href {https://ui.adsabs.harvard.edu/abs/2019ApJ...884L..50M} {884, L50}

\bibitem[\protect\citeauthoryear{{McKernan}, {Ford}, {O'Shaugnessy}  \& {Wysocki}}{{McKernan} et~al.}{2020}]{McKerna2020}
{McKernan} B.,  {Ford} K.~E.~S.,  {O'Shaugnessy} R.,   {Wysocki} D.,  2020, \mn@doi [\mnras] {10.1093/mnras/staa740}, \href {https://ui.adsabs.harvard.edu/abs/2020MNRAS.494.1203M} {494, 1203}

\bibitem[\protect\citeauthoryear{{Mehta}, {Buonanno}, {Gair}, {Miller}, {Farag}, {deBoer}, {Wiescher}  \& {Timmes}}{{Mehta} et~al.}{2022}]{Mehta2022}
{Mehta} A.~K.,  {Buonanno} A.,  {Gair} J.,  {Miller} M.~C.,  {Farag} E.,  {deBoer} R.~J.,  {Wiescher} M.,   {Timmes} F.~X.,  2022, \mn@doi [\apj] {10.3847/1538-4357/ac3130}, \href {https://ui.adsabs.harvard.edu/abs/2022ApJ...924...39M} {924, 39}

\bibitem[\protect\citeauthoryear{Mennekens \& Vanbeveren}{Mennekens \& Vanbeveren}{2014}]{Mennekens_2014}
Mennekens N.,  Vanbeveren D.,  2014, \mn@doi [Astronomy & Astrophysics] {10.1051/0004-6361/201322198}, 564, A134

\bibitem[\protect\citeauthoryear{{Menon} et~al.,}{{Menon} et~al.}{2021}]{Menon2021}
{Menon} A.,  et~al., 2021, \mn@doi [\mnras] {10.1093/mnras/stab2276}, \href {https://ui.adsabs.harvard.edu/abs/2021MNRAS.507.5013M} {507, 5013}

\bibitem[\protect\citeauthoryear{{Michaely} \& {Perets}}{{Michaely} \& {Perets}}{2014}]{MichaelyPerets2014}
{Michaely} E.,  {Perets} H.~B.,  2014, \mn@doi [\apj] {10.1088/0004-637X/794/2/122}, \href {https://ui.adsabs.harvard.edu/abs/2014ApJ...794..122M} {794, 122}

\bibitem[\protect\citeauthoryear{{Miller} \& {Hamilton}}{{Miller} \& {Hamilton}}{2002}]{MillerHamiltion2002}
{Miller} M.~C.,  {Hamilton} D.~P.,  2002, \mn@doi [\apj] {10.1086/341788}, \href {https://ui.adsabs.harvard.edu/abs/2002ApJ...576..894M} {576, 894}

\bibitem[\protect\citeauthoryear{{Misner}, {Thorne}  \& {Wheeler}}{{Misner} et~al.}{1973}]{MisnerThorneWheeler_Gravitation1_973}
{Misner} C.~W.,  {Thorne} K.~S.,   {Wheeler} J.~A.,  1973, {Gravitation}.
W. H. Freeman Princeton University Press

\bibitem[\protect\citeauthoryear{{Moe} \& {Di Stefano}}{{Moe} \& {Di Stefano}}{2017}]{MoeDiStefano2017}
{Moe} M.,  {Di Stefano} R.,  2017, \mn@doi [\apjs] {10.3847/1538-4365/aa6fb6}, \href {https://ui.adsabs.harvard.edu/abs/2017ApJS..230...15M} {230, 15}

\bibitem[\protect\citeauthoryear{{Moody}, {Shi}  \& {Stone}}{{Moody} et~al.}{2019}]{Moody2019}
{Moody} M. S.~L.,  {Shi} J.-M.,   {Stone} J.~M.,  2019, \mn@doi [\apj] {10.3847/1538-4357/ab09ee}, \href {https://ui.adsabs.harvard.edu/abs/2019ApJ...875...66M} {875, 66}

\bibitem[\protect\citeauthoryear{{Moreno M{\'e}ndez}, {De Colle}, {L{\'o}pez C{\'a}mara}  \& {Vigna-G{\'o}mez}}{{Moreno M{\'e}ndez} et~al.}{2022}]{2022arXiv220703514M}
{Moreno M{\'e}ndez} E.,  {De Colle} F.,  {L{\'o}pez C{\'a}mara} D.,   {Vigna-G{\'o}mez} A.,  2022, arXiv e-prints, \href {https://ui.adsabs.harvard.edu/abs/2022arXiv220703514M} {p. arXiv:2207.03514}

\bibitem[\protect\citeauthoryear{{Mu{\~n}oz}, {Miranda}  \& {Lai}}{{Mu{\~n}oz} et~al.}{2019}]{Munoz2019}
{Mu{\~n}oz} D.~J.,  {Miranda} R.,   {Lai} D.,  2019, \mn@doi [\apj] {10.3847/1538-4357/aaf867}, \href {https://ui.adsabs.harvard.edu/abs/2019ApJ...871...84M} {871, 84}

\bibitem[\protect\citeauthoryear{{Naoz}}{{Naoz}}{2016}]{Naoz2016}
{Naoz} S.,  2016, \mn@doi [\araa] {10.1146/annurev-astro-081915-023315}, \href {https://ui.adsabs.harvard.edu/abs/2016ARA&A..54..441N} {54, 441}

\bibitem[\protect\citeauthoryear{{Naoz}, {Farr}, {Lithwick}, {Rasio}  \& {Teyssandier}}{{Naoz} et~al.}{2013}]{Naoz2013}
{Naoz} S.,  {Farr} W.~M.,  {Lithwick} Y.,  {Rasio} F.~A.,   {Teyssandier} J.,  2013, \mn@doi [\mnras] {10.1093/mnras/stt302}, \href {https://ui.adsabs.harvard.edu/abs/2013MNRAS.431.2155N} {431, 2155}

\bibitem[\protect\citeauthoryear{{Nieuwenhuijzen} \& {de Jager}}{{Nieuwenhuijzen} \& {de Jager}}{1990}]{Nieuwenhuijzen1990}
{Nieuwenhuijzen} H.,  {de Jager} C.,  1990, \aap, \href {https://ui.adsabs.harvard.edu/abs/1990A&A...231..134N} {231, 134}

\bibitem[\protect\citeauthoryear{{{\"O}pik}}{{{\"O}pik}}{1924}]{OpikLaw1924}
{{\"O}pik} E.,  1924, Publications of the Tartu Astrofizica Observatory, \href {https://ui.adsabs.harvard.edu/abs/1924PTarO..25f...1O} {25, 1}

\bibitem[\protect\citeauthoryear{Paczynski}{Paczynski}{1976}]{Paczynski1976}
Paczynski B.,  1976, \mn@doi [Symposium - International Astronomical Union] {10.1017/S0074180900011864}, 73, 75–80

\bibitem[\protect\citeauthoryear{{Paxton}, {Bildsten}, {Dotter}, {Herwig}, {Lesaffre}  \& {Timmes}}{{Paxton} et~al.}{2011}]{Paxton_2010}
{Paxton} B.,  {Bildsten} L.,  {Dotter} A.,  {Herwig} F.,  {Lesaffre} P.,   {Timmes} F.,  2011, \mn@doi [\apjs] {10.1088/0067-0049/192/1/3}, \href {https://ui.adsabs.harvard.edu/abs/2011ApJS..192....3P} {192, 3}

\bibitem[\protect\citeauthoryear{{Perets} \& {Fabrycky}}{{Perets} \& {Fabrycky}}{2009}]{PeretsFabrycky2009}
{Perets} H.~B.,  {Fabrycky} D.~C.,  2009, \mn@doi [\apj] {10.1088/0004-637X/697/2/1048}, \href {https://ui.adsabs.harvard.edu/abs/2009ApJ...697.1048P} {697, 1048}

\bibitem[\protect\citeauthoryear{{Perets} \& {Kratter}}{{Perets} \& {Kratter}}{2012}]{PeretsKratter2012}
{Perets} H.~B.,  {Kratter} K.~M.,  2012, \mn@doi [\apj] {10.1088/0004-637X/760/2/99}, \href {https://ui.adsabs.harvard.edu/abs/2012ApJ...760...99P} {760, 99}

\bibitem[\protect\citeauthoryear{Peters}{Peters}{1964}]{Peters64}
Peters P.~C.,  1964, \mn@doi [Phys. Rev.] {10.1103/PhysRev.136.B1224}, 136, B1224

\bibitem[\protect\citeauthoryear{{Petrovich}}{{Petrovich}}{2015}]{Petrovich2015}
{Petrovich} C.,  2015, \mn@doi [\apj] {10.1088/0004-637X/799/1/27}, \href {https://ui.adsabs.harvard.edu/abs/2015ApJ...799...27P} {799, 27}

\bibitem[\protect\citeauthoryear{{Pijloo}, {Caputo}  \& {Portegies Zwart}}{{Pijloo} et~al.}{2012}]{Pijloo2012}
{Pijloo} J.~T.,  {Caputo} D.~P.,   {Portegies Zwart} S.~F.,  2012, \mn@doi [\mnras] {10.1111/j.1365-2966.2012.21431.x}, \href {https://ui.adsabs.harvard.edu/abs/2012MNRAS.424.2914P} {424, 2914}

\bibitem[\protect\citeauthoryear{{Pols}, {Schr{\"o}der}, {Hurley}, {Tout}  \& {Eggleton}}{{Pols} et~al.}{1998}]{Pols1998}
{Pols} O.~R.,  {Schr{\"o}der} K.-P.,  {Hurley} J.~R.,  {Tout} C.~A.,   {Eggleton} P.~P.,  1998, \mn@doi [\mnras] {10.1046/j.1365-8711.1998.01658.x}, \href {https://ui.adsabs.harvard.edu/abs/1998MNRAS.298..525P} {298, 525}

\bibitem[\protect\citeauthoryear{{Portegies Zwart} \& {Verbunt}}{{Portegies Zwart} \& {Verbunt}}{1996}]{Portegies_Zwart1996}
{Portegies Zwart} S.~F.,  {Verbunt} F.,  1996, \aap, \href {https://ui.adsabs.harvard.edu/abs/1996A&A...309..179P} {309, 179}

\bibitem[\protect\citeauthoryear{{Rakavy} \& {Shaviv}}{{Rakavy} \& {Shaviv}}{1967}]{Rakavy1967}
{Rakavy} G.,  {Shaviv} G.,  1967, \mn@doi [\apj] {10.1086/149204}, \href {https://ui.adsabs.harvard.edu/abs/1967ApJ...148..803R} {148, 803}

\bibitem[\protect\citeauthoryear{{Reimers}}{{Reimers}}{1975}]{Reimers1975}
{Reimers} D.,  1975, Memoires of the Societe Royale des Sciences de Liege, \href {https://ui.adsabs.harvard.edu/abs/1975MSRSL...8..369R} {8, 369}

\bibitem[\protect\citeauthoryear{{Renzo}, {Farmer}, {Justham}, {G{\"o}tberg}, {de Mink}, {Zapartas}, {Marchant}  \& {Smith}}{{Renzo} et~al.}{2020}]{Renzo2020}
{Renzo} M.,  {Farmer} R.,  {Justham} S.,  {G{\"o}tberg} Y.,  {de Mink} S.~E.,  {Zapartas} E.,  {Marchant} P.,   {Smith} N.,  2020, \mn@doi [\aap] {10.1051/0004-6361/202037710}, \href {https://ui.adsabs.harvard.edu/abs/2020A&A...640A..56R} {640, A56}

\bibitem[\protect\citeauthoryear{{Renzo} et~al.,}{{Renzo} et~al.}{2021}]{Renzo_2021}
{Renzo} M.,  et~al., 2021, \mn@doi [\apj] {10.3847/1538-4357/ac1110}, \href {https://ui.adsabs.harvard.edu/abs/2021ApJ...919..128R} {919, 128}

\bibitem[\protect\citeauthoryear{{Riley}, {Mandel}, {Marchant}, {Butler}, {Nathaniel}, {Neijssel}, {Shortt}  \& {Vigna-G{\'o}mez}}{{Riley} et~al.}{2021}]{Riley2021MNRAS.505..663R}
{Riley} J.,  {Mandel} I.,  {Marchant} P.,  {Butler} E.,  {Nathaniel} K.,  {Neijssel} C.,  {Shortt} S.,   {Vigna-G{\'o}mez} A.,  2021, \mn@doi [\mnras] {10.1093/mnras/stab1291}, \href {https://ui.adsabs.harvard.edu/abs/2021MNRAS.505..663R} {505, 663}

\bibitem[\protect\citeauthoryear{{Riley} et~al.,}{{Riley} et~al.}{2022}]{Riley2022:compas_paper}
{Riley} J.,  et~al., 2022, \mn@doi [\apjs] {10.3847/1538-4365/ac416c}, \href {https://ui.adsabs.harvard.edu/abs/2022ApJS..258...34R} {258, 34}

\bibitem[\protect\citeauthoryear{{Rodriguez} \& {Antonini}}{{Rodriguez} \& {Antonini}}{2018}]{RodriguezAntonini2018}
{Rodriguez} C.~L.,  {Antonini} F.,  2018, \mn@doi [\apj] {10.3847/1538-4357/aacea4}, \href {https://ui.adsabs.harvard.edu/abs/2018ApJ...863....7R} {863, 7}

\bibitem[\protect\citeauthoryear{{Rozner} \& {Perets}}{{Rozner} \& {Perets}}{2022}]{Rozner2022}
{Rozner} M.,  {Perets} H.~B.,  2022, \mn@doi [\apj] {10.3847/1538-4357/ac6d55}, \href {https://ui.adsabs.harvard.edu/abs/2022ApJ...931..149R} {931, 149}

\bibitem[\protect\citeauthoryear{{Sabhahit}, {Vink}, {Higgins}  \& {Sander}}{{Sabhahit} et~al.}{2021}]{Sabhahit2021}
{Sabhahit} G.~N.,  {Vink} J.~S.,  {Higgins} E.~R.,   {Sander} A. A.~C.,  2021, \mn@doi [\mnras] {10.1093/mnras/stab1948}, \href {https://ui.adsabs.harvard.edu/abs/2021MNRAS.506.4473S} {506, 4473}

\bibitem[\protect\citeauthoryear{{Sana} et~al.,}{{Sana} et~al.}{2012}]{Sana2012}
{Sana} H.,  et~al., 2012, \mn@doi [Science] {10.1126/science.1223344}, \href {https://ui.adsabs.harvard.edu/abs/2012Sci...337..444S} {337, 444}

\bibitem[\protect\citeauthoryear{{Sana} et~al.,}{{Sana} et~al.}{2014}]{Sana2014}
{Sana} H.,  et~al., 2014, \mn@doi [\apjs] {10.1088/0067-0049/215/1/15}, \href {https://ui.adsabs.harvard.edu/abs/2014ApJS..215...15S} {215, 15}

\bibitem[\protect\citeauthoryear{Sander \& Vink}{Sander \& Vink}{2020}]{Sander_2020}
Sander A. A.~C.,  Vink J.~S.,  2020, \mn@doi [\mnras] {10.1093/mnras/staa2712}, 499, 873–892

\bibitem[\protect\citeauthoryear{{Schootemeijer}, {Langer}, {Grin}  \& {Wang}}{{Schootemeijer} et~al.}{2019}]{Schootemeijer2019}
{Schootemeijer} A.,  {Langer} N.,  {Grin} N.~J.,   {Wang} C.,  2019, \mn@doi [\aap] {10.1051/0004-6361/201935046}, \href {https://ui.adsabs.harvard.edu/abs/2019A&A...625A.132S} {625, A132}

\bibitem[\protect\citeauthoryear{{Schr{\o}der}, {MacLeod}, {Ramirez-Ruiz}, {Mandel}, {Fragos}, {Loeb}  \& {Everson}}{{Schr{\o}der} et~al.}{2021}]{Schroder2021}
{Schr{\o}der} S.~L.,  {MacLeod} M.,  {Ramirez-Ruiz} E.,  {Mandel} I.,  {Fragos} T.,  {Loeb} A.,   {Everson} R.~W.,  2021, arXiv e-prints, \href {https://ui.adsabs.harvard.edu/abs/2021arXiv210709675S} {p. arXiv:2107.09675}

\bibitem[\protect\citeauthoryear{{Shappee} \& {Thompson}}{{Shappee} \& {Thompson}}{2013}]{ShappeeThompson2013}
{Shappee} B.~J.,  {Thompson} T.~A.,  2013, \mn@doi [\apj] {10.1088/0004-637X/766/1/64}, \href {https://ui.adsabs.harvard.edu/abs/2013ApJ...766...64S} {766, 64}

\bibitem[\protect\citeauthoryear{{Silsbee} \& {Tremaine}}{{Silsbee} \& {Tremaine}}{2017}]{SilsbeeTremaine2017}
{Silsbee} K.,  {Tremaine} S.,  2017, \mn@doi [\apj] {10.3847/1538-4357/aa5729}, \href {https://ui.adsabs.harvard.edu/abs/2017ApJ...836...39S} {836, 39}

\bibitem[\protect\citeauthoryear{{Siwek}, {Weinberger}  \& {Hernquist}}{{Siwek} et~al.}{2023}]{Siwek2023}
{Siwek} M.,  {Weinberger} R.,   {Hernquist} L.,  2023, \mn@doi [\mnras] {10.1093/mnras/stad1131}, \href {https://ui.adsabs.harvard.edu/abs/2023MNRAS.522.2707S} {522, 2707}

\bibitem[\protect\citeauthoryear{{Smeyers} \& {Willems}}{{Smeyers} \& {Willems}}{2001}]{SmeyersWillems2001}
{Smeyers} P.,  {Willems} B.,  2001, \mn@doi [\aap] {10.1051/0004-6361:20010563}, \href {https://ui.adsabs.harvard.edu/abs/2001A&A...373..173S} {373, 173}

\bibitem[\protect\citeauthoryear{{Smith}}{{Smith}}{2014}]{Smith2004}
{Smith} N.,  2014, \mn@doi [\araa] {10.1146/annurev-astro-081913-040025}, \href {https://ui.adsabs.harvard.edu/abs/2014ARA&A..52..487S} {52, 487}

\bibitem[\protect\citeauthoryear{{Soberman}, {Phinney}  \& {van den Heuvel}}{{Soberman} et~al.}{1997}]{Soberman1997}
{Soberman} G.~E.,  {Phinney} E.~S.,   {van den Heuvel} E.~P.~J.,  1997, \aap, \href {https://ui.adsabs.harvard.edu/abs/1997A&A...327..620S} {327, 620}

\bibitem[\protect\citeauthoryear{{Soker} \& {Bear}}{{Soker} \& {Bear}}{2021}]{Soker2021}
{Soker} N.,  {Bear} E.,  2021, \mn@doi [\mnras] {10.1093/mnras/stab1561}, \href {https://ui.adsabs.harvard.edu/abs/2021MNRAS.505.4791S} {505, 4791}

\bibitem[\protect\citeauthoryear{{Song}, {Meynet}, {Maeder}, {Ekstr{\"o}m}  \& {Eggenberger}}{{Song} et~al.}{2016}]{Song2016}
{Song} H.~F.,  {Meynet} G.,  {Maeder} A.,  {Ekstr{\"o}m} S.,   {Eggenberger} P.,  2016, \mn@doi [\aap] {10.1051/0004-6361/201526074}, \href {https://ui.adsabs.harvard.edu/abs/2016A&A...585A.120S} {585, A120}

\bibitem[\protect\citeauthoryear{{Stegmann}, {Antonini}, {Schneider}, {Tiwari}  \& {Chattopadhyay}}{{Stegmann} et~al.}{2022a}]{Stegmann2022}
{Stegmann} J.,  {Antonini} F.,  {Schneider} F. R.~N.,  {Tiwari} V.,   {Chattopadhyay} D.,  2022a, \mn@doi [\prd] {10.1103/PhysRevD.106.023014}, \href {https://ui.adsabs.harvard.edu/abs/2022PhRvD.106b3014S} {106, 023014}

\bibitem[\protect\citeauthoryear{{Stegmann}, {Antonini}  \& {Moe}}{{Stegmann} et~al.}{2022b}]{Stegman2022}
{Stegmann} J.,  {Antonini} F.,   {Moe} M.,  2022b, \mn@doi [\mnras] {10.1093/mnras/stac2192}, \href {https://ui.adsabs.harvard.edu/abs/2022MNRAS.516.1406S} {516, 1406}

\bibitem[\protect\citeauthoryear{{Stevenson}, {Sampson}, {Powell}, {Vigna-G{\'o}mez}, {Neijssel}, {Sz{\'e}csi}  \& {Mandel}}{{Stevenson} et~al.}{2019}]{Stevenson2019}
{Stevenson} S.,  {Sampson} M.,  {Powell} J.,  {Vigna-G{\'o}mez} A.,  {Neijssel} C.~J.,  {Sz{\'e}csi} D.,   {Mandel} I.,  2019, \mn@doi [\apj] {10.3847/1538-4357/ab3981}, \href {https://ui.adsabs.harvard.edu/abs/2019ApJ...882..121S} {882, 121}

\bibitem[\protect\citeauthoryear{{Stone}, {Metzger}  \& {Haiman}}{{Stone} et~al.}{2017}]{Stone2017}
{Stone} N.~C.,  {Metzger} B.~D.,   {Haiman} Z.,  2017, \mn@doi [\mnras] {10.1093/mnras/stw2260}, \href {https://ui.adsabs.harvard.edu/abs/2017MNRAS.464..946S} {464, 946}

\bibitem[\protect\citeauthoryear{{Swaruba Rajamuthukumar}, {Hamers}, {Neunteufel}, {Pakmor}  \& {de mink}}{{Swaruba Rajamuthukumar} et~al.}{2022}]{Rajamuthukumar2022}
{Swaruba Rajamuthukumar} A.,  {Hamers} A.,  {Neunteufel} P.,  {Pakmor} R.,   {de mink} S.~E.,  2022, arXiv e-prints, \href {https://ui.adsabs.harvard.edu/abs/2022arXiv221104463S} {p. arXiv:2211.04463}

\bibitem[\protect\citeauthoryear{{Sz{\'e}csi}, {Langer}, {Yoon}, {Sanyal}, {de Mink}, {Evans}  \& {Dermine}}{{Sz{\'e}csi} et~al.}{2015}]{Szecsi2015}
{Sz{\'e}csi} D.,  {Langer} N.,  {Yoon} S.-C.,  {Sanyal} D.,  {de Mink} S.,  {Evans} C.~J.,   {Dermine} T.,  2015, \mn@doi [\aap] {10.1051/0004-6361/201526617}, \href {https://ui.adsabs.harvard.edu/abs/2015A&A...581A..15S} {581, A15}

\bibitem[\protect\citeauthoryear{{Takahashi}, {Yoshida}  \& {Umeda}}{{Takahashi} et~al.}{2018}]{Takahashi2018A}
{Takahashi} K.,  {Yoshida} T.,   {Umeda} H.,  2018, \mn@doi [\apj] {10.3847/1538-4357/aab95f}, \href {https://ui.adsabs.harvard.edu/abs/2018ApJ...857..111T} {857, 111}

\bibitem[\protect\citeauthoryear{{The LIGO Scientific Collaboration} et~al.,}{{The LIGO Scientific Collaboration} et~al.}{2021}]{Abbott2021GWTC-3}
{The LIGO Scientific Collaboration} et~al., 2021, \mn@doi [arXiv e-prints] {10.48550/arXiv.2111.03606}, \href {https://ui.adsabs.harvard.edu/abs/2021arXiv211103606T} {p. arXiv:2111.03606}

\bibitem[\protect\citeauthoryear{{Thompson}}{{Thompson}}{2011}]{Thompson2011}
{Thompson} T.~A.,  2011, \mn@doi [\apj] {10.1088/0004-637X/741/2/82}, \href {https://ui.adsabs.harvard.edu/abs/2011ApJ...741...82T} {741, 82}

\bibitem[\protect\citeauthoryear{{Tiede}, {Zrake}, {MacFadyen}  \& {Haiman}}{{Tiede} et~al.}{2020}]{Tiede2020}
{Tiede} C.,  {Zrake} J.,  {MacFadyen} A.,   {Haiman} Z.,  2020, \mn@doi [\apj] {10.3847/1538-4357/aba432}, \href {https://ui.adsabs.harvard.edu/abs/2020ApJ...900...43T} {900, 43}

\bibitem[\protect\citeauthoryear{{Tokovinin}}{{Tokovinin}}{2010}]{Tokovinin_2010}
{Tokovinin} A.,  2010, VizieR Online Data Catalog, \href {https://ui.adsabs.harvard.edu/abs/2010yCat..73890925T} {p. J/MNRAS/389/925}

\bibitem[\protect\citeauthoryear{{Tokovinin}}{{Tokovinin}}{2014}]{Tokovinin2014}
{Tokovinin} A.,  2014, \mn@doi [\aj] {10.1088/0004-6256/147/4/87}, \href {https://ui.adsabs.harvard.edu/abs/2014AJ....147...87T} {147, 87}

\bibitem[\protect\citeauthoryear{{Tokovinin}, {Thomas}, {Sterzik}  \& {Udry}}{{Tokovinin} et~al.}{2006}]{Tokovinin2006}
{Tokovinin} A.,  {Thomas} S.,  {Sterzik} M.,   {Udry} S.,  2006, \mn@doi [\aap] {10.1051/0004-6361:20054427}, \href {https://ui.adsabs.harvard.edu/abs/2006A&A...450..681T} {450, 681}

\bibitem[\protect\citeauthoryear{{Toonen}, {Nelemans}  \& {Portegies Zwart}}{{Toonen} et~al.}{2012}]{Toonen_2012}
{Toonen} S.,  {Nelemans} G.,   {Portegies Zwart} S.,  2012, \mn@doi [\aap] {10.1051/0004-6361/201218966}, \href {https://ui.adsabs.harvard.edu/abs/2012A&A...546A..70T} {546, A70}

\bibitem[\protect\citeauthoryear{{Toonen}, {Hamers}  \& {Portegies Zwart}}{{Toonen} et~al.}{2016}]{Toonen2016}
{Toonen} S.,  {Hamers} A.,   {Portegies Zwart} S.,  2016, \mn@doi [Computational Astrophysics and Cosmology] {10.1186/s40668-016-0019-0}, \href {https://ui.adsabs.harvard.edu/abs/2016ComAC...3....6T} {3, 6}

\bibitem[\protect\citeauthoryear{{Toonen}, {Perets}  \& {Hamers}}{{Toonen} et~al.}{2018}]{Toonen2018}
{Toonen} S.,  {Perets} H.~B.,   {Hamers} A.~S.,  2018, \mn@doi [\aap] {10.1051/0004-6361/201731874}, \href {https://ui.adsabs.harvard.edu/abs/2018A&A...610A..22T} {610, A22}

\bibitem[\protect\citeauthoryear{{Toonen}, {Portegies Zwart}, {Hamers}  \& {Bandopadhyay}}{{Toonen} et~al.}{2020}]{Toonen2020}
{Toonen} S.,  {Portegies Zwart} S.,  {Hamers} A.~S.,   {Bandopadhyay} D.,  2020, \mn@doi [\aap] {10.1051/0004-6361/201936835}, \href {https://ui.adsabs.harvard.edu/abs/2020A&A...640A..16T} {640, A16}

\bibitem[\protect\citeauthoryear{{Toonen}, {Boekholt}  \& {Portegies Zwart}}{{Toonen} et~al.}{2022}]{Toonen2022}
{Toonen} S.,  {Boekholt} T.~C.~N.,   {Portegies Zwart} S.,  2022, \mn@doi [\aap] {10.1051/0004-6361/202141991}, \href {https://ui.adsabs.harvard.edu/abs/2022A&A...661A..61T} {661, A61}

\bibitem[\protect\citeauthoryear{{Tutukov} \& {Yungelson}}{{Tutukov} \& {Yungelson}}{1979}]{Tutukov1979}
{Tutukov} A.,  {Yungelson} L.,  1979, in {Conti} P.~S.,  {De Loore} C.~W.~H.,  eds, ~ Vol. 83, Mass Loss and Evolution of O-Type Stars. pp 401--406

\bibitem[\protect\citeauthoryear{{Ulrich} \& {Burger}}{{Ulrich} \& {Burger}}{1976}]{UlrichBurger1976}
{Ulrich} R.~K.,  {Burger} H.~L.,  1976, \mn@doi [\apj] {10.1086/154406}, \href {https://ui.adsabs.harvard.edu/abs/1976ApJ...206..509U} {206, 509}

\bibitem[\protect\citeauthoryear{{Vassiliadis} \& {Wood}}{{Vassiliadis} \& {Wood}}{1993}]{Vassiliadis1993}
{Vassiliadis} E.,  {Wood} P.~R.,  1993, \mn@doi [\apj] {10.1086/173033}, \href {https://ui.adsabs.harvard.edu/abs/1993ApJ...413..641V} {413, 641}

\bibitem[\protect\citeauthoryear{{Veras}, {Wyatt}, {Mustill}, {Bonsor}  \& {Eldridge}}{{Veras} et~al.}{2011}]{Veras2011}
{Veras} D.,  {Wyatt} M.~C.,  {Mustill} A.~J.,  {Bonsor} A.,   {Eldridge} J.~J.,  2011, \mn@doi [\mnras] {10.1111/j.1365-2966.2011.19393.x}, \href {https://ui.adsabs.harvard.edu/abs/2011MNRAS.417.2104V} {417, 2104}

\bibitem[\protect\citeauthoryear{Verbunt, Igoshev  \& Cator}{Verbunt et~al.}{2017}]{Verbunt2017}
Verbunt F.,  Igoshev A.,   Cator E.,  2017, \mn@doi [Astronomy & Astrophysics] {10.1051/0004-6361/201731518}, 608, A57

\bibitem[\protect\citeauthoryear{{Vigna-G{\'o}mez}, {Toonen}, {Ramirez-Ruiz}, {Leigh}, {Riley}  \& {Haster}}{{Vigna-G{\'o}mez} et~al.}{2021}]{VignaGomez2021}
{Vigna-G{\'o}mez} A.,  {Toonen} S.,  {Ramirez-Ruiz} E.,  {Leigh} N. W.~C.,  {Riley} J.,   {Haster} C.-J.,  2021, \mn@doi [\apjl] {10.3847/2041-8213/abd5b7}, \href {https://ui.adsabs.harvard.edu/abs/2021ApJ...907L..19V} {907, L19}

\bibitem[\protect\citeauthoryear{{Vigna-G{\'o}mez}, {Liu}, {Aguilera-Dena}, {Grishin}, {Ramirez-Ruiz}  \& {Soares-Furtado}}{{Vigna-G{\'o}mez} et~al.}{2022}]{VignaGomez2022}
{Vigna-G{\'o}mez} A.,  {Liu} B.,  {Aguilera-Dena} D.~R.,  {Grishin} E.,  {Ramirez-Ruiz} E.,   {Soares-Furtado} M.,  2022, \mn@doi [\mnras] {10.1093/mnrasl/slac067}, \href {https://ui.adsabs.harvard.edu/abs/2022MNRAS.515L..50V} {515, L50}

\bibitem[\protect\citeauthoryear{{Vigna-G{\'o}mez} et~al.,}{{Vigna-G{\'o}mez} et~al.}{2023}]{Vigna-Gomez2023}
{Vigna-G{\'o}mez} A.,  et~al., 2023, arXiv e-prints, \href {https://ui.adsabs.harvard.edu/abs/2023arXiv231001509V} {p. arXiv:2310.01509}

\bibitem[\protect\citeauthoryear{{Vink} \& {Sabhahit}}{{Vink} \& {Sabhahit}}{2023}]{VinkSabhahit2023}
{Vink} J.~S.,  {Sabhahit} G.~N.,  2023, \mn@doi [\aap] {10.1051/0004-6361/202347801}, \href {https://ui.adsabs.harvard.edu/abs/2023A&A...678L...3V} {678, L3}

\bibitem[\protect\citeauthoryear{{Vink} \& {de Koter}}{{Vink} \& {de Koter}}{2005}]{VinkdeKoter05}
{Vink} J.~S.,  {de Koter} A.,  2005, \mn@doi [\aap] {10.1051/0004-6361:20052862}, \href {https://ui.adsabs.harvard.edu/abs/2005A&A...442..587V} {442, 587}

\bibitem[\protect\citeauthoryear{{Vink}, {de Koter}  \& {Lamers}}{{Vink} et~al.}{2001}]{Vink2001}
{Vink} J.~S.,  {de Koter} A.,   {Lamers} H.~J.~G.~L.~M.,  2001, \mn@doi [\aap] {10.1051/0004-6361:20010127}, \href {https://ui.adsabs.harvard.edu/abs/2001A&A...369..574V} {369, 574}

\bibitem[\protect\citeauthoryear{{Wen}}{{Wen}}{2003}]{Wen2003}
{Wen} L.,  2003, \mn@doi [\apj] {10.1086/378794}, \href {https://ui.adsabs.harvard.edu/abs/2003ApJ...598..419W} {598, 419}

\bibitem[\protect\citeauthoryear{{Woods} \& {Ivanova}}{{Woods} \& {Ivanova}}{2011}]{Woods2011}
{Woods} T.~E.,  {Ivanova} N.,  2011, \mn@doi [\apjl] {10.1088/2041-8205/739/2/L48}, \href {https://ui.adsabs.harvard.edu/abs/2011ApJ...739L..48W} {739, L48}

\bibitem[\protect\citeauthoryear{{Woosley}}{{Woosley}}{2017}]{Woosley2017}
{Woosley} S.~E.,  2017, \mn@doi [\apj] {10.3847/1538-4357/836/2/244}, \href {https://ui.adsabs.harvard.edu/abs/2017ApJ...836..244W} {836, 244}

\bibitem[\protect\citeauthoryear{{Woosley} \& {Heger}}{{Woosley} \& {Heger}}{2021}]{WoosleyHeger:2021:MassGap}
{Woosley} S.~E.,  {Heger} A.,  2021, \mn@doi [\apjl] {10.3847/2041-8213/abf2c4}, \href {https://ui.adsabs.harvard.edu/abs/2021ApJ...912L..31W} {912, L31}

\bibitem[\protect\citeauthoryear{{Woosley} \& {Weaver}}{{Woosley} \& {Weaver}}{1982}]{Woosley1982}
{Woosley} S.~E.,  {Weaver} T.~A.,  1982, in {Rees} M.~J.,  {Stoneham} R.~J.,  eds,  NATO Advanced Study Institute (ASI) Series C Vol. 90, Supernovae: A Survey of Current Research. p.~79

\bibitem[\protect\citeauthoryear{{Yoon} \& {Langer}}{{Yoon} \& {Langer}}{2005}]{Yoon+2005}
{Yoon} S.-C.,  {Langer} N.,  2005, \aap, 443, 643

\bibitem[\protect\citeauthoryear{{Yoon}, {Langer}  \& {Norman}}{{Yoon} et~al.}{2006}]{Yoon+2006}
{Yoon} S.-C.,  {Langer} N.,   {Norman} C.,  2006, \aap, 460, 199

\bibitem[\protect\citeauthoryear{{Yoshida}, {Umeda}, {Maeda}  \& {Ishii}}{{Yoshida} et~al.}{2016}]{Yoshida2016}
{Yoshida} T.,  {Umeda} H.,  {Maeda} K.,   {Ishii} T.,  2016, \mn@doi [\mnras] {10.1093/mnras/stv3002}, \href {https://ui.adsabs.harvard.edu/abs/2016MNRAS.457..351Y} {457, 351}

\bibitem[\protect\citeauthoryear{{de Kool}, {van den Heuvel}  \& {Pylyser}}{{de Kool} et~al.}{1987}]{deKool1987}
{de Kool} M.,  {van den Heuvel} E.~P.~J.,   {Pylyser} E.,  1987, \aap, \href {https://ui.adsabs.harvard.edu/abs/1987A&A...183...47D} {183, 47}

\bibitem[\protect\citeauthoryear{{de Mink} \& {King}}{{de Mink} \& {King}}{2017}]{deMinkKing2017}
{de Mink} S.~E.,  {King} A.,  2017, \mn@doi [\apjl] {10.3847/2041-8213/aa67f3}, \href {https://ui.adsabs.harvard.edu/abs/2017ApJ...839L...7D} {839, L7}

\bibitem[\protect\citeauthoryear{de Mink \& Mandel}{de~Mink \& Mandel}{2016}]{de_Mink_2016}
de Mink S.~E.,  Mandel I.,  2016, \mn@doi [\mnras] {10.1093/mnras/stw1219}, 460, 3545–3553

\bibitem[\protect\citeauthoryear{{de Mink}, {Cantiello}, {Langer}, {Pols}, {Brott}  \& {Yoon}}{{de Mink} et~al.}{2009}]{deMink2009}
{de Mink} S.~E.,  {Cantiello} M.,  {Langer} N.,  {Pols} O.~R.,  {Brott} I.,   {Yoon} S.~C.,  2009, \mn@doi [\aap] {10.1051/0004-6361/200811439}, \href {https://ui.adsabs.harvard.edu/abs/2009A&A...497..243D} {497, 243}

\bibitem[\protect\citeauthoryear{de Vries, Portegies~Zwart  \& Figueira}{de~Vries et~al.}{2014}]{deVries2014}
de Vries N.,  Portegies~Zwart S.,   Figueira J.,  2014, \mn@doi [\mnras] {10.1093/mnras/stt1688}, 438, 1909

\bibitem[\protect\citeauthoryear{deBoer et~al.,}{deBoer et~al.}{2017}]{deBoer2017}
deBoer R.~J.,  et~al., 2017, \mn@doi [Rev. Mod. Phys.] {10.1103/RevModPhys.89.035007}, 89, 035007

\bibitem[\protect\citeauthoryear{{du Buisson} et~al.,}{{du Buisson} et~al.}{2020}]{duBuisson2020}
{du Buisson} L.,  et~al., 2020, \mn@doi [\mnras] {10.1093/mnras/staa3225}, \href {https://ui.adsabs.harvard.edu/abs/2020MNRAS.499.5941D} {499, 5941}

\bibitem[\protect\citeauthoryear{{van den Heuvel}, {Portegies Zwart}  \& {de Mink}}{{van den Heuvel} et~al.}{2017}]{vandenheuvel17}
{van den Heuvel} E.~P.~J.,  {Portegies Zwart} S.~F.,   {de Mink} S.~E.,  2017, \mn@doi [\mnras] {10.1093/mnras/stx1430}, \href {https://ui.adsabs.harvard.edu/abs/2017MNRAS.471.4256V} {471, 4256}

\bibitem[\protect\citeauthoryear{{von Zeipel}}{{von Zeipel}}{1910}]{1910AN....183..345V}
{von Zeipel} H.,  1910, \mn@doi [Astronomische Nachrichten] {10.1002/asna.19091832202}, \href {https://ui.adsabs.harvard.edu/abs/1910AN....183..345V} {183, 345}

\makeatother
\end{thebibliography}




\appendix

\section{Additional figures}
\label{sec:additional_figures}

In this section, we present the low metallicity model  (i.e. Z = 0.0005) counterparts of some of the figures presented in the main text. In Fig. \ref{fig:m_transferred_ext_z_high}, we show the mass ratios at the onset of TMT and the amount of (relative) mass transferred towards the inner binary. In Fig. \ref{fig:inner_ecc_teritary_rlof_z_low}, we show the cumulative distribution of eccentricities for CHE triples that experience TMT at the onset of the mass transfer phase. In \ref{fig:m2_bh_frac_z_low}, we show the $M_{\rm 2,ZAMS}$ distribution of TMT sources, distinguishing them based on the evolutionary phase of the inner binary. In Fig. \ref{fig:uter_perti_types_z_low} we show the distribution of initial inner pericenters of CHE triples, distinguishing systems based on the maxium inner eccentricity reached during evolution (left panel), and the based on the evolutionary channel (right panel). In Fig \ref{fig:a_outer_evol_zlow}, we show the outer pericentre before and after the TMT episode for systems with MS-MS inner binaries (upper panel) and BH-BH inner binaries (lower panel) at the onset of the mass transfer phase.

\begin{figure*}
\includegraphics[trim=1.5cm 0.5cm 2.5cm 0cm, clip, width=\textwidth]{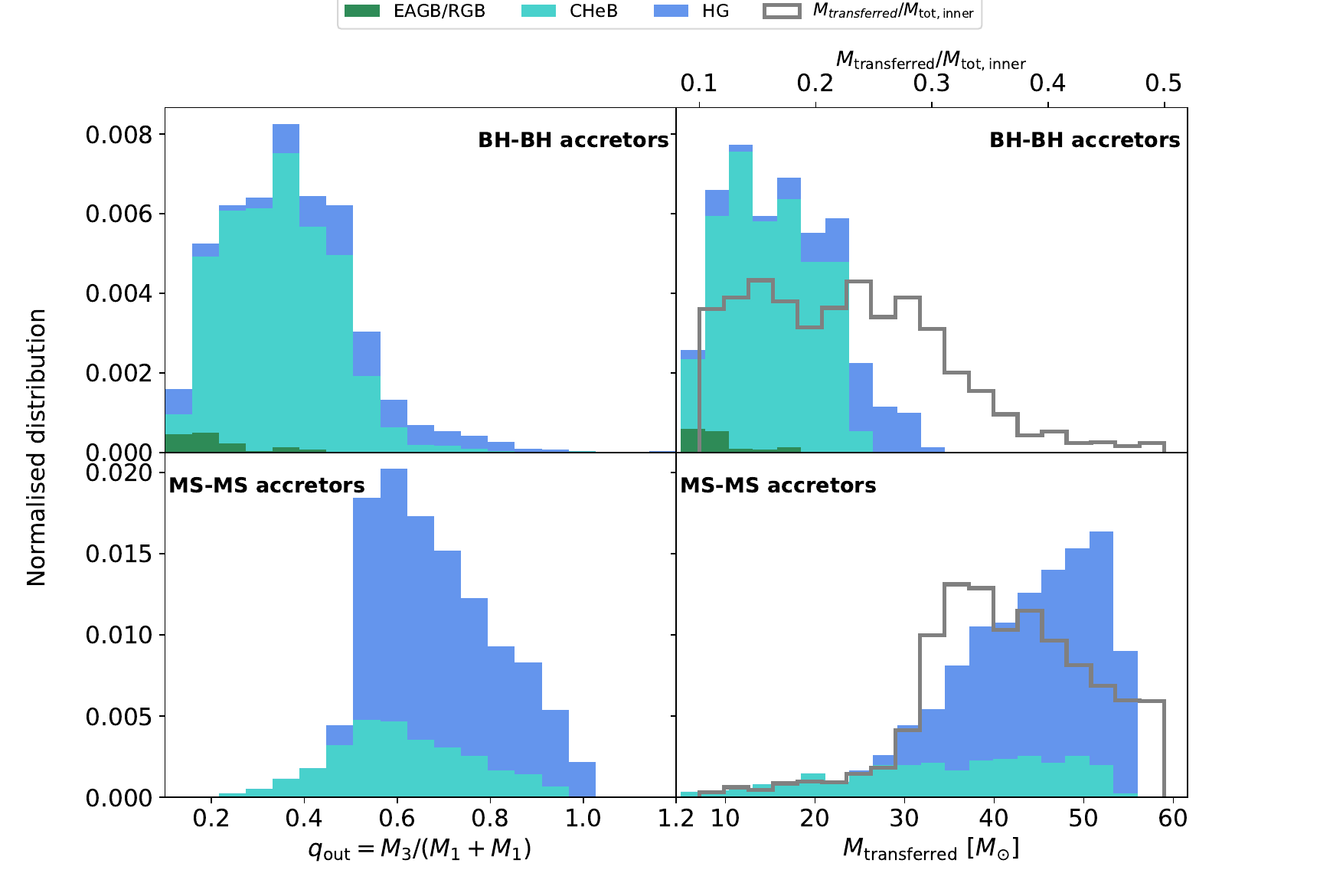}
\caption{The same figure as Fig. \ref{fig:m_transferred_ext_z_high} but for our model at Z = 0.0005.
}
\label{fig:m_transferred_ext_z_low}
\end{figure*}

\begin{figure}
\includegraphics[width=\columnwidth]{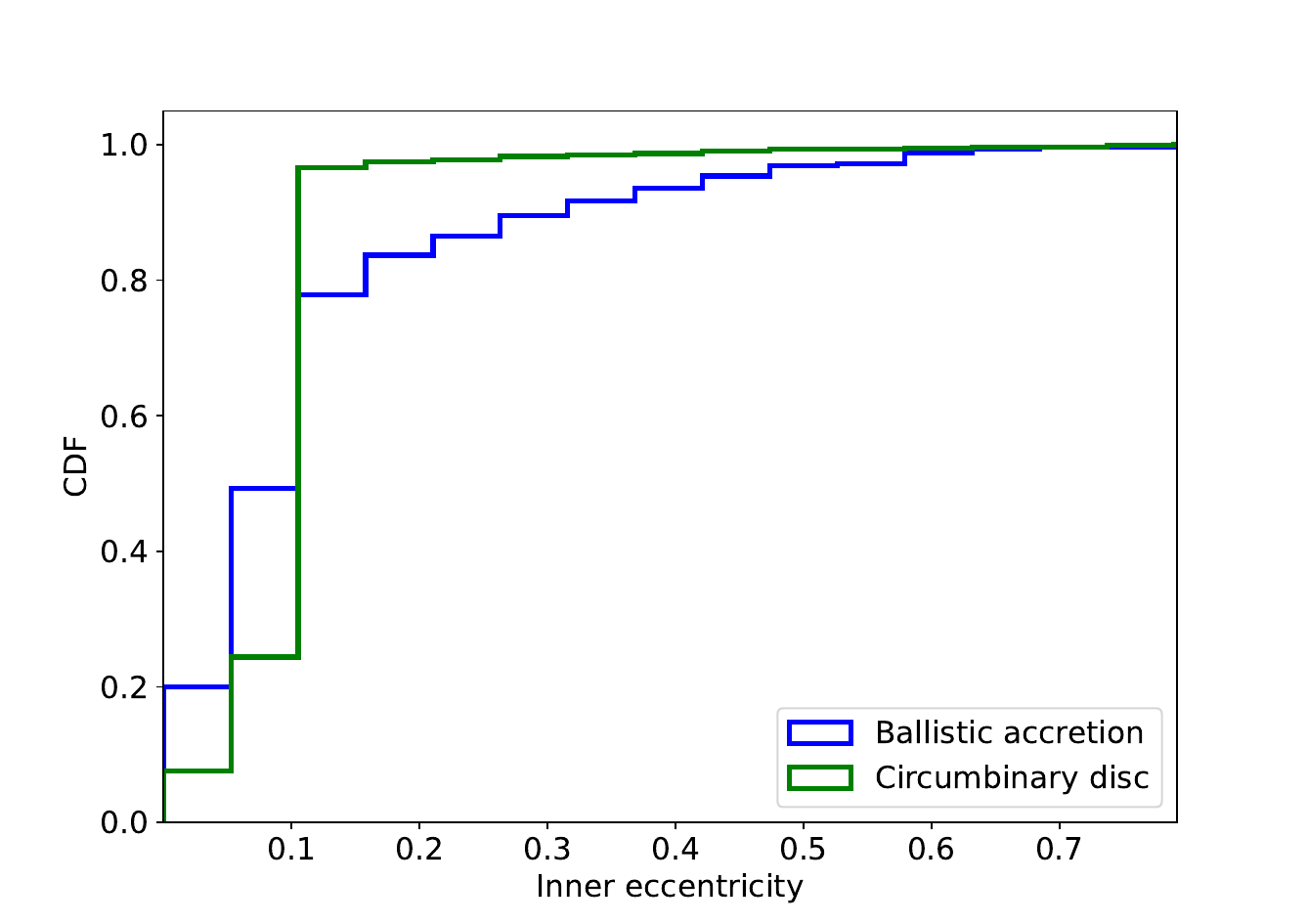}
\caption{The same figure as Fig. \ref{fig:inner_ecc_teritary_rlof} but for our model at Z = 0.0005.
}
\label{fig:inner_ecc_teritary_rlof_z_low}
\end{figure}

\begin{figure}
\includegraphics[trim=0 0 0.0cm 0, clip,width=\columnwidth]{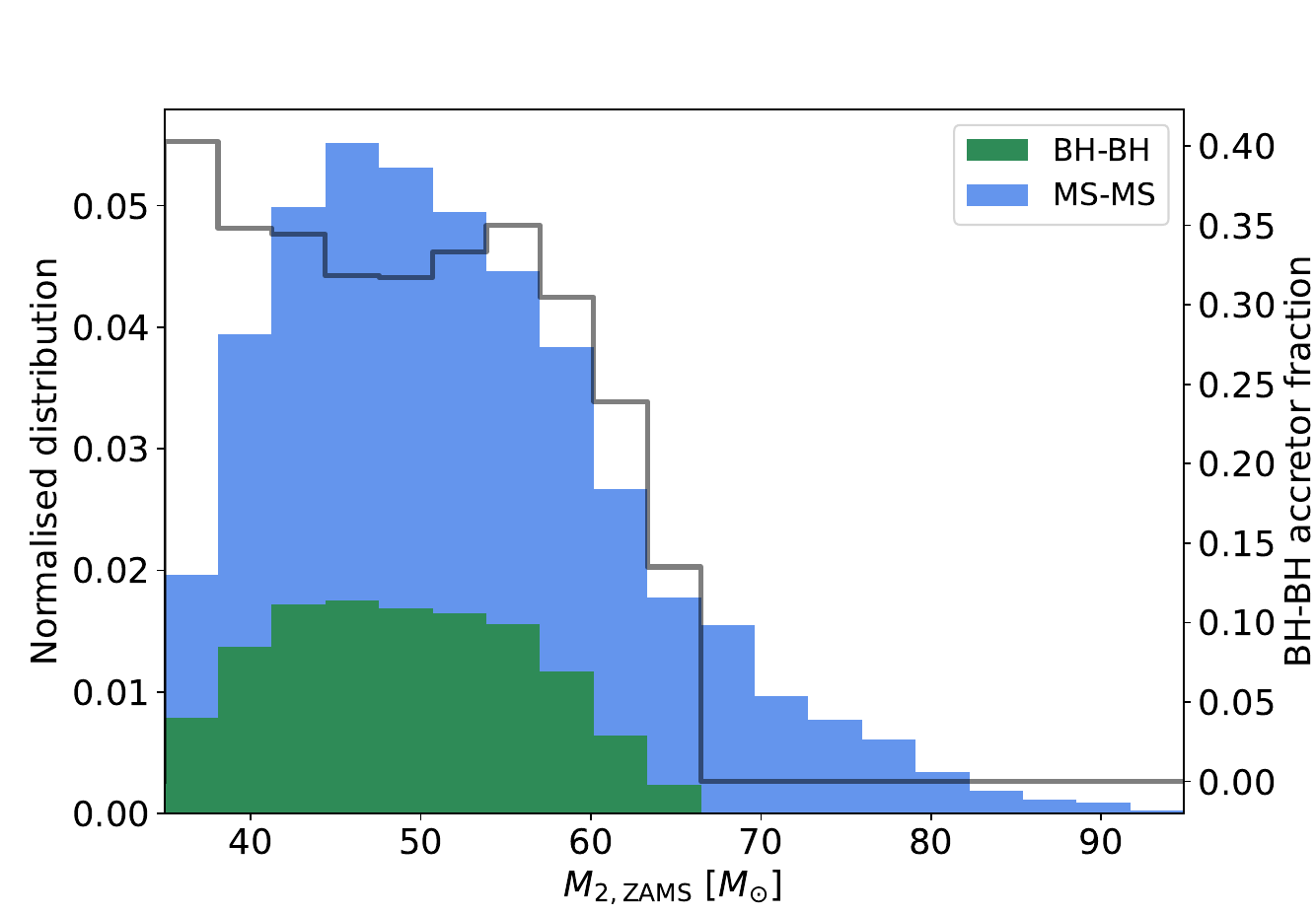}
\caption{The same as Fig. \ref{fig:m2_bh_frac_z_high} but for our model at Z = 0.0005 }
\label{fig:m2_bh_frac_z_low}
\end{figure}

\begin{figure*}
\includegraphics[trim=0 0 1cm 0, clip,width=\columnwidth]
{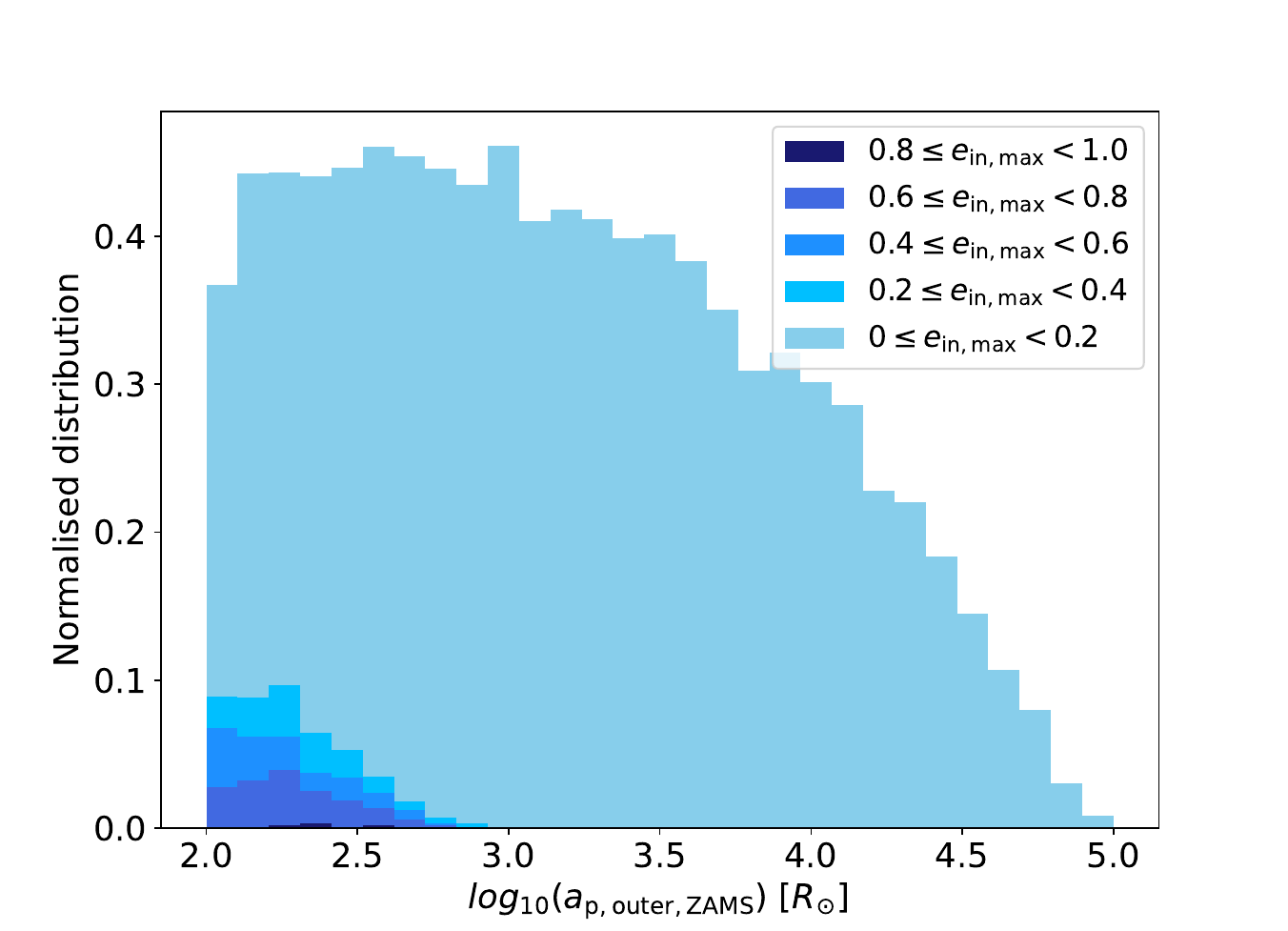} \hfill
\includegraphics[trim=0 0 1cm 0, clip,width=\columnwidth]
{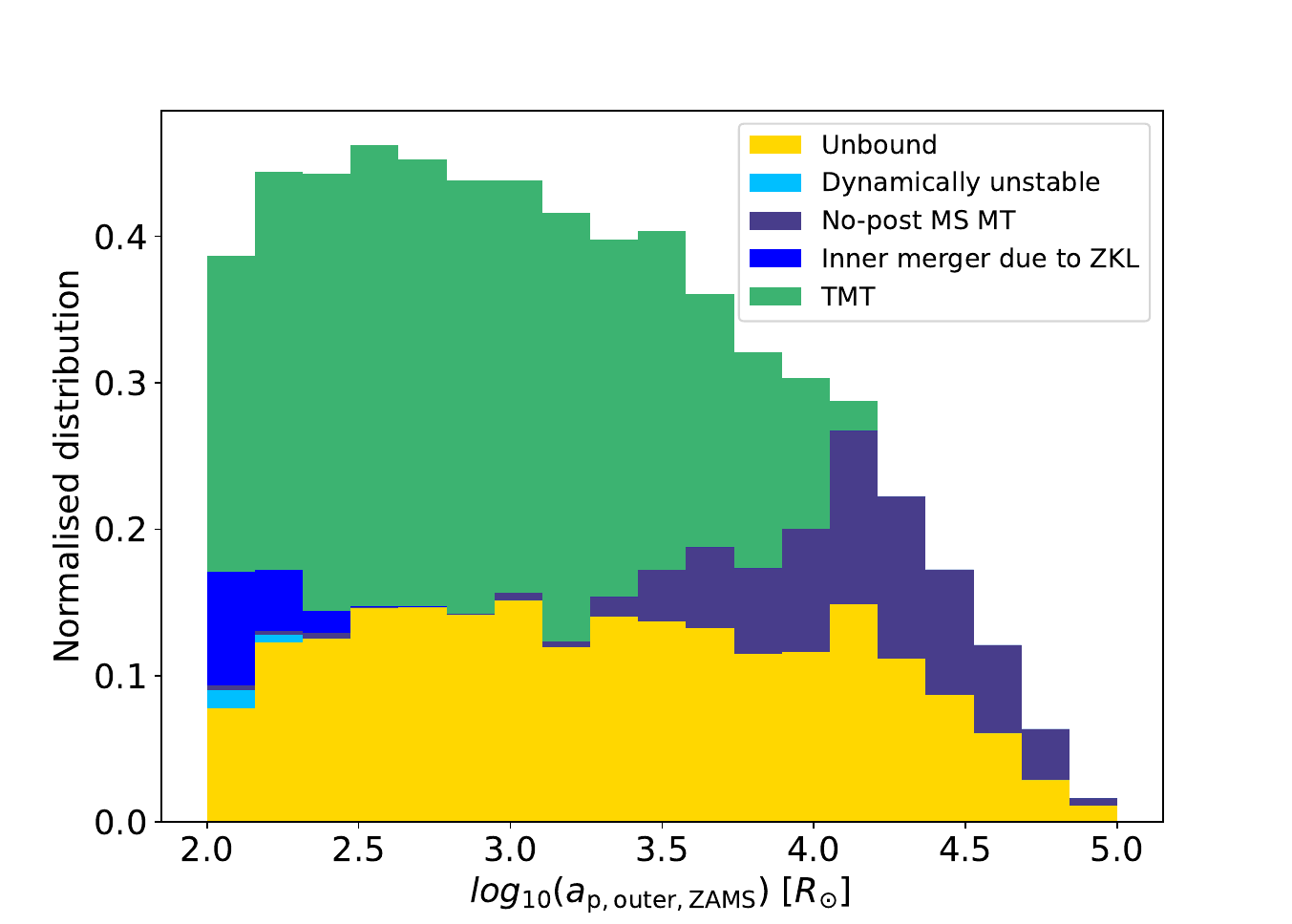}
\caption{The same as Fig. \ref{fig:uter_perti_types_z_high} but at Z = 0.0005} 
\label{fig:uter_perti_types_z_low}
\end{figure*}

\begin{figure}
    \includegraphics[width=\columnwidth]{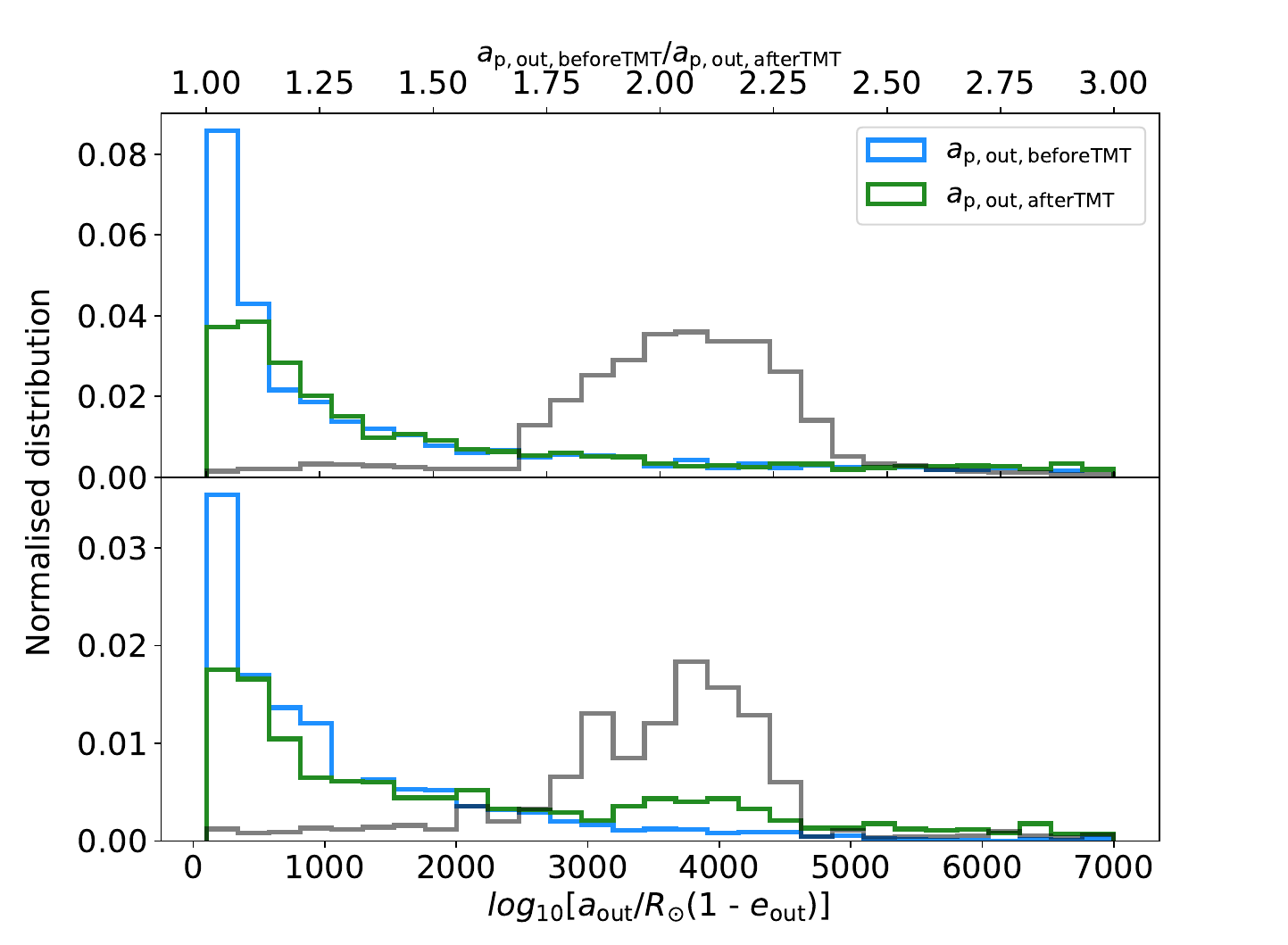}
  \caption{The orbital separation of the outer orbit before and at the onset of TMT for MS-MS inner  (upper panel) and BH-BH inner binary accretors (lower panel) as calculated with equation \ref{eq:iso} at Z = 0.0005.}
\label{fig:a_outer_evol_zlow}
\end{figure}



\section{Calculation of birth and merger rate density}
\label{subsec:app_norm}

Throughout the paper, we estimate the:
\begin{enumerate}
    \item Formation efficiency (equation \ref{eq:formation_efficiency})
    \item Birth rate density (equation \ref{eq:birth_rate})  
    \item Merger rate density (equation \ref{eq:event_rate})
\end{enumerate}
for each identified evolutionary channels. In this section, we discuss in detail how we determine these quantities.

\textit{(i) Formation efficiency:} The formation efficiency expresses the number of ZAMS stellar systems formed that will evolve according to a specific evolutionary channel as a fraction of all ZAMS stellar systems formed. We calculate this quantity as:
\begin{equation}
\label{eq:formation_efficiency}
    \epsilon_{\rm formation} =  f_{\rm pm} \cdot \frac{N_{\rm channel}}{N_{\rm simulated}},
\end{equation}
where $N_{\rm channel}$ is the number simulated systems that evolves according to the channel of interest, $N_{\rm \rm simulated}$ the total number of sampled systems, and $f_{\rm pm}$ is the portion of the simulated parameter space with respect to the complete parameter space, that is:
\begin{equation}
\label{eq:fpm}
    f_{\rm pm} =  f_{\rm triple}\cdot f_{\rm M_{\rm 1, ZAMS}} \cdot f_{\rm q, in} \cdot f_{\rm q, out} \cdot f_{\rm a, in} \cdot f_{\rm a, out},
\end{equation}
where $f_{\rm triple}$ is the assumed triple fraction, $f_{\rm M_{\rm 1, ZAMS}}$ is the fraction of the simulated parameter space of primary masses:
\begin{equation}
    f_{\rm M_{\rm 1, ZAMS}} = \frac{\int_{20\,M_{\odot}}^{100\,M_{\odot}} M_{\rm 1, ZAMS}^{-2.3} dm}{\int_{0.08\,M_{\odot}}^{0.5\,M_{\odot}} M_{\rm 1, ZAMS}^{-1.3} dm + \int_{0.5\,M_{\odot}}^{100\,M_{\odot}} M_{\rm 1, ZAMS}^{-2.3} dm},
\end{equation}
where we assumed that the absolute minimum stellar mass is $M_{\rm ZAMS,min} = 0.08\,M_{\odot}$ and the absolute maximum stellar mass is $M_{\rm ZAMS,max} = 100\,M_{\odot}$, and as explained in section \ref{subsec:init}, we sample primary masses in the range of 20-100$\,M_{\odot}$.
The fraction of the simulated parameter space of inner mass ratios is:
\begin{equation}
\label{eq:f_qin}
    f_{\rm q, in} = \frac{1.0-0.7}{1.0-0.0},
\end{equation}
since the distribution of (inner and outer) mass ratios is assumed to be uniform. In equation \ref{eq:f_qin}, we assume that inner mass ratios of hierarchical triples have an interval of (0,1] and we sample from the interval of [0.7,1].
The fraction of the simulated parameter space of outer mass ratios is
\begin{equation}
    f_{\rm q, out} = \frac{1.0-0.1}{1.0-0.0},
\end{equation}
where we assume that outer mass ratios triples have an interval of (0,1] and we sample from the interval of [0.1,1].
The fraction of the simulated parameter space of inner semimajor axis is:
\begin{equation}
    f_{\rm a, in} = \frac{\rm{log}_{10}(40\,R_{\odot})-\rm{log}_{10}(14\,R_{\odot})}{\rm{log}_{10}(10^5\,R_{\odot})-\rm{log}_{10}(14\,R_{\odot})},
\end{equation}
since the distribution of (inner and outer) semimajor axis is assumed to be uniform in a logarithmic space. We assume that inner mass semimajor axes of all triples range from 14$\,R_{\odot}$ to $10^5\,R_{\odot}$ and we sample from the interval of [14,40] $R_{\odot}$.
Finally, the fraction of the simulated parameter space of outer semimajor axis is:
\begin{equation}
    f_{\rm a, out} = \frac{\rm{log}_{10}(10^5\,R_{\odot})-\rm{log}_{10}(10^2\,R_{\odot})}{\rm{log}_{10}(10^5\,R_{\odot})-\rm{log}_{10}(10^2\,R_{\odot})},
\end{equation}
where we assume that inner mass semimajor axes of all triples range from $10^2\,R_{\odot}$ to $10^5\,R_{\odot}$ and we sample from the enitre interval. Equation \ref{eq:fpm} for channels involving isolated binaries reduces to $f_{\rm pm} = f_{\rm binary}\cdot f_{\rm M_{\rm 1, ZAMS}} \cdot f_{\rm q, in} \cdot f_{\rm a, in}$.

\textit{ii) Birth rate density:} The birth rate density gives the number density of ZAMS stellar systems in the local universe (that is at redshift $z\approx0$), which will evolve according to a specific channel. 
We calculate the birth rate of systems in a certain channel as:
\begin{equation}
\label{eq:birth_rate}
    R_{\rm birth} = \sum_{Z_i}\frac{\textrm{SFRd}^{*}(Z_i,z_{\rm ZAMS} = 0)}{\Tilde{M}}\cdot \epsilon_{\rm formation},
\end{equation}
where we sum over the two metallicity values, at which we performed our simulations; Z = 0.005 and Z = 0.0005.  $\rm{SFRd}^{*}(Z, z)$ is defined as the metallicity-specific star formation rate density, and it gives the stellar mass formed within a metallicity range $Z_{\rm low} \leq Z \leq Z_{\rm high}$ at redshift $z$:
\begin{equation}
\label{eq:SFRdz}
    \rm{SFRd}^{*}(Z, z) = \int_{Z_{\rm low}}^{Z_{\rm high}} f_{\rm met}(Z, z) \rm{SFRd}(z) dZ,
\end{equation}
where $Z_{\rm low}$ and $Z_{\rm high}$ are 0.0015 ($10^{-10}$) and 0.01 (0.0015), respectively, for our model with Z = 0.005 (Z = 0.0005). Here, Z = 0.0015 is the midpoint between Z = 0.005 and Z = 0.0005 in logarithmic space, Z = 0.01 is the highest metallicity at which CHE binaries can still form GW sources at appreciable numbers and Z = $10^{-10}$ is an arbitrarily chosen, extremely low metallicity value. In equation \ref{eq:SFRdz}, $\rm{SFRd}(z)$ is the star formation rate density, and we use the model from \citet{MadauDickinson2014}:
\begin{equation}
\label{eq:sfrd}
    \textrm{SFRd}(z) = \frac{0.01\cdot(1 + z)^{2.6}}{1 + ((1+z)/3.2)^{6.2}} \, M_{\odot}\rm{yr^{-1}}Mpc^{-3},
\end{equation}
and $f_{\rm met}(Z,z)$ is the metallicity distribution of the stellar mass formed. This quantity is also redshift dependent and assumed to follow a log-normal distribution \citep{Madau_2017}:
\begin{equation}
    \label{eq:fmet}
    f_{\rm met}(Z,z) = \frac{1}{\sigma \sqrt{2\pi}}\exp\left(\frac{(\log_{10}(Z) - \mu(z))^2}{2\sigma^2}\right),
\end{equation}
with a standard deviation of $\sigma = 0.5$ and with a redshift-dependent mean metallicity $\mu(z) = \log_{10}(Z_{\odot}\cdot 10 ^{0.153 - 0.074z^{1.34}}) - 0.5\rm{ln}(10)\sigma^2$.
Finally, the term $\tilde{M}$ in equation \ref{eq:birth_rate} is the average mass of all stellar systems and we calculate this as:
\begin{equation}
\label{eq:avgmass}
\begin{split}
&\tilde{M}=f_{\rm single}\cdot \tilde{M}_{\rm 1,ZAMS}  +\\ 
&f_{\rm binary}\cdot \int_{0}^{1} (1 + q_{\rm in}) \tilde{M}_{\rm 1,ZAMS}  dq_{\rm in}  +\\
&f_{\rm triple}\cdot \int_{0}^{1} \int_{0}^{1}  (1 + q_{\rm in}) (1 + q_{\rm out}) \tilde{M}_{\rm 1,ZAMS}  dq_{\rm in} dq_{\rm out},
\end{split}
\end{equation}
where we have defined $\tilde{M}_{\rm 1,ZAMS}$, as the average mass of the primary, i.e.:
\begin{equation}
    \tilde{M}_{\rm 1,ZAMS} = \int_{0.08\,M_{\odot}}^{100\,M_{\odot}} M_{\rm 1,ZAMS} f_{\rm IMF} dM_{\rm 1,ZAMS}
\end{equation}
where $f_{\rm IMF}$ is the normalised, piecewise continuous initial mass function of \citet{Kroupa},  $f_{\rm single}$ and $f_{\rm binary}$ are the single and binary fractions, respectively. We neglect higher order systems, such that $f_{\rm triple} = 1 -f_{\rm single} - f_{\rm binary}$. 

We note that we also assume that the binary and triple fractions are independent on the primary mass of the system \citep[which is clearly not consistent with observations, see e.g.][but commonly assumed in population synthesis studies as a simplification]{MoeDiStefano2017}. Assuming flat mass ratio distributions for both the inner and outer binary, equation \ref{eq:avgmass} becomes:
\begin{equation}
\label{eq:avgmass_simp}
    \tilde{M} = \left(f_{\rm single} + \frac{3}{2} f_{\rm binary} + \frac{9}{4}  \cdot f_{\rm triple}\right) \cdot \tilde{M}_{\rm 1,ZAMS},
\end{equation}
The term $\textrm{SFRd}^{*}(Z_i,z_{\rm ZAMS} = 0)/\tilde{M}$ in equation \ref{eq:birth_rate} then gives the average number of stars formed at redshift $z = 0$ in a metallicity range of $Z_{\rm i,low} \leq Z \leq Z_{\rm i, high}$. Multiplying this term with $\epsilon_{\rm formation}$ gives the number of systems formed in a given formation channel as a fraction of all systems formed in the above mentioned metallicity range for a given star formation history model. Summing these values over all of our metallicity bins therefore yields the total birth rate of systems in a specific channel.

\textit{iii) Merger rate density:} The merger rate density gives the rate density of a given astrophysical event (such as GW transients from coalescing double compact objects) in the local universe. The main difference between the birth and merger rate is due to the considerable delay time between the formation of the stellar system and the occurrence of the GW merger. For example, if the delay time for a GW source at $z = 0$ is $t_{\rm delay} = $ 10.5 Gyr, then the redshift at ZAMS of its progenitor systems is $z_{\rm ZAMS} \approx 2$, at which the star formation rate density is an order of magnitude higher with respect to its value at $z = 0$ \citep[e.g. see models of][]{MadauDickinson2014}.
 We determine the merger rate  density at $z = 0$ as:
\begin{equation}
\label{eq:event_rate}
    R_{\rm event} = \sum_{Z_i}  \int_{0\,\rm{Gyr}}^{13.5\,\rm{Gyr}}\frac{\textrm{SFRd}^{*}(Z_i,z_{\rm ZAMS}(t_{\rm delay}))}{\Tilde{M}}\cdot \tilde{\epsilon}(t_{\rm delay})  dt_{\rm delay},
\end{equation}
where $z_{\rm ZAMS}$ is the redshift at which the progenitor of a given astrophysical event is formed (and therefore it is a function of delay time), $\tilde{\epsilon}$ is the number of astrophysical events occurring at z = 0 with a delay time of $t_{\rm delay}$ as a fraction of all ZAMS stellar
systems formed at z = $z_{\rm ZAMS}$. We determine $z_{\rm ZAMS}$ for a given delay time via the standard relation for lookback time:
\begin{equation}
    t_{\rm delay} = \frac{1}{H_0}\int_{z = 0}^{z_{\rm ZAMS}} \frac{dz'}{(1 + z') E(z')},
\end{equation}
where $E(z) = \sqrt{\Omega_m(1+z)^3 + \Omega_{\lambda}}$, with $\Omega_M = 0.3$, $\Omega_{\lambda} = 0.7$ and $H_{0} = 70 \rm{kms^{-1}}\rm{Mpc}^{-1}$.


 We note that our merger rate density should be only considered as an order of magnitude estimate at best. This imprecision is due to several uncertainties in stellar physics and, notably, the limited density of our metallicity grid. We performed simulations only at two metallicities to determine the merger rate density. However, the formation efficiency and delay times of GW sources originating from CHE systems is expected to be sensitively dependent on metallicity.

In particular, we overestimate the delay times for GW sources formed at $0.001<Z\leq0.005$, which in turn leads to an overestimation of the merger rate density at z = 0. This is because, we represent all systems formed in this metallicity range with our models at Z = 0.005, at which the stellar winds are stronger and therefore lead to wider BH-BH binaries. The longer time delays imply that GW sources merging at z = 0 are predicted to have formed at a larger redshift, at which the star formation rate is higher. In particular, \citealt{MadauDickinson2014} predicts that the cosmic star formation rate montonically increases up to $z\sim2$. This could also explain why our merger rate is a factor of two higher than predicted by \citep{Riley2021MNRAS.505..663R}. Similarly, we underestimate the delay times for GW sources formed at $0.0005<Z\leq0.001$, and therefore we might underestimate the merger rate densities for such systems. In particular, this could mean that the merger rate density of the TMT with a MS-MS accretor channel (discussed in section \ref{subsubsec:gw_TMT_with_MSMS_accretors}) could be significantly higher than predicted (shown in Table \ref{tab:gwnumbers}).

\bsp	
\label{lastpage}
\end{document}